\documentclass[usenatbib]{mnras}
\usepackage[english]{babel}

\pdfoutput=1
\usepackage{amsfonts}
\usepackage{amsmath}
\usepackage{amssymb}
\usepackage{xcolor}
\usepackage{enumitem}
\usepackage{graphicx}
\usepackage{gensymb}
\usepackage{fancyhdr}
\usepackage{times}
\usepackage{widetext}
\usepackage{ulem}
\usepackage{caption}
\usepackage{subcaption} 
\usepackage{float} 

\graphicspath{{./Figures/}}

\begin{document}

\title{KiDS-450: Enhancing cosmic shear with clipping transformations}

\author[Giblin et al.]
{Benjamin Giblin$^1$\thanks{bengib@roe.ac.uk}, Catherine Heymans$^1$, Joachim Harnois-D{\'e}raps$^1$, Fergus Simpson$^2$, \newauthor J{\"o}rg P. Dietrich$^{3,4}$, Ludovic Van Waerbeke$^5$, Alexandra Amon$^1$, Marika Asgari$^1$,\newauthor Thomas Erben$^6$, Hendrik Hildebrandt$^6$, Benjamin Joachimi$^7$, Konrad Kuijken$^8$,\newauthor Nicolas Martinet$^6$, Peter Schneider$^6$ and Tilman Tr{\"o}ster$^1$   \\
$^1$ Scottish Universities Physics Alliance, Institute for Astronomy, University of Edinburgh, Blackford Hill, Edinburgh, EH9 3HJ, UK \\ $^2$ Instituto de Ciencias del Cosmos, University of Barcelona, UB-IEEC, Marti i Franques 1, E08028, Barcelona, Spain \\ $^3$ Faculty of Physics, Ludwig-Maximilians-Universit\"at, Scheinerstr.\ 1, 81679 M\"unchen, Germany \\ $^4$ Excellence Cluster Universe, Boltzmannstr.\ 2, 85748 Garching
b. M\"unchen, Germany  \\ $^5$ Department of Physics and Astronomy, University of British Columbia, BC V6T 1Z1, Canada \\ $^6$ Argelander-Institut f{\"u}r Astronomie, Universit{\"a}t Bonn, Auf dem H{\"u}gel 71, D-53121 Bonn, Germany \\ $^7$ Department of Physics and Astronomy, University College London, Gower Street, London WC1E 6BT, UK \\ $^8$ Leiden Observatory, Leiden University, P.O.Box 9513, 2300RA Leiden, The Netherlands \\ }

\newcommand{\Master}{\textit{Master{\_}CorrFun}}

\maketitle
\begin{abstract}
We present the first ``clipped" cosmic shear measurement using data from the Kilo-Degree Survey (KiDS-450).  ``Clipping" transformations suppress the signal from the highest density, non-linear regions of cosmological fields. We demonstrate that these transformations improve constraints on $S_8=\sigma_8(\Omega_{\rm{m}}/0.3)^{0.5}$ when used in combination with conventional two-point statistics. For the KiDS-450 data, we find that the combined measurements improve the constraints on $S_8$ by 17\%, compared to shear correlation functions alone. We determine the expectation value of the clipped shear correlation function using a suite of numerical simulations, and develop methodology to mitigate the impact of masking and shot noise.  Future improvements in numerical simulations and mass reconstruction methodology will permit the precise calibration of clipped cosmic shear statistics such that clipping can become a standard tool in weak lensing analyses.  
\end{abstract} 
\begin{keywords}
   Gravitational lensing: weak -- Cosmology: observations -- Cosmology: cosmological parameters -- Surveys
\end{keywords}


\section{Introduction}\label{sec:introduction}

The use of two-point statistics in extracting information from cosmological fields has been eminently successful to date. Observations of the CMB temperature and polarisation power spectra \citep{Planck_2018}, weak lensing shear-shear correlation functions \citep{Hildebrandtetal_2016,DES_Cosmic_Shear_Yr1}
and shear-shear/convergence power spectra \citep{Kohlinger_etal_2017, vanUitert_etal_2017}, for example, have placed meaningful constraints on the cosmological model, helping forge our current understanding of the Universe. However, some degree of tension has emerged between state-of-the-art results from the weak lensing and CMB cosmological probes. Constraints from the Kilo Degree Survey \citep[KiDS;][]{Hildebrandtetal_2016} and the Canada France Hawaii Telescope Lensing Survey \citep[CFHTLenS;][]{Heymans_etal_2013}, whilst consistent with each other are in some tension with those of the Planck Collaboration \citep{Planck_2018}. The Year 1 cosmology results from the Dark Energy Survey \citep{DES_Cosmic_Shear_Yr1, DES_Clustering_Cosmic_Shear_Yr1} ``bridge the gap" between the aforementioned studies, being broadly in agreement with all, as is also the case with the Nine-Year Wilkinson Microwave Anisotropy Probe \citep[WMAP9;][]{Hinshaw_etal_2013}. On the other hand, the cosmic shear measurements from the Deep Lens Survey \citep[DLS;][]{Yoon_etal_2018} are fully consistent with Planck and are in some tension with KiDS and CFHTLenS. The range of results on this subject highlights the necessity for more precise and accurate cosmological parameter constraints, thereby affirming whether or not the existing tension is a signature of an exotic form of dark energy or new physics within our Universe \citep[see for example][]{Joudaki_etal_2016}. It is with regards to this necessity that we review our employment of two-point statistics for cosmology.

\medskip

When considering alternatives to two-point statistics, the computational- and time-intensiveness of collecting and reducing observations in the era of precision cosmology must also be considered. Two-point statistics alone fail to exploit the full wealth of information within these expensive datasets, on account of the presence of regions of non-linear gravitational collapse. Consequently, it is crucial that we employ all possible statistical tools to capitalise on the available datasets. 

\medskip

Indeed, the sub-optimality of two-point statistics has driven research involving non-Gaussian statistics. Counting the abundance of convergence peaks, known as ``peak statistics" \citep{Jain_LvW_2000}, as well as extending the cosmological analysis to third and higher order statistics \citep{Takada_Jain_2002, Bernardeau_2005, Kilbinger_Schneider_2005, Semboloni_etal_2011, Fu_etal_2014} have been shown to yield improved constraints on cosmology. In addition, one can perform transformations to enhance the linearity of the cosmological field in question, improving the capacity of two-point statistics to contrain cosmology. For example, \cite{Neyrinck_etal_2009} and \cite{Seo_et_al_2011} found various logarithmic transformations are sufficient for this purpose. 

\medskip

In particular, ``clipping" transformations have been shown to be beneficial to a number of analyses. Clipping truncates the peaks above a given threshold within a density field, thereby suppressing the contributions of high-density regions to the power spectrum. This methodology was successfully applied to galaxy number counts within numerical simulations, and found to increase the range of Fourier modes in which the power spectrum and bispectrum can be related with tree-level perturbation theory, leading to precise determination of the galaxy bias and the amplitude of matter perturbations $\sigma_8$ \citep{Simpson_etal_2011, Simpson_etal_2013}. Furthermore, \cite{Simpson_etal_2016} clip galaxy number counts from the Galaxy and Mass Assembly Survey (GAMA), to reduce the impact of non-linear processes and galaxy bias on the analysis, allowing for reliable constraints on the rate of growth of structure in the Universe. \citet{Wilson_2016} employed clipping in estimating the growth rate of structure from the VIMOS Public Extragalactic Redshift Survey as part of a redshift-space distortion analysis. \cite{Lombriser_etal_2015} also demonstrate that clipping density fields allows for modified gravity models to more easily be distinguished from concordance cosmology.

\medskip

Clipping can also be combined with standard cosmological statistics, as demonstrated by \citet[][henceforth `S15']{Simpson_et_al_2015} in a weak lensing analysis. They truncate the peaks in simulated fields of the projected surface density, i.e. the convergence, and measure the effect on the convergence power spectrum. The objective of clipping in this context is to reduce the correlations between the Fourier modes in the convergence power spectrum in order to unlock previously inaccessible cosmological information. An alternative interpretation of the information gain in clipping, is that it is analogous to that which is found in peak statistics analyses, since both methods selectively target high-density regions. Via a Fisher matrix analysis, S15 predict the constraints on the amplitude of matter perturbations, $\sigma_8$, and the matter density parameter, $\Omega_{\rm{m}}$, one would obtain from the ``clipped" and the conventional ``unclipped" convergence power spectra. They find that clipping engenders a small clockwise rotation of the clipped contours relative to the unclipped, breaking the degeneracy in the $\Omega_{\rm{m}}$-$\sigma_8$ parameter space (see Figure 2 of S15). The consequence of this is that when the contours from the two power spectra are combined (taking into account the cross-covariance of the clipped and unclipped statistics, so as to avoid double-counting) the constraints on $\Omega_{\rm{m}}$ and $\sigma_8$ are increased overall by more than a factor of three. Moreover, clipping is found to be more constraining than the alternative logarithmic transforms proposed by \cite{Neyrinck_etal_2009}.

\medskip

A crucial aspect of clipping convergence fields containing regions of non-linear gravitational collapse, is the fact that there currently exists no analytical prescription for the clipped statistics one will subsequently measure. This means that numerical simulations are necessary for establishing their cosmological dependence. This is not a disadvantage specific to clipping, given that peak statistics \citep{Jain_LvW_2000, Kacprzak_etal_2016, Martinet_etal_2018} and higher order statistics \citep{Takada_Jain_2002,Semboloni_etal_2011}, similarly necessitate simulations for calibration. What is more, simulations are also required for investigating the behaviour of standard cosmological statistics on non-linear scales \citep{Smith_etal_2003, Takahashi_etal_2012}.

\medskip

In this work we apply clipping to weak lensing convergence fields measured from the first 450 square degrees of $r$-band data from the Kilo-Degree Survey (hereafter `KiDS-450'). In contrast to S15, rather than determine the effect of clipping on the convergence power spectrum, we investigate for the first time the properties of the clipped two-point shear correlation functions. This is to facilitate a direct comparison of the clipped statistics to the conventional shear correlation functions used in constraining cosmology in the \citet{Hildebrandtetal_2016} analysis. By exploring the cosmological dependence of clipping with the \citet[][hereafter `DH10']{Dietrich_Hartlap_2010} simulations, and by measuring the covariance of these new statistics using the Scinet Light Cone Simulations (SLICS) from \citet{Harnois_etal_2018}, we constrain the cosmology of the KiDS-450 data. We also characterise how clipping is affected by masking and shape noise, and demonstrate how these can be accounted for. The format of this paper is as follows; in Section \ref{sec:KiDS_Sims} we discuss the KiDS-450 data and the $N$-body simulations at our disposal, in Section \ref{sec:method} we explain our methodology for measuring the clipped shear correlation functions and discuss calibration corrections, in Section \ref{sec:results} we present our results, and finally we conclude in Section \ref{sec:conclusions}.


\section{Data and Simulations} \label{sec:KiDS_Sims}

\begin{table*}
  \centering
    \caption{A comparison of the specifications of the SLICS and DH10 suites used in this paper. These simulations are used for estimation of the covariance, and the dependence on cosmological parameters, of the clipped shear correlation functions, $\xi_\pm^{\rm{clip}}$, respectively.} \label{tab:Mocks_Table}
  \begin{tabular}{lcr}
      &     SLICS    & DH10 \\
  \hline
   Science case & Covariance Matrices & Cosmological Dependence \\
  Cosmologies & 1 & 158 \\
  Realisations per cosmology & 932 & 35(Fiducial)$+$1(Other)\\
  Lightcone area [deg$^2$] & 100 & 36 \\
  Box size [Mpc$/h$]$^3$ & $505^3$ & $140^3$ \\  
  Particles & $1536^3$ & $256^3$\\
  Particle Mass [$\rm{M}_\odot$] & $4.17 \times 10^9$ & $9.3\times 10^9$--$8.2\times 10^{10}$\\
  \hline
  \end{tabular}
\end{table*}

\medskip

The Kilo Degree Survey (KiDS) is an ESO public survey which will span 1350 square degrees upon completion. KiDS observes with the VLT Survey Telescope (VST) in the $ugri$ bands, with science goals pertaining to cosmology and galaxy evolution. In this paper we focus on the KiDS-450 data release, containing the first 450 square degrees of four-band coverage \citep[][hereafter `H17']{Hildebrandtetal_2016}. The KiDS-450 data is divided between five patches, G9, G12, G15, G23 and GS \citep{deJong_etal_2017} and consists of \textit{lens}fit \citep{Miller_et_al_2013} shear estimates for $\sim$15 million galaxies. The effective number of galaxies per square arcminute in the data is 8.53 and the galaxy ellipticities have a dispersion of $\sigma_e=0.29$ per component. The photometric redshifts of the background galaxies are estimated from the four-band photometry using the Bayesian photometric redshift BPZ code from \cite{Benitez_2000}, as described in \cite{Hildebrandt_et_al_2012}. In addition, three different techniques for calibrating the effective redshift distribution $n(z)$ are investigated in H17 and found to produce consistent cosmic shear results. In constraining the KiDS-450 cosmology in this analysis, we adopt the method favoured in H17 -- the weighted direct calibration (``DIR"). This follows the methodology of \citet{Lima_etal_2008}, where a subsample of galaxies with spectroscopic redshifts are reweighted such that the photometric observables (e.g. colours, magnitudes) of the reweighted sample match the larger sample of galaxies with photometric redshifts only. The reweighted spectroscopic redshift distribution is then taken to be representative of the whole sample. We refer the reader to \citet{Kuijken_etal_2015} for more technical discussion of the survey.

\medskip

The shapes of galaxies in KiDS-450, characterised by two ellipticity components, are measured with the \textit{lens}fit algorithm \citep{Miller_et_al_2013} from the $r$-band data, as described in \cite{Fenech_Conti_et_al_2017}. \textit{Lens}fit models the point spread function (PSF) at the pixel level for individual exposures, and then measures the ellipticity components by fitting a PSF-convolved disc and bulge model to each galaxy via a likelihood-based method. Weights for the shape measurement are then derived from the likelihood surface. We calibrate the shape measurements with the additive and multiplicative corrections detailed in Appendix D of H17. The former correction is determined empirically by averaging the observed ellipticities in the data, whereas the latter is quantified with image simulations resembling the KiDS-450 $r$-band.

\medskip

\begin{figure}
\begin{center}
\includegraphics[scale=0.4]{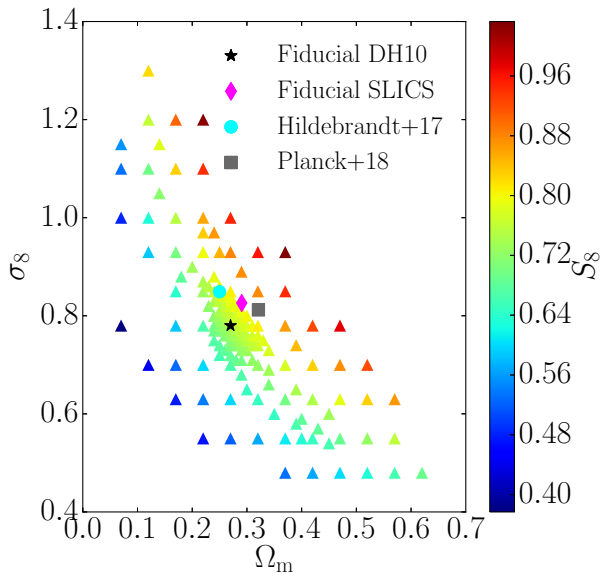}
\caption{The 158 cosmologies of the DH10 simulations in the $\Omega_{\rm{m}}$-$\sigma_8$ plane (triangles), colour coded by $S_8=\sigma_8(\Omega_{\rm{m}}/0.3)^{0.5}$. The fiducial cosmologies of DH10 and SLICS are shown by the black star and magenta diamond, respectively. The cyan circle and grey square designate the best-fit $(\Omega_{\rm{m}},\sigma_8)$ determined from the KiDS-450 data in H17, and from the TT$+$lowE analysis of the Planck data in \citet{Planck_2018}, respectively. } \label{fig:DH10_Cosmologies}
\end{center}
\end{figure}

\medskip

The absence of an analytical prescription for clipped statistics means that in order to use clipping to constrain cosmological parameters, we require a suite of numerical simulations for various cosmologies to determine how clipping responds to changes in said parameters. In addition, this task requires that the covariance of our clipped statistic is accurately measured, which necessitates a large number of independent realisations for a given cosmology. These requirements are at odds with one another; given the computational expense, simulators typically must choose between producing simulations for a large range of cosmological configurations, or producing many realisations for a single cosmology. Therefore we are compelled to use two different simulation suites to satisfy these two criteria -- DH10 and SLICS.

\medskip  

The DH10 suite \citep{Dietrich_Hartlap_2010} consists of numerical $N$-body simulations ran with the TREEPM code GADGET-2 \citep{Springel2005} and initial conditions generated with the \cite{Eisenstein_Hu_1998} transfer function. There are 192 DH10 simulations spanning 158 different flat $\Lambda$CDM cosmologies. Each simulation has 256$^3$ dark matter particles in a box with sides of length 140 $h^{-1}$Mpc, evolved from $z=50$ to $z=0$. The lightcone area per simulation is 6$\times$6 square-degrees, and the particle mass varies from $\rm{m}_{\rm{p}}=9.3\times10^9\rm{M}_{\odot}$ for $\Omega_{\rm{m}}=0.07$, to $\rm{m}_{\rm{p}}=8.2\times10^{10}\rm{M}_{\odot}$ for $\Omega_{\rm{m}}=0.62$. 35 of the simulations have the fiducial cosmological parameters given by $\boldsymbol{\pi_0} = (\Omega_{\rm{m}}=0.27, \Omega_\Lambda = 0.73, \Omega_{\rm{b}}=0.04, \sigma_8=0.78, n_{\rm{s}}=1.0, h=0.7)$. The remaining 157 cosmologies, each of which comprise a single $N$-body simulation, differ only in $\Omega_{\rm{m}}$ and $\sigma_8$, the range of which is displayed in Figure \ref{fig:DH10_Cosmologies}. Hence, in this work we only demonstrate the power of clipping in constraining $S_8=\sigma_8(\Omega_{\rm{m}}/0.3)^{0.5}$, which probes the $\Omega_{\rm{m}}$-$\sigma_8$ parameter space in the direction approximately perpendicular to the degeneracy between these parameters, for a flat $\Lambda$CDM Universe. These constraints are obtained with the other cosmological parameters fixed to their fiducial values.

\medskip

Catalogues of the noise-free shear components for galaxies are produced by ray-tracing through each DH10 $N$-body simulation. This consists of propagating light rays through the matter distribution constructed by the $N$-body simulation, from galaxies with a given distribution in redshift. The matter distribution exists in the form of mass snapshots at various redshifts; the deflection of light rays by these mass planes determines the shear of the mock galaxies. Five pseudo-independent shear catalogues are obtained for a given simulation by ray-tracing through five different random angles. Thus, in this work we are using $35\times5$ shear catalogues for the fiducial cosmological parameters, and $1\times5$ shear catalogues for the remaining 157 cosmologies.

\medskip

In order to measure the covariance of clipped statistics, we employ the public\footnote{SLICS $N$-body simulations; \url{http://slics.roe.ac.uk}} Scinet Light Cone Simulations (SLICS) of \citet{Harnois_etal_2018}. The SLICS suite evolved 1536$^3$ particles of mass $m_{\rm{p}}=4.17 \times10^9\rm{M}_{\odot}$, from $z=120$ to $z=0$ in a box with sides of length 505 $h^{-1}$Mpc. They were created using the CUBEP$^3$M $N$-body code \citep{Harnois_etal_2013}, with initial conditions selected from the Zel'dovich displacement of particles based on a transfer function from CAMB \citep{Lewis_etal_2000}. The SLICS consist of just three cosmologies and are therefore unable to determine the cosmological dependence of clipping. However, on account of there being 932 realisations of 100 deg$^2$ light cones for the fiducial cosmology ($\Omega_{\rm{m}}=0.2905 $, $\Omega_{\Lambda}=0.7095$, $\Omega_{b}=0.0473$, $h=0.6898$, $\sigma_8=0.826$ and $n_{\rm{s}}=0.969$), SLICS are very well suited to covariance estimation. In this work we use only the SLICS with the fiducial cosmology, and assume that the covariance measured from these realisations is robust to changes in cosmology. This is a commonly made approximation, as neglecting the cosmological dependence of the covariance has been shown to have little effect on the best-fit value of $S_8$ if the fiducial cosmology is sufficiently close to that of the best-fit \citep{Eifler_etal_2009}. In our case, the SLICS cosmological parameters are close to the best-fit from the H17 analysis of the KiDS-450 data, the fiducial cosmology of DH10, and the best-fit from \citet{Planck_2018}, as is shown in Figure \ref{fig:DH10_Cosmologies}. Thus our approximation of a cosmology-independent covariance matrix is reasonable given the data we are working with. A comparison of the DH10 and SLICS specifications is presented in Table \ref{tab:Mocks_Table}. Both suites consist of dark matter particles only.

\medskip

The fact that galaxies can be intrinsically aligned through gravitational interaction, rather than have their alignments induced by weak gravitational lensing, poses a systematic bias to cosmological inference \citep{Bridle_King_2007}. In order to reduce the influence of intrinsic alignments in this work, we follow \cite{Benjamin_et_al_2013} and restrict our analysis to the 0.5--0.9 photometric redshift range in the KiDS-450 data. Within this tomographic interval, the density of source galaxies is 3.32 gal/arcmin$^2$ and the galaxy ellipticities have a dispersion of $\sigma_e=0.28$ per component. We downsample the SLICS and DH10 mock catalogues so as to have the same source density and redshift distribution of the data, which we take to be the KiDS-450 DIR-calibrated redshift distribution (H17), which has mean and standard deviation of 0.76 and 0.29, respectively, in our chosen redshift bin. We also introduce Gaussian-distributed galaxy ellipticities to the mocks, with standard deviation, $\sigma_e$, equal to that of the KiDS-450 data. We do not truncate the Gaussian distribution to ellipticites between -1 and 1, since less than 0.05\% of mock galaxies are allocated ellipticities outside of this range, and their contributions to the correlation functions are negligible. We also verified that using ellipticities directly sampled from the distribution in the data, instead of from a Gaussian, does not affect our results. Matching the shape noise (which in this work we use to refer to all factors contributing to the measured galaxy shape, bar the shear itself) and source densities, means that the noise in the covariance matrices and the clipped predictions from the mocks reflect that of KiDS-450. The effect of baryonic physics on the shear correlation functions is another source of bias in weak lensing analyses \citep{Semboloni_etal_2011b}, and could in principle affect clipped statistics differently than the unclipped. For this first proof-of-concept analysis however, we do not contend with baryonic effects in this work.


\section{Methodology} \label{sec:method}

In this Section, we describe the pipeline in which we apply clipping transformations to the mocks and KiDS-450 data, and subsequently measure the ``clipped" two-point shear correlation functions $\xi_\pm^{\rm{clip}}$. Measuring these statistics allows for a comparison to the conventional ``unclipped" shear correlation functions, which are directly calculated from the observed galaxy ellipticities in the data. We begin with a very brief summary of the key steps in our method for easy referral. We discuss these steps in greater detail in the Sections that follow.

\medskip

\begin{itemize}

\item Our pipeline takes as input catalogues of the ellipticities and positions of galaxies. We project these onto a Cartesian grid of pixels with a resolution of 5 arcseconds, smooth these maps with a Gaussian filter and reconstruct the projected surface mass density, i.e. the convergence, $\kappa$, following \citet{KSB}.

\item We subject these convergence maps to clipping; anywhere the convergence exceeds a certain threshold value, we set the convergence equal to that threshold. 

\item The resulting ``clipped" convergence map is subtracted from the ``unclipped" thereby generating a map containing the projected surface density exceeding the threshold, and zeroes elsewhere. On this ``residual" convergence map, we invert the mass reconstruction process and recover the shear corresponding to these projected peaks.

\item This ``residual" shear is subtracted from the original shear values yielding the ``clipped" shear. From the clipped shear, we calculate the clipped shear correlation functions, $\xi_\pm^{\rm{clip}}$, using {\sc{TreeCorr}} \citep{TreeCorr}. To measure the unclipped shear correlation functions, $\xi_\pm^{\rm{unclip}}$, we feed the catalogues of the observed ellipticities to {\sc{TreeCorr}} directly.  

\item We repeat this process for successive SLICS realisations to measure the covariance of the $\xi_\pm^{\rm{clip}}$ and $\xi_\pm^{\rm{unclip}}$ statistics, and for successive DH10 realisations to determine the cosmological dependence of the $\xi_\pm^{\rm{clip}}$.

\end{itemize}

\subsection{Mass reconstruction} \label{subsec:Mass_Recon}

In order to clip the densest non-linear regions from our analysis, we first produce maps of the projected surface mass density, or convergence, $\kappa$, using the methodology of \citet[][`KS93' hereafter]{KSB}. In this analysis, the process of ``mass reconstruction" begins with the observed ellipticities, which can be written in the complex form $\epsilon^{\rm{obs}}=\epsilon_1^{\rm{obs}} + {\rm{i}}\epsilon_2^{\rm{obs}}$ \citep{Seitz_Schneider_1996}. The observed ellipticities have contributions from the reduced shear $g$, the intrinsic ellipticity $\epsilon^{\rm{int}}$ and the shape measurement noise $\eta$ via

\medskip

\begin{equation} \label{eqn:eobs}
\epsilon^{\rm{obs}} = \frac{g+\epsilon^{\rm{int}}}{1+g^{*}\epsilon^{\rm{int}} } + \eta \,,
\end{equation}

\noindent where $g^*$ is the complex conjugate of $g$. The reduced shear is related to the shear $\gamma$ and the convergence $\kappa$ by $g=\gamma/(1-\kappa)$. In a weak lensing analysis, we assume that the magnitudes of both the shear and the convergence are much smaller than unity, such that the average of the observed ellipticities $\langle \epsilon^{\rm{obs}} \rangle \simeq g \simeq \gamma$. In this case, it is possible to reconstruct the convergence from the observed ellipticities via the KS93 inversion method. We begin with the gravitational deflection potential $\Psi(\boldsymbol{\theta})$. This is related to the convergence $\kappa$ for a particular source redshift  and angular coordinate on the sky $\boldsymbol{\theta}=(\theta_1,\theta_2)$, via Poisson's equation,

\begin{equation} \label{eqn:Poisson}
\nabla^2 \Psi(\boldsymbol{\theta}) = 2\kappa(\boldsymbol{\theta})\,,
\end{equation}

\noindent where $\Psi(\boldsymbol{\theta})$ is given by the line of sight integral over the 3D matter gravitational potential $\Phi$,

\begin{equation}
\Psi(\boldsymbol{\theta}) = \int_0^{\chi_s} d\chi^\prime \frac{f_{\rm{K}}(\chi-\chi^\prime)}{f_{\rm{K}}(\chi)f_{\rm{K}}(\chi^\prime)}\Phi \left[ f_{\rm{K}}(\chi^\prime)\boldsymbol{\theta},\chi^\prime \right] .
\end{equation}

\noindent Here $\chi$ is the comoving radial distance, $\chi_s$ is the comoving radial distance to the source, and $f_{\rm{K}}(\chi)$ is the comoving angular diameter distance. The potential $\Psi(\boldsymbol{\theta})$ is related to the shear components $\gamma_i(\boldsymbol{\theta})$ via

\begin{equation} \label{eqn:eitheta}
\gamma_i(\boldsymbol{\theta}) = D_i \Psi(\boldsymbol{\theta})  \,,
\end{equation}

\noindent where

\begin{equation}
\begin{pmatrix}
D_1 \\ D_2
\end{pmatrix}
= 
\frac{1}{2}\begin{pmatrix}
\partial^2/\partial\theta_1 \partial\theta_1 - \partial^2/\partial\theta_2 \partial\theta_2 \\ 2\partial^2/\partial\theta_1 \partial\theta_2
\end{pmatrix} \,,
\end{equation}

\noindent and $\partial$ denotes partial derivatives. Combining equations \ref{eqn:Poisson} and \ref{eqn:eitheta} and taking the Fourier transform yields

\begin{equation} \label{eqn:eik}
\tilde{\gamma_i}(\boldsymbol{\ell}) = ^{•}F_i(\boldsymbol{\ell}) \tilde{\kappa}(\boldsymbol{\ell}) \,,
\end{equation}

\noindent where

\begin{equation}
\begin{pmatrix}
F_1 \\ F_2
\end{pmatrix}
 \equiv
\begin{pmatrix}
({\ell_1}^2 - {\ell_2}^2)/\ell^2 \\ 2\ell_1\ell_2/\ell^2
\end{pmatrix}\,,
\end{equation}

\noindent and $\boldsymbol{\ell}=(\ell_1,\ell_2)$ is the 2D Fourier conjugate of $\boldsymbol{\theta}$.

\medskip

From equation \ref{eqn:eik} we see that, in principle, either $\tilde{\gamma_1}(\boldsymbol{\ell})/F_1(\boldsymbol{\ell})$ or $\tilde{\gamma_2}(\boldsymbol{\ell})/F_2(\boldsymbol{\ell})$ would suffice to give an estimate of $\tilde{\kappa}(\boldsymbol{\ell})$, which can then be inverse-Fourier transformed to recover $\kappa(\boldsymbol{\theta})$. Both $F_1(\boldsymbol{\ell})$ and $F_2(\boldsymbol{\ell})$ vanish for particular directions however, so instead we sum over the $\tilde{\gamma_i}(\boldsymbol{\ell})$ components weighted by $F_i(\boldsymbol{\ell})$ to obtain the convergence,   

\begin{equation} \label{eqn:Sigma_Tilde}
\sum_{i=1}^2  F_i(\boldsymbol{\ell})\tilde{\gamma_i}(\boldsymbol{\ell}) = \sum_{i=1}^2|F_i(\boldsymbol{\ell})|^2\tilde{\kappa}(\boldsymbol{\ell}) = \tilde{\kappa}(\boldsymbol{\ell}) \,,
\end{equation} 

\noindent where we have employed the fact that $\sum_{i=1}^2|F_i(\boldsymbol{\ell})|^2$ is equal to unity \citep{Kaiser92}. An inverse-Fourier transform is performed to reconstruct the $\kappa(\boldsymbol{\theta})$ map, the real part of which contains the E-modes, whereas the imaginary part contains the B-modes\footnote{\citet{Hildebrandtetal_2016} and \citet{vanUitert_etal_2017} report significant B-modes within the KiDS-450 data but as these are at such a low-level in comparison to the E-mode signal we do not consider them in this analysis.} \citep{Schneider_etal_2002b}.

\medskip

The KS93 mass reconstruction can be summarised in the following:

\begin{itemize}
\item The shear is projected onto a Cartesian grid and smoothed with a Gaussian filter with width $\sigma_{\rm{s}}$ to reduce the impact of mask features (which removes artefacts) on the reconstruction.  
\item A border of zero values is added to the smoothed shear map, increasing the dimensions by 1 deg on each side, before Fourier transforming the field. The border serves to reduce edge effects in the transform \citep{vanWaerbeke_et_al_2013}.
\item $\tilde{\kappa}(\boldsymbol{\ell})$ is computed via equation \ref{eqn:Sigma_Tilde}.
\item An inverse-Fourier transform is performed to reconstruct the $\kappa(\boldsymbol{\theta})$ map. 
\end{itemize}

\noindent The steps we take in mass reconstruction follow this recipe. However, in this analysis we are working with real data and simulations tailored to the data in terms of the redshift distribution, source density and galaxy shape noise. Our observed ellipticities (see equation \ref{eqn:eobs}), smoothed with the Gaussian filter, are treated as an unbiased estimator for the shear and take the place of $\gamma$ in the above equations. Furthermore, the KiDS-450 data has masked regions leading to gaps in the observed patches. The Gaussian smoothing accounts for the number of masked pixels within the smoothing window, to minimise the bias in the resultant smoothed ellipticity \citep[see][for more details]{vanWaerbeke_et_al_2013}. The effect of masking on the clipped shear correlation functions $\xi_\pm^{\rm{clip}}$ is discussed in Section \ref{subsec:Mask_Correction}. We refer to the width of the Gaussian smoothing filter as the \textit{smoothing scale}, $\sigma_{\rm{s}}$, hereafter. 

\medskip
 
The KS93 methodology has been shown to be accurate for relatively small fields ($\lesssim 100$ deg$^2$) which may be approximated as flat \citep{vanWaerbeke_et_al_2013}. Other mass reconstruction methods do exist; for example \citet{Seitz_Schneider_1996} generalise the KS93 technique into the lensing regime where the $\kappa \ll 1$ approximation no longer holds, whereas \citet{Chang_etal_2017} conduct curved-sky mass reconstruction with a spherical harmonic formalism. The KS93 methodology is sufficiently accurate for our purposes however, since the KiDS-450 patches, DH10 mocks and SLICS are well described by the flat-sky approximation, and the convergence is sufficiently small (see Section \ref{subsec:Choose_SS_Kc}). Future clipping analyses, especially those involving datasets with larger sky coverage, will require these improved methodologies. Convergence maps for the KiDS-450 patches created following KS93 are presented in Appendix \ref{sec:Appendix_MassMaps}.  

\medskip

\subsection{Clipping methodology} \label{subsec:Clip_Meth}

After the convergence field is generated it is clipped if above a given threshold $\kappa^c$ according to

\begin{equation}
\kappa^{\rm{clip}}_s(\boldsymbol{\theta}) =  
\begin{cases}
\kappa^c, & \text{if } \kappa_s(\boldsymbol{\theta}) \geq \kappa^c  \\ \kappa_s(\boldsymbol{\theta}), & \text{otherwise}  
\end{cases}\,,
\end{equation}

\noindent where the `s' subscript is used to denote fields either directly smoothed with the Gaussian filter, or those derived from fields which have been directly smoothed. We calculate the ``residual" convergence $\Delta\kappa_s$, given by

\begin{equation} \label{eqn:deltakappa}
\Delta\kappa_s(\boldsymbol{\theta}) = \kappa_s(\boldsymbol{\theta}) - \kappa^{\rm{clip}}_s(\boldsymbol{\theta}).
\end{equation}

\noindent The $\Delta\kappa_s$ map features the projected surface density exceeding the threshold $\kappa^c$, and zeroes elsewhere. We subject this map to an inversion of the mass reconstruction process following equation \ref{eqn:eik}. This generates the ``residual" ellipticity maps $\Delta \epsilon_s$, which exhibit the strongest signal around the positions of the peaks, and weaker signal elsewhere. The residual ellipticities are defined on a grid; in order to obtain $\Delta \epsilon_s$ at the locations of the galaxies in the original, ``unclipped" ellipticity catalogue, $\boldsymbol{\theta}_{\rm{g}}$, we perform 2D linear interpolation from the $\Delta \epsilon_s$ maps. The clipped ellipticity ${\epsilon}^{\rm{clip}}_s$ is the difference between the observed (unclipped) ellipticity $\epsilon^{\rm{obs}}$ and the residual ellipticity $\Delta \epsilon_s$,

\begin{equation} \label{eqn:clipped_e}
{\epsilon}^{\rm{clip}}_s(\boldsymbol{\theta}_{\rm{g}}) = \epsilon^{\rm{obs}}(\boldsymbol{\theta}_{\rm{g}}) - \Delta \epsilon_s(\boldsymbol{\theta}_{\rm{g}}).
\end{equation}

\medskip

It is inadvisable to recover the clipped ellipticity, ${\epsilon}^{\rm{clip}}_s$, by conducting inverse mass reconstruction \textit{directly} on the clipped convergence map, $\kappa^{\rm{clip}}_s$. This is because $\kappa^{\rm{clip}}_s$ has been affected by smoothing in all regions where the convergence is below the clipping threshold $\kappa^c$ (those regions with convergence above $\kappa^c$ are set to the constant threshold itself), and smoothing incurs a loss of signal. This corresponds to $\sim$90\% of the area of  $\kappa^{\rm{clip}}_s$ being affected by smoothing, for the $\kappa^c$ and smoothing scale, $\sigma_{\rm{s}}$, values we identify in Section \ref{subsec:Choose_SS_Kc}. In contrast, if we invert the mass reconstruction on the $\Delta\kappa_s$, only $\sim$10\% of the area of which is smoothed, and subtract the $\Delta \epsilon_s$ from the unsmoothed observed ellipticities, $\epsilon^{\rm{obs}}$, we minimise the impact of smoothing on our overall signal.

\medskip

After computing the clipped ellipticity components via equation \ref{eqn:clipped_e}, using {\sc{TreeCorr}} \citep{TreeCorr} we calculate estimators for the clipped and unclipped angular shear correlation functions in nine logarithmically spaced angular bins, $\theta$, with bin centres from 0.78 to 219 arcmin. We define these estimators, within a single tomographic bin, accordingly 

\begin{equation} \label{eqn:xi+-}
\widehat{\xi}_{\pm}(\theta) = \frac{\sum_{\rm{ab}} w_{\rm{a}} w_{\rm{b}} \left[ \epsilon_\text{t} (\boldsymbol{\theta}_{\rm{g},\rm{a}}) \epsilon_\text{t} (\boldsymbol{\theta}_{\rm{g},\rm{b}}) \, \pm \, \epsilon_\times (\boldsymbol{\theta}_{\rm{g},\rm{a}}) \epsilon_\times (\boldsymbol{\theta}_{\rm{g},\rm{b}})
\right]}{
\sum_{\rm{ab}} w_{\rm{a}} w_{\rm{b}} } \,,
\end{equation}

\noindent where the summation is over pairs of galaxies $\rm{a}$ and $\rm{b}$ positioned at angular coordinates $\boldsymbol{\theta}_{\rm{g},\rm{a}/\rm{b}}$, within an interval $\Delta \theta$ about the angular separation $\theta$ \citep{Bartelmann_Schneider_2001}. The $\epsilon_\text{t}$ and $\epsilon_\times$ terms designate the tangential- and cross- components of the clipped ellipticities (in the case of the $\widehat{\xi}_{\pm}^{\rm{clip}}$ estimator) or the observed ellipticities (in the case of the unclipped estimator $\widehat{\xi}_{\pm}^{\rm{unclip}}$) measured relative to the vector $\boldsymbol{\theta}_{\rm{g},\rm{a}} - \boldsymbol{\theta}_{\rm{g},\rm{b}}$ connecting the galaxy pairs. $w$ is the weight ascribed to the measurement of the ellipticity components, which comes from the \textit{lens}fit algorithm in the case of KiDS-450 (refer to Section \ref{sec:KiDS_Sims} for more details) or takes the value of unity in the case of the mocks. We treat the observed ellipticities, a combination of the shear and shape noise via equation \ref{eqn:eobs}, in the mocks and data as unbiased estimators for the shear. Accordingly we treat $\widehat{\xi}_\pm^{\rm{unclip}}$ as an unbiased estimator of the theoretical unclipped shear correlation functions, $\xi_\pm^{\rm{unclip}}$, defined in equation \ref{eqn:xi+-theory}. Consequently, in this work we follow H17 and refer to the \textit{estimators} for the unclipped shear correlation functions simply as the unclipped shear correlation functions, and omit the $\widehat{}$ notation. There is currently no established theoretical prediction for $\xi_\pm^{\rm{clip}}$. Thus it is not meaningful to include the $\widehat{}$ notation nor ``estimator" prefix for our measured clipped statistics, and we similarly drop this nomenclature henceforth. However, we encourage the reader not to regard the clipped statistics measured from the mocks as unbiased estimators of the clipped measurement made in the absence of shape noise (as we do with the unclipped statistic). The clipped statistics we measure not only depend on the level of shape noise, but also the clipping threshold and level of smoothing applied in the analysis (see Section \ref{subsec:Choose_SS_Kc}).

\medskip

The theoretical unclipped shear correlation functions $\xi_\pm^{\rm{unclip}}$  are related to the convergence power spectrum $P_\kappa(\ell)$ via

\begin{equation} \label{eqn:xi+-theory}
\xi_\pm^{\rm{unclip}}(\theta) = \frac{1}{2\pi}\int \text{d}\ell \, \ell \,P_\kappa(\ell) \, J_{0,4}(\ell \theta) \, , 
\end{equation}

\medskip

\noindent where the zeroth $J_{0}(\ell \theta)$ and fourth $J_{4}(\ell \theta)$ order Bessel functions of the first kind are used for $\xi_+^{\rm{unclip}}$ and $\xi_-^{\rm{unclip}}$ respectively. The convergence power spectrum $P_\kappa(\ell)$ is in turn related to the matter power spectrum $P_\delta(\ell)$ via 

\medskip

\begin{equation} \label{eqn:Pkappa}
P_\kappa(\ell) = \int_0^{\chi_{\rm H}} \text{d} \chi \, \frac{q(\chi)^2}{f_{\rm{K}}(\chi)^2} \, P_\delta \left( k=\frac{[\ell+1/2]}{f_{\rm{K}}(\chi)},\chi \right),
\end{equation}

\medskip

\noindent where $\chi_{\rm{H}}$ is the comoving radial distance to the horizon and $k$ is the Fourier conjugate of $\chi$. Here we have used the flat-sky first-order extended Limber approximation, which is sufficiently accurate for the KiDS-450 data \citep[see][]{Kilbinger_etal_2018}. The lensing efficiency, $q(\chi)$, is defined as

\medskip

\begin{equation} \label{eqn:lensing_efficiency}
q(\chi) = \frac{3 H_0^2 \Omega_{\rm m}}{2c^2} \frac{f_{\rm{K}}(\chi)}{a(\chi)}\int_\chi^{\chi_{\rm H}}\, \text{d} \chi^\prime\ n(\chi^\prime) 
\frac{f_{\rm{K}}(\chi^\prime-\chi)}{f_{\rm{K}}(\chi^\prime)}\,,
\end{equation}

\noindent where $a$ is the scale factor, $n(\chi)$ is the probability density of galaxies as a function of $\chi$, $H_0$ is the Hubble constant and $c$ is the speed of light.

\medskip

Constraining the cosmology of the KiDS-450 data requires covariance matrices for the clipped and unclipped $\xi_\pm$. We measure the covariance of these statistics across $N \sim$900 independent SLICS realisations. The $i^{\text{th}}$ and $j^{\text{th}}$ elements of the covariance matrices are given by

\medskip

\begin{equation} \label{eqn:CovMat}
C_\pm(\theta_i, \theta_j) = \sum_k^N  \frac{({\xi_\pm^k(\theta_i)} - \overline{\xi_\pm}(\theta_i))({\xi_\pm^k(\theta_j)} -  \overline{\xi_\pm}(\theta_j))}{N-1} \,,
\end{equation}

\noindent where $\overline{\xi_\pm}(\theta_i)$ refers to either the mean clipped or mean unclipped $\xi_\pm$, across $N$ realisations each numerated by $k$, within the $i^{\text{th}}$ angular separation bin, given by $\sum_k^N \xi_\pm^k(\theta_i) / N$. When computing the auto-covariance of the clipped (or unclipped) statistic, all correlation functions in equation \ref{eqn:CovMat} correspond to $\xi_\pm^{\rm{clip}}$ (or $\xi_\pm^{\rm{unclip}}$). When computing the cross-covariance between the clipped and unclipped, the $\xi_\pm$ correspond to clipped in one bracket, and to unclipped in the other. In order to constrain the cosmology of KiDS-450, we scale the covariance matrices measured from SLICS by the ratio of the areas of SLICS and KiDS-450 \citep{Schneider_et_al_2002}. We note that this is an approximation and does not account for the survey geometry, as is discussed in \citet{Troxel_etal_2018}. Correlation coefficient matrices, calculated from the SLICS covariance matrices, are present in Appendix \ref{sec:Appendix_Covariance}.

\subsection{Choosing the clipping threshold and smoothing scale} \label{subsec:Choose_SS_Kc}

In a clipping analysis, the values of the convergence threshold, $\kappa^c$, at which peaks are truncated and the width of the Gaussian with which the ellipticity maps are smoothed, i.e. the smoothing scale $\sigma_{\rm{s}}$, are free parameters. Thus an important aspect of clipping is to identify values which are appropriate for the data one wishes to analyse. Suitable choices of these parameters depend on the depth and resolution of the data. These parameters are also degenerate with one another; for a given value of $\kappa^c$, a lower level of smoothing results in more of the convergence field exceeding the clipping threshold. Similarly, for a fixed $\sigma_{\rm{s}}$, lesser values of $\kappa^c$ correspond to more aggressive clipping. The interplay of these parameters means that the optimal values for constraining cosmology are costly to determine. Consequently, in this work we only determine values which are well suited to the KiDS-450 data. We also investigate the effect of different choices of the smoothing scale and clipping threshold on the clipped correlation functions.     

\medskip

\begin{figure}
\begin{center}
\includegraphics[width=0.48\textwidth]{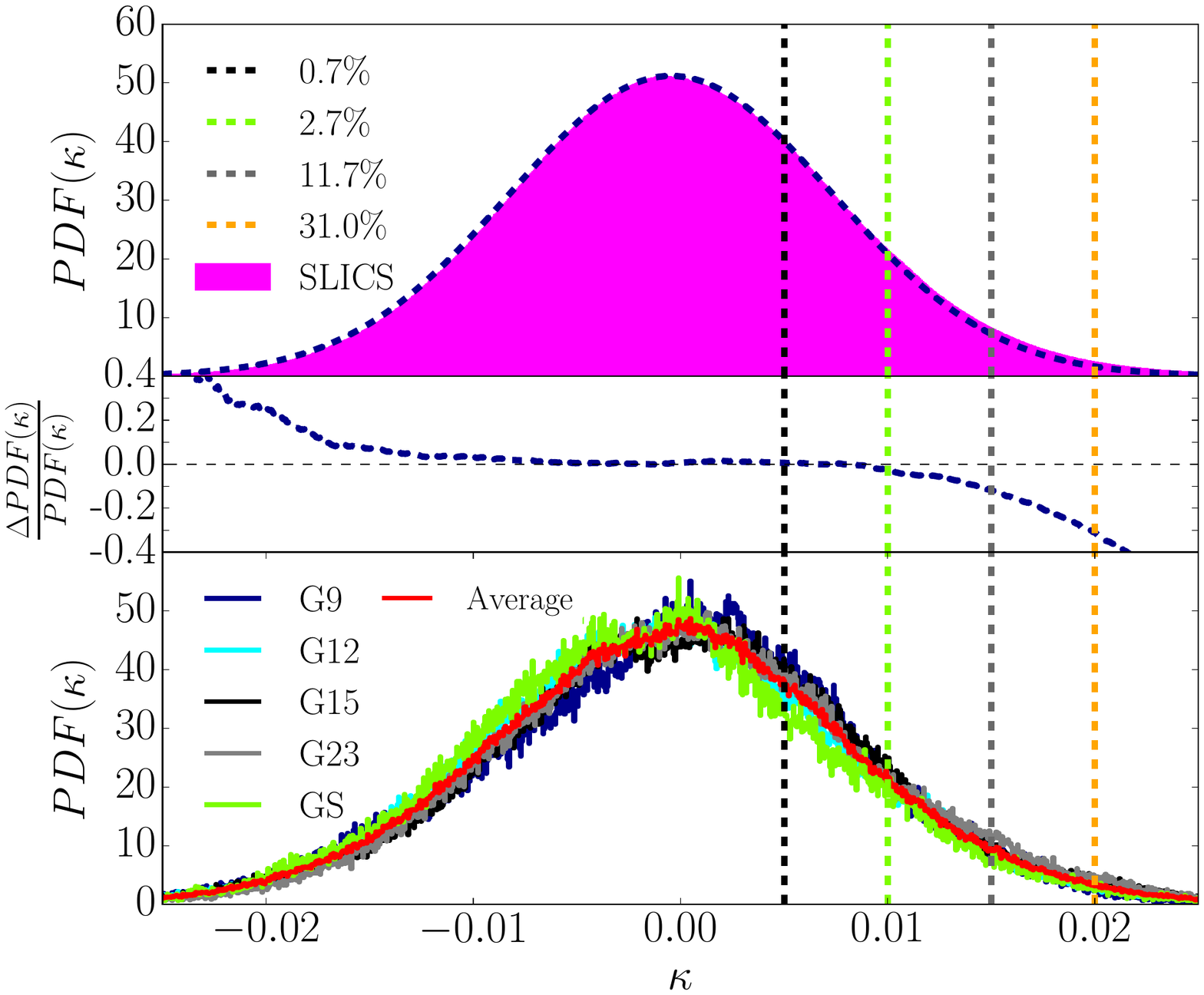} 
\caption{\textit{Upper:} PDF of the convergence $\kappa$ from 50 SLICS realisations in magenta and a Gaussian fit in dashed blue. The percentage deviations between the Gaussian fit and the PDF($\kappa$) at $\kappa=(0.005,0.010,0.015,0.020)$, shown by the dashed lines, are detailed in the legend. \textit{Middle:} the fractional difference between the Gaussian fit and the SLICS PDF($\kappa$). \textit{Lower:} The PDFs of the five KiDS-450 patches and their average.} \label{fig:Kappa_PDFs}
\end{center}
\end{figure}

\medskip

We first establish a clipping threshold which targets the most non-linear regions of the field, without over-clipping the linear field. An intuitive way of doing this is to first fix the smoothing scale and determine where the PDF of the convergence deviates from Gaussian. However, we find that even for relatively large values of the smoothing scale, the KiDS-450 PDF$(\kappa)$ is too noisy for this test. We therefore use the SLICS, the fiducial $\Omega_{\rm{m}}$ and $\sigma_8$ of which are similar to the best-fit values from the H17 analysis of the KiDS-data (see Figure \ref{fig:DH10_Cosmologies}). In Figure \ref{fig:Kappa_PDFs} we compare the PDF$(\kappa)$ measured from 50 SLICS with a smoothing scale of 6.6 arcmin (upper panel), to those from the five KiDS-450 patches (lower panel). We overplot vertical dashed lines at $\kappa=0.005, 0.010, 0.015$ and $0.020$ and detail the deviations between a Gaussian fit and the SLICS PDF$(\kappa)$ at these convergence values in the legend. The middle panel shows the fractional difference between the Gaussian fit and the SLICS PDF$(\kappa)$. We find that in the range $-0.005 \leq \kappa \leq 0.005$, the PDF of the SLICS convergence is well described by the Gaussian, but deviations of a few percent arise at $\kappa \gtrsim 0.010$. At the high-end tail of the convergence, the SLICS PDF is considerably non-Gaussian, differing by $\gtrsim 30\%$. This suggests that a clipping threshold $\kappa^c \gtrsim 0.010$ is appropriate for isolating non-linear features of the field.

\medskip

\begin{figure}
\begin{center}
\includegraphics[width=0.5\textwidth]{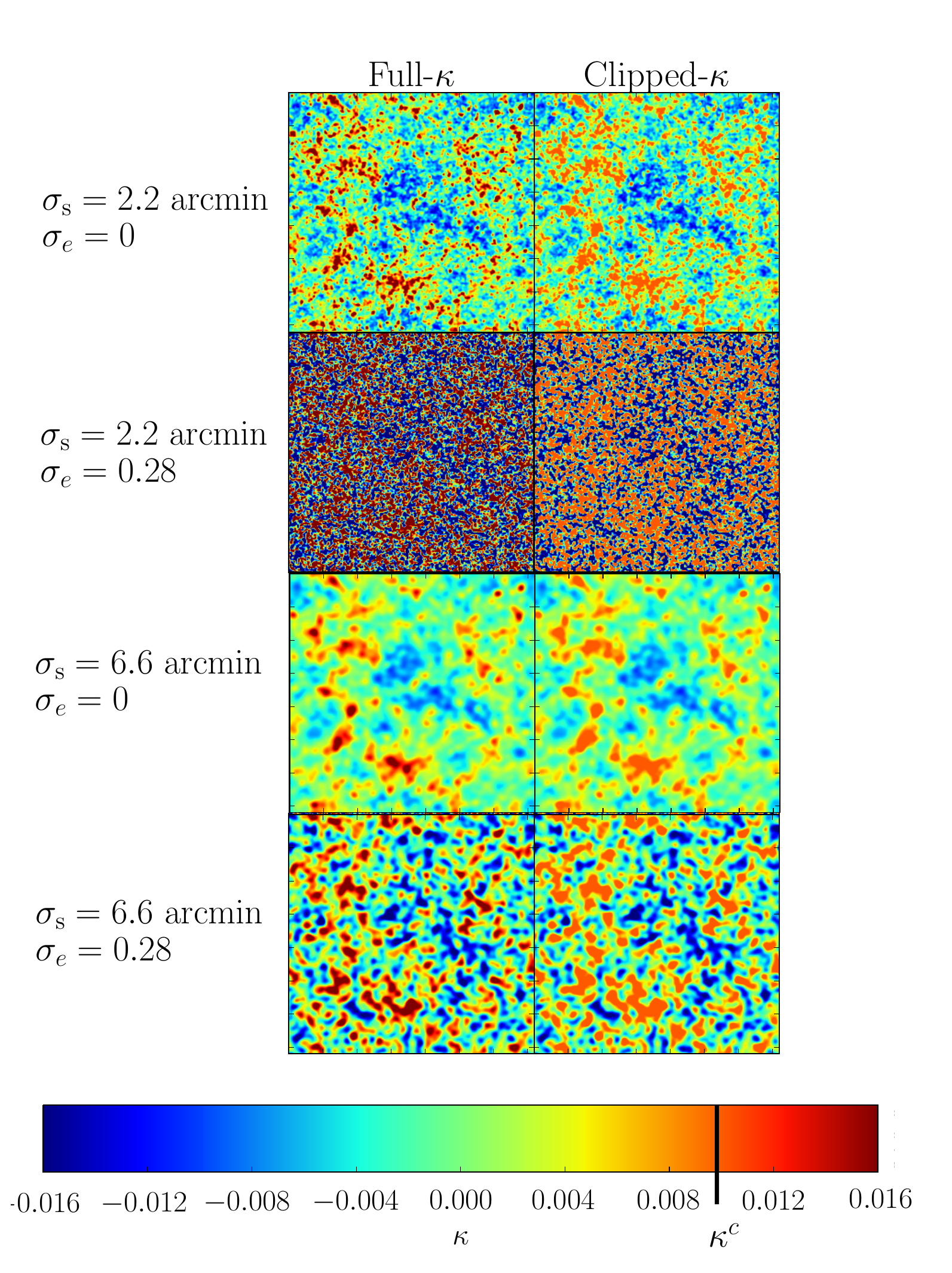}
\caption{Unclipped (left hand panels) and clipped (right hand panels; $\kappa^c = 0.010$) convergence maps for a single 100 deg$^2$ SLICS. For the upper two panels, the smoothing scale, $\sigma_{\rm{s}}$, is equal to 2.2 arcmin. Comparison of these panels shows that the features in both the clipped and unclipped convergence maps for a noise-free field ($\sigma_e=0$) change dramatically with the inclusion of KiDS-450 level shape noise (Gaussian distributed with width $\sigma_e=0.28$). The lower two panels however have $\sigma_{\rm{s}}=6.6$ arcmin. Comparison of these panels shows that the clipped/unclipped maps change less dramatically with the inclusion of shape noise if the smoothing scale is set to the higher level. This suggests that using $\sigma_{\rm{s}}=2.2$ arcmin results in the clipping of mainly pure noise features, and that $\sigma_{\rm{s}}=6.6$ arcmin is a more appropriate level of smoothing for clipping the KiDS-450 cosmological signal.} \label{fig:MassMapsMocks}
\end{center}
\end{figure}

\medskip

\begin{figure*}
\begin{center}
\includegraphics[width=\textwidth]{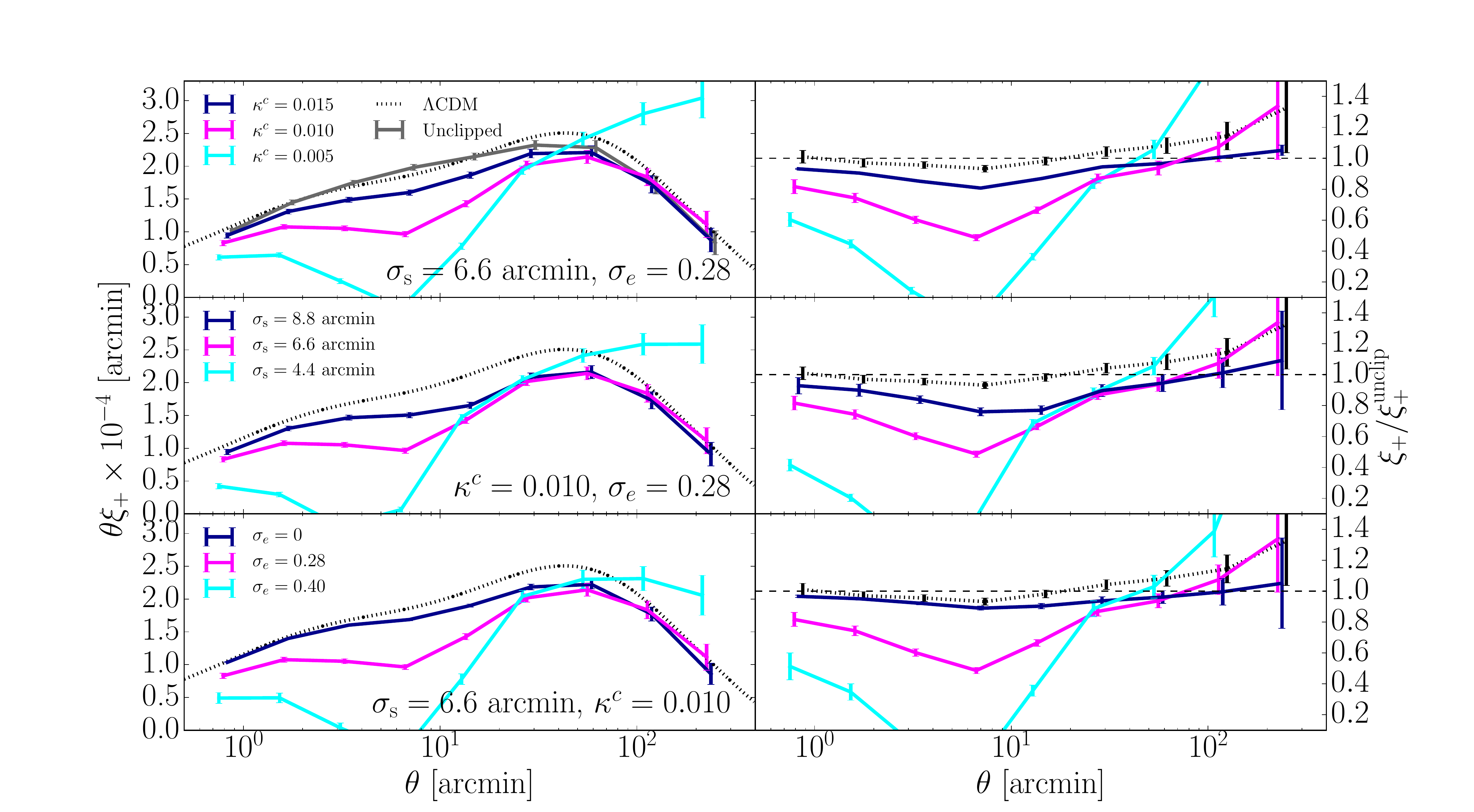}
\caption{The mean unclipped (solid grey) and clipped (other solid colours) $\xi_+$ correlation functions measured from the SLICS realisations. The dashed black line is the theoretical unclipped prediction from equation \ref{eqn:xi+-theory}. The left hand panels display $\theta \xi_+$, the right hand the measurements normalised to the unclipped statistic from SLICS. The annotation in the lower right hand corner of each panel specifies which of the parameters are held constant in the calculations. The upper panel is concerned with variations in the clipping threshold, $\kappa^c$, with fixed smoothing scale, $\sigma_{\rm{s}}$, and shape noise characteristics, $\sigma_e$.
The middle and lower panels present variations in the smoothing scale and shape noise respectively. The magenta line in all cases depicts the measurement for the fiducial parameters: $\kappa^c=0.010, \sigma_{\rm{s}}=6.6$ arcmin and $\sigma_e=0.28$. The error bars are the error on the mean measurement.} \label{fig:BiggerPlot}
\end{center}
\end{figure*}

\medskip

In setting the value of $\sigma_{\rm{s}}$, one should aim to reduce the prominency of peaks caused solely by noise fluctuations, but not to the extent that we lose a significant amount of the cosmological information. A comparison of the SLICS convergence maps when clipped at different smoothing scales, with and without intrinsic galaxy shape noise, serves as a useful visual indicator of whether $\sigma_{\rm{s}}$ is appropriate for the data. Figure \ref{fig:MassMapsMocks} illustrates the unclipped (left column) and clipped (right column) convergence fields from a single 100 deg$^2$ SLICS realisation, with a smoothing scale of 2.2 arcmin (upper two panels) and 6.6 arcmin (lower two panels). We chose these values of $\sigma_{\rm{s}}$, simply to illustrate the substantial differences in the clipped convergence fields these scales facilitate. The first and third panels have no shape noise ($\sigma_e=0$), whereas the second and fourth panels have shape noise at the level of KiDS-450 (Gaussian distributed with mean zero and $\sigma_e=0.28$). The clipped fields here have a convergence threshold of $\kappa^c=0.010$. Comparing the first and second panels, smoothed with $\sigma_{\rm{s}}=2.2$ arcmin, we see that the features within the clipped and unclipped maps change dramatically when shape noise is introduced. The third and fourth panels however show that the maps change less dramatically with the inclusion of shape noise when the smoothing scale is set to 6.6 arcmin. This indicates that the higher of the two smoothing scales is better suited to SLICS and by extension the data.  

\medskip

\medskip

An additional test of whether the chosen $(\kappa^c,\sigma_{\rm{s}})$ combination is suitable comes from inspection of the clipped and unclipped correlation functions. The optimal choices for these parameters will facilitate clipping of the non-linear regions exclusively, leaving the linear signal untouched. In this case, the unclipped and clipped $\xi_+$ should converge on the larger, linear angular scales. In Figure \ref{fig:BiggerPlot}, we present how the $\xi_+^{\rm{clip}}$ measured from the SLICS are affected by variations in the clipping threshold, smoothing scale and the galaxy shape noise. Similar trends are seen for the $\xi_-^{\rm{clip}}$ statistic at higher angular scales (we refer the reader to Section \ref{sec:results}). The left hand panels in this figure display $\theta \xi_+$, where $\xi_+$ is the mean unclipped (in solid grey) or clipped (other colours) correlation function measured from the SLICS realisations. The right hand panels display the various correlation functions normalised to that of the unclipped. In calculating the error on the ratios, we take into account the cross-covariance between the clipped and unclipped statistics. The magenta line on all panels is the same and corresponds to $\kappa^c=0.010, \sigma_{\rm{s}}=6.6$ arcmin with KiDS-450 level shape noise.

\medskip

The upper panel of Figure \ref{fig:BiggerPlot} illustrates the effect of increasing the clipping threshold from $\kappa^c=0.005$ to 0.010 to 0.015, whilst the smoothing scale is fixed to 6.6 arcmin and the shape noise is fixed to the KiDS-450 level. On average, $26 \pm 3\%$ of the area of the field is clipped in the case of the most aggressive clipping threshold, $\kappa^c=0.005$, and $3 \pm 1\%$ is clipped in the case of the least aggressive, $\kappa^c=0.015$. We see that when adopting $\kappa^c=0.005$, the clipped signal exhibits a large reduction  in power at angular scales around 6 arcmin and a failure to converge with the unclipped at the larger angular scales. The power deprecation is caused by overly aggressive clipping; subtracting too much of the shear signal engenders anticorrelations in the $\xi_+^{\rm{clip}}$. The excess power at large $\theta$ is caused by the smoothing transferring small-scale power to larger scales. This effect is illustrated by considering the convolution of a single $\delta$-function with a Gaussian smoothing kernel; the signal is spread by an extent given by the width of the Gaussian. This panel suggests that $\kappa^c=0.010$ and 0.015 are more appropriate thresholds as they better recover the large scale behaviour of the $\xi_+^{\rm{unclip}}$.

\medskip

The variations in the $\xi_+^{\rm{clip}}$ when the smoothing scale is altered, whilst $\kappa^c$ is fixed to 0.010 and the shape noise is fixed to KiDS-450 level, are shown in the middle panel of Figure \ref{fig:BiggerPlot}. We note the lack of convergence between the unclipped and the clipped signal with $\sigma_{\rm{s}}=4.4$ arcmin, indicating over-clipping of the convergence field. We also see that the angular scale at which the loss of power in the $\xi_+^{\rm{clip}}$ is maximised translates right with increasing smoothing scale. This is due to the loss of signal incurred from smoothing over features of this angular size. The upper and middle panels of Figure \ref{fig:BiggerPlot} illustrate the importance of identifying a clipping threshold and smoothing scale which are high enough to diminish the clipping of pure noise features, but low enough to avoid smoothing out the cosmological content in the clipped statistic.

\medskip 
 
The lower panel of Figure \ref{fig:BiggerPlot} illustrates the sensitivity of the $\xi_+^{\rm{clip}}$ to the shape noise, whilst $\kappa^c$ and $\sigma_{\rm{s}}$ are fixed to $0.010$ and 6.6 arcmin respectively. Where $\sigma_e > 0$ the shape noise is sampled from a Gaussian distribution with width equal to $\sigma_e$, whereas $\sigma_e=0$ refers to a measurement made in the absence of shape noise. Shape noise sampled from the broader Gaussian with $\sigma_e=0.4$, causes greater proportions of the convergence map to exceed the clipping threshold and hence we see a greater reduction in the power after clipping. This demonstrates the importance of matching the shape noise properties of galaxies in the mocks to the data in order to get a simulated model of the clipped correlation functions. We also note that we see only a small reduction in the power in the shape-noise-free clipped relative to the unclipped, suggesting that most of the clipped content is shape noise rather than non-linear regions. Nevertheless, we find that this small amount of clipping of non-linear cosmological signal, is sufficient for informing the parameter inference with some independent information, as evidenced by the constraints obtained in Section \ref{sec:results} and the cross-correlation coefficient matrices in Appendix \ref{sec:Appendix_Covariance}.

\medskip

Having quantified the effect of different choices of the clipping threshold and smoothing scale with the SLICS, in clipping the KiDS-450 data we adopt the most aggressive clipping parameters that satisfy our requirement that the clipped and unclipped $\xi_+$ converge within 1$\sigma_{\rm{mean}}$, where $\sigma_{\rm{mean}}$ is the error on the mean measurement, on large angular scales. This is in order to maximise the difference between the clipped and unclipped statistics and thus enhance the cosmological parameter constraints. Henceforth we set $\kappa^c=0.010$ and $\sigma_{\rm{s}}=6.6$ arcmin, and conduct clipping with these parameters on the KiDS-450 data and all simulations.

\subsection{Calibration Corrections}

In this section we discuss various calibration corrections which are necessary in order to use clipping to constrain cosmological parameters in this proof-of-concept study. These corrections, necessitated by the imperfect mass reconstruction due to the presense of masks, as well as the the finite box size and low level bias in the simulations, are not intrinsic to the clipping methodology. 

\subsubsection{Mask bias} \label{subsec:Mask_Correction}

Real data is subjected to masking, which complicates all methods seeking to transform the density field. This is because it is unclear how to interpret regions where the density field is unknown. In order to investigate how masking affects the clipped correlation functions, we take a $5 \times 10$ deg$^2$ section of the G9 mask (H17) and concatenate it with a copy of itself, in order to fit the $10 \times 10$ deg$^2$ field of view of SLICS. We apply the resultant mask to each of the realisations.

\medskip

\begin{figure}
\includegraphics[width=0.5\textwidth]{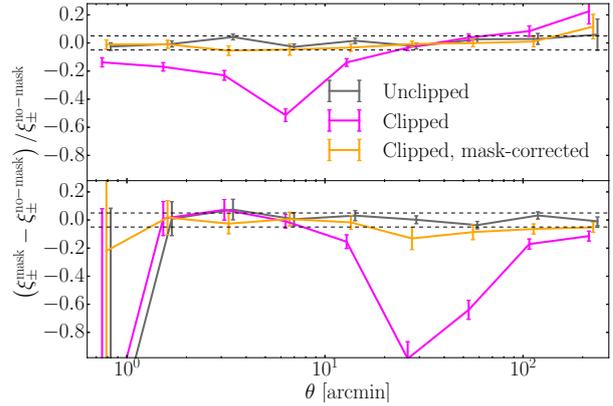} 
\caption{The effect of the mask bias for the clipped and unclipped $\xi_+$ (\textit{upper}) and $\xi_-$ (\textit{lower}) from SLICS. The grey curve shows the fractional difference between the masked and the unmasked $\xi_\pm^{\rm{unclip}}$ -- the fact that this curve has a 5\% consistency with zero across all angular scales illustrates that the $\xi_\pm^{\rm{unclip}}$ is fairly unaffected by masking. The magenta curve shows the fractional difference between the masked and unmasked $\xi_\pm^{\rm{clip}}$ -- the significant deviation from zero illustrates the biasing caused by the mask. The orange curve displays the fractional difference between the masked $\xi_\pm^{\rm{clip}}$, once corrected for the bias with 100 noise realisations via the methodology discussed in the text, and the unmasked $\xi_\pm^{\rm{clip}}$. The correction reduces the mask bias to $\lesssim 5\%$ in the case of the $\xi_+^{\rm{clip}}$; the $\xi_-^{\rm{clip}}$ however still suffers residual mask-bias at a level of $\sim$10\% between 20 and 50 arcmin after we employ our masking correction. The clipped measurements were made with $\kappa^c=0.010$ and $\sigma_{\rm{s}}=$6.6 arcmin, and the error bars are measured from the SLICS realisations. } \label{fig:Mask_Correction}
\end{figure}

As expected, the change in the $\xi_\pm^{\rm{unclip}}$ from SLICS  when a mask is applied is small, in line with the sampling variance on the measurement. However we find considerable deviations between the measurements of $\xi_\pm^{\rm{clip}}$ from the masked and unmasked SLICS. Figure \ref{fig:Mask_Correction} shows the fractional difference between the masked and unmasked clipped and unclipped $\xi_\pm$ measured from the SLICS with $\kappa^c=0.010$ and smoothing scale of 6.6 arcmin. The fractional difference for the $\xi_\pm^{\rm{unclip}}$ (in grey) differs from zero by less than 5\% across all angular scales whereas that of the clipped (magenta) features considerable deviations at angular scales below 20 arcmin. Deviations of similar magnitude and shape arise when we use masks which have different geometry but reduce the field area by similar amounts. We refer to the influence which the mask has on the clipped measurements as the \textit{mask bias}.

\medskip

The mask bias arises from the way we handle masks and edge effects in mass reconstruction. We follow the methodology of \cite{vanWaerbeke_et_al_2013} by setting the convergence to zero in regions where more than 50\% of the volume of the Gaussian smoothing window is centred on masked pixels. Where masked regions coincide with high convergence regions, this process causes the convergence surrounding the masked regions to be underestimated, and the overall power in the $\xi_\pm^{\rm{clip}}$ statistics to be diminished. This does not affect the $\xi_\pm^{\rm{unclip}}$ since no mass reconstruction is performed in arriving at these measurements. This issue is not a problem intrinsic to clipping, so much as it is a general issue with mass reconstruction methodology in the presence of masks. This is an active topic of research \citep[see for example][]{VanderPlas_etal_2012, Liu_etal_2014, Jullo_etal_2014, Chang_etal_2017} and rigorously solving this problem is beyond the scope of this paper. We instead opt to numerically calibrate and correct for the effect of the mask on our clipped statistics.

\medskip

We find that for $\kappa^c=0.010$ and $\sigma_{\rm{s}}=6.6$ arcmin, mask bias is negligible in the absence of galaxy shape noise. Consequently, we assume that for our chosen clipping threshold and smoothing scale, the mask bias is dependent on the level of shape noise and the mask geometry, and independent of the cosmology. This is a reasonable assumption given the statistical power of KiDS-450. Our assumption prompts us to investigate the effect of the mask on fields consisting of pure galaxy shape noise and zero lensing. We model the mask bias correction to the clipped correlation function as 

\begin{equation} \label{maskbias2}
\text{mask bias} =  \xi{_\pm}^{\text{mask, noise}} - \xi{_\pm}^{\text{no-mask, noise}}\,,
\end{equation}

\noindent where $\xi{_\pm}^{\text{mask, noise}}$ and $\xi{_\pm}^{\text{no-mask, noise}}$ are the average measurements from fields of pure Gaussian shape noise, with mean zero and $\sigma_e=0.28$, which are masked/unmasked respectively. By subtracting the mask bias correction from the clipped correlation functions calculated with a mask applied, we find that the influence of the mask can be mostly corrected for.

\medskip

Figure \ref{fig:Mask_Correction} displays the $\xi_\pm^{\rm{clip}}$ corrected for the mask bias (modelled in equation \ref{maskbias2}), using 100 noise fields, in orange. The mask applied here is that of the G9 patch reformatted to fit the SLICS lightcone, but we verify that we obtain the same results for the corrected $\xi_\pm^{\rm{clip}}$ if we apply a different mask to the SLICS and recompute the correction specific to said mask. We find that the corrected $\xi_+^{\rm{clip}}$ is consistent with the measurement made in the absence of masking to within 5\%. Although the corrected $\xi_-^{\rm{clip}}$ is much closer to the unmasked than the masked measurement, we find that the mask bias remains present at a $\sim$10\% level at angular scales of $\sim$30 arcmin. A larger number of noise realisations does not reduce the mask bias further, implying that a more sophisticated treatment of the masks is critical if clipping is to be used in future cosmological analyses. The residual mask bias affecting the $\xi_-^{\rm{clip}}$ measurement, combined with the fact that $\xi_-^{\rm{unclip}}$ is the least powerful shear correlation function in terms of constraining cosmological parameters, motivates us to continue in this analysis using the $\xi_+^{\rm{clip}}$ and $\xi_+^{\rm{unclip}}$ statistics only. 

\medskip

We proceed to compute and correct the mask bias for each of the KiDS-450 patches individually. The corrections for each of the patches are similar, which is expected given the masks cause a similar reduction in effective area per patch. All clipped correlation functions from KiDS-450, presented in this paper and used in the likelihood analysis in Section \ref{sec:results}, have been corrected for mask bias, whereas all those from the simulations were computed without masks applied. As this is a proof-of-concept, we do not propagate the error on the mask bias through to the cosmological constraints with KiDS-450, as we want to see the improvement obtained through clipping in a scenario where the mask bias is under control.

\subsubsection{Finite box effects} \label{subsec:DH10_finite_box}

\begin{figure}
\begin{center}
\includegraphics[width=0.5\textwidth]{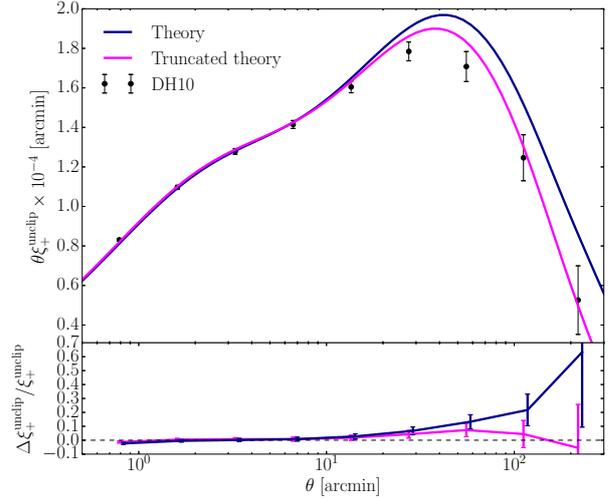}
\end{center}
\caption{\textit{Upper:} the noise-free $\xi_+^{\rm{unclip}}$ measured for the fiducial cosmology of the DH10 simulations (data points), the theoretical prediction from a non-truncated box (dark blue; equation \ref{eqn:xi+-theory}), and the theoretical prediction from a truncated box of size $L_{\rm{box}}=250 \, h^{-1}$Mpc (magenta). The error bars on the data points come from the dispersion across the 175 realisations (35 simulations $\times$ 5 ray-tracing angles) for this cosmology. The difference between the dark blue and magenta lines is the finite box correction we apply to the DH10 measurement. \textit{Lower:} the fractional difference between the theoretical $\xi_+^{\rm{unclip}}$ predictions and the DH10 measurement.} \label{fig:Fitting_kmin}
\end{figure}

The DH10 simulations span a broad range in the $\Omega_{\rm{m}}$-$\sigma_8$ parameter space at the cost of having a small number of realisations per cosmology and a small box size relative to the SLICS (see Figure \ref{fig:DH10_Cosmologies} and Table \ref{tab:Mocks_Table} for details). In simulations, the finite size of the box means that the matter power spectrum $P_\delta(k,\chi)$, appearing in equation \ref{eqn:Pkappa}, is limited by two scales: $k_{\rm{min}}=2\pi/L_{\rm{box}}$, where $L_{\rm{box}}$ is the size of the simulation box, and $k_{\rm{max}}=2\pi/L_{\rm{res}}$, where $L_{\rm{res}}$ is the smallest scale which can be resolved in the simulation. The missing modes with $k<k_{\rm{min}}$ cause the unclipped shear correlation functions expressed in equation \ref{eqn:xi+-theory} to lose power at large angular scales  \citep[see for example][]{Harnois_Waerbeke_2015}. Similarly, the missing modes with $k>k_{\rm{max}}$ engender a loss of power at small $\theta$. The effect of the resolution of DH10 is not prominent at the angular scales probed by our measured shear correlation functions, as is evidenced by the consistency between the theoretical and mock $\xi_+^{\rm{unclip}}$ at angular scales $<10$ arcmin, shown in Figure \ref{fig:Fitting_kmin}. On the other hand, the $k$-modes absent due to the box size do cause the DH10 $\xi_+^{\rm{unclip}}$ to be underestimated on angular scales $>10$ arcmin. We therefore need to correct for the effect of the finite box in order to constrain the cosmology of the real Universe using the DH10 $\xi_+^{\rm{clip}}$ measurements.

\medskip

We obtain cosmology- and angular-scale-dependent corrections for the finite box effect on $\xi_+^{\rm{unclip}}$ by measuring the difference between the theoretical prediction from equation \ref{eqn:xi+-theory} in a non-truncated box, and the prediction within a box of size $L_{\rm{box}}$. For these predictions we use the {\sc{Nicaea}} code from \citet{Kilbinger_etal_2009} with the {\sc{Halofit}} model from \citet{Smith_etal_2003}, since it is a better match to the DH10 $\xi_+^{\rm{unclip}}$ than that of \citet{Takahashi_etal_2012}. The correction for the loss of power at large angular separations, due to the finite box, is robust to the choice of {\sc{Halofit}} model however, since \citet{Smith_etal_2003} and \citet{Takahashi_etal_2012} converge at these scales. 

\medskip

The obvious choice for the size of the truncated box used in calibrating the finite box effect is that in which the DH10 were created, 140 $h^{-1}\rm{Mpc}$. The theoretical $\xi_+^{\rm{unclip}}$ from a box of this size however overestimates the loss of power at large $\theta$ seen in DH10. This is because the simulations are constructed as a lightcone through the box, resulting in a smooth decay in the power, in contrast to a sharp cutoff at the $L_{\rm{box}}$ scale. We follow \cite{Sellentin_etal_2017}, by modelling the finite box effect with an effective cutoff, performing a $\chi^2$  fitting of the theoretical $\xi_+^{\rm{unclip}}$ for different values of the box size to the shape-noise-free mean measurement from the fiducial DH10 cosmology. We fit the box size for the fiducial cosmology only, on account of there being the largest number of realisations and thus the lowest sampling variance overall (though we stress that the corrections we apply are specific to each cosmology). We use the covariance matrix measured from the 175 realisations for the fiducial DH10 cosmology, rather than the one from SLICS, since the former will better describe the uncertainty on DH10. Furthermore, we use only the five angular separation points $>10$ arcmin in the fitting as we are most interested in finding the effective box size that best describes the large-scale behaviour of the mocks where the effect of the finite box size becomes relevant. We find that the $\xi_+^{\rm{unclip}}$ for the fiducial cosmology of DH10 is best described by the prediction in an \textit{effective} box size of 250 $h^{-1}$Mpc. This prediction, shown by the magenta curve in Figure \ref{fig:Fitting_kmin}, fits the DH10 measurement well, with a $\chi^2$ of 4.99 for the 4 degrees of freedom. The correction for the finite box size for this cosmology is the difference between the theoretical prediction from the non-truncated box (shown in dark blue), and the truncated box prediction.

\medskip

The lack of a theoretical prediction for $\xi_+^{\rm{clip}}$ limits our inference of the finite box effect for this statistic. We assume therefore that the loss of power in the clipped correlation functions due to the finite box effect is equal to that of the unclipped, and so the calibration correction we derive for the unclipped correlation functions per cosmology, is applicable also to the clipped. This assumption is likely to be valid since the effect of the finite box is most prominent on scales where $\xi_+^{\rm{unclip}}$ and $\xi_+^{\rm{clip}}$ converge. We also test how much the marginalised means and 68\% confidence intervals on the cosmological parameters change when the finite box correction is included/omitted and find that the effect is small and does not change our conclusions. This approach is suitable for this proof-of-concept analysis and the correction can easily be circumvented in the future with the use of larger simulations such as the Mira Titan suite \citep{Heitmann_etal_2016}. 

\medskip

We compute individual calibration corrections for each of the 158 DH10 cosmologies, using the box size fit to the fiducial cosmology. We then additively scale up the whole angular separation range of the clipped and unclipped $\xi_+$ from the simulations (the small scales remaining practically unchanged by the calibration). An additive, rather than a multiplicative, correction is appropriate for accounting for the missing $k$-modes in the integration over $P_\delta(k,\chi)$ in equation \ref{eqn:Pkappa}. The correction we apply also has the benefit of not inflating the noise in the DH10 predictions.

\medskip

The SLICS are also affected by the limitations of a finite box, though the box size is larger than that of DH10, engendering a loss of power at the largest angular scales that is of order 10-30\% (we refer the reader to the ratio of the theoretical and SLICS $\xi_+^{\rm{unclip}}$ shown in Figure \ref{fig:BiggerPlot}). In general the covariance that we calculate from SLICS will be affected by the loss of power in the correlation functions, but since the correction for the finite box in DH10 has a very small impact on the cosmological parameter constraints, and this effect is much smaller for SLICS, we therefore treat the SLICS covariance matrices as unbiased by the box size. We note however that returning to the effect of the finite box on covariance estimation is an important topic for future work.

\subsubsection{Cosmological bias} \label{subsec:DH10_Bias}

\begin{figure}
\begin{center}
\includegraphics[width=0.5\textwidth]{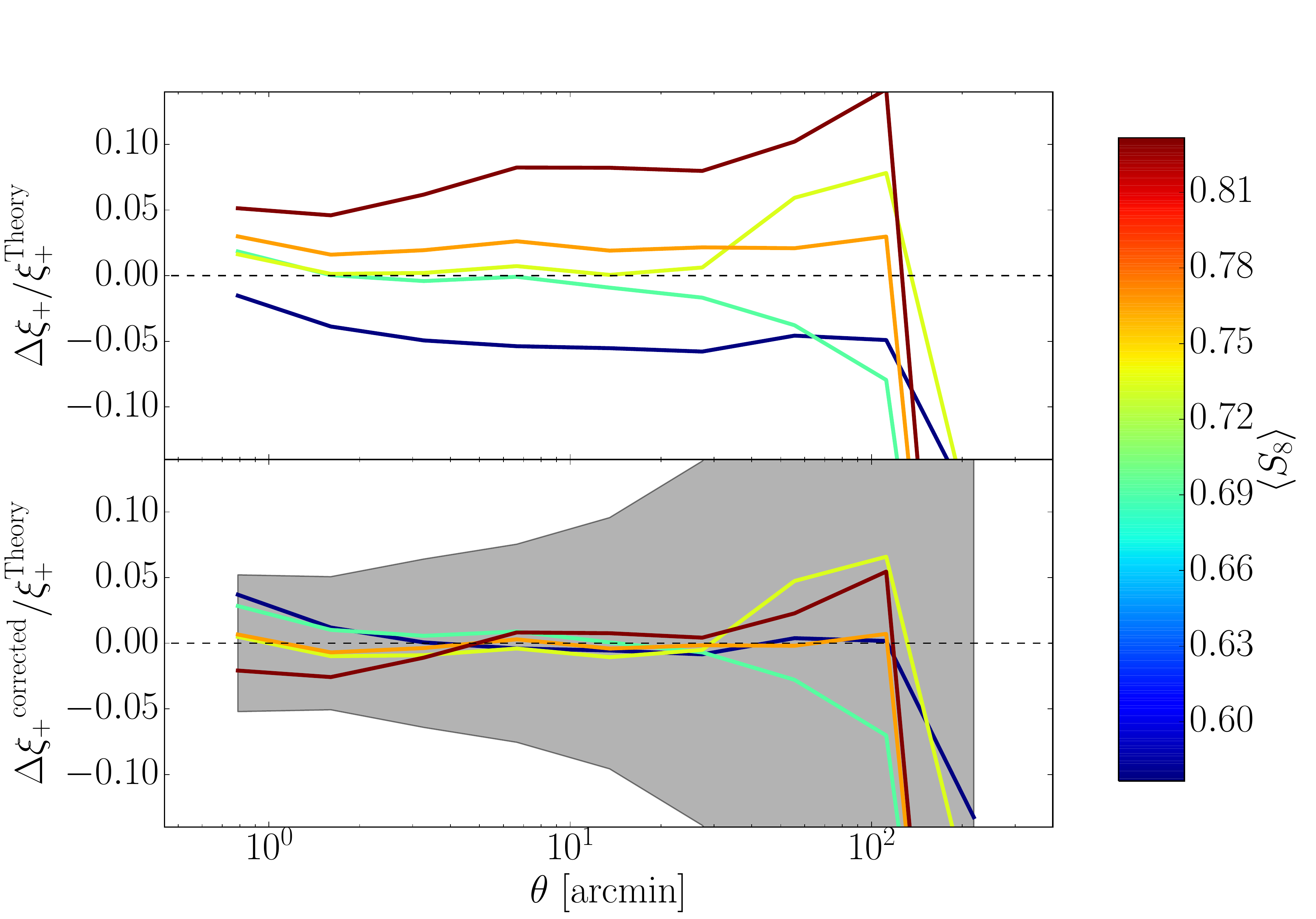}
\end{center}
\caption{\textit{Upper panel:} the fractional difference between the 158 shape-noise-free DH10 $\xi_+^{\rm{unclip}}$ measurements and the theoretical predictions (equation \ref{eqn:xi+-theory}) binned in terms of $S_8=\sigma_8(\Omega_{\rm{m}}/0.3)^{0.5}$, with the colours designating the mean $S_8$ in each bin. We see that the low $S_8$ measurements underestimate $\xi_+^{\rm{unclip}}$, whereas the high $S_8$ measurements overestimate. \textit{Lower panel:} the same measurements but corrected for the cosmological bias via the methodology discussed in the text. Any remaining bias can be compared to the uncertainty on the clipped predictions (shaded grey) that is included in our analysis when using the DH10 simulations.} \label{fig:DH10_Bias}
\end{figure}

In Figure \ref{fig:Fitting_kmin} we show that the fiducial DH10 cosmology reproduces the expected $\xi_+^{\rm unclip}$, modulo a small correction for the finite box effect on large scales. In the upper panel of Figure \ref{fig:DH10_Bias} we compare noise-free measurements of $\xi_+^{\rm unclip}$, corrected for the box size,  with theoretical predictions (equation \ref{eqn:xi+-theory}), now for the full range of 158 cosmologies spanned by the DH10 simulations. Binning the relative difference by the input cosmology $S_8$ (see colour-bar) we see a trend where the low $S_8$ simulations tend to underestimate $\xi_+^{\rm unclip}$ by $\sim$7\% between angular separations of 1 and 110 arcmin, whereas high $S_8$ simulations overestimate by $\sim$10\% in this range. The cause of this cosmological bias, which is present irrespective of whether the finite box correction is applied, is currently unknown. Uncovering its origin is part of an on-going analysis where we are building a next generation of varying cosmology lensing simulations. However, the bias is less than the level of uncertainty due to shot noise and sampling variation in the DH10 $\xi_+^{\rm clip}$ predictions (see Section \ref{subsec:Cosmol_Depend}) that increases from 5 to $\sim$100\% over the full angular range, shown by the grey shaded  region in the lower panel. It is therefore accounted for, to some extent, in our clipped analysis that includes an error budget to account for this level of uncertainty in the DH10 predictions. Nevertheless, we employ a correction scheme to ensure that this systematic does not artificially contribute to the improvements yielded by the combined clipped-and-unclipped analysis. 

\medskip

We determine a cosmological bias correction by averaging the relative difference between the shape-noise-free DH10 and theoretical $\xi_+^{\rm{unclip}}$ between 1 and 60 arcmin, where the bias varies slowly, in each of the five $S_8$ bins shown in Figure \ref{fig:DH10_Bias}. This produces a smooth mean-bias function which monotonically increases from $-5\%$ in the lowest $S_8$ bin to $+8\%$ in the highest. We obtain the mean-bias for each of the 158 DH10 cosmologies by linearly interpolating/extrapolating from this function for the simulation $S_8$ values. The corrected $\xi_+^{\rm{unclip}}$ is obtained by multiplicatively scaling the DH10 measurements by $1/[1+b(S_8)]$, where $b(S_8)$ is the mean-bias corresponding to the $S_8$ value of a given simulation. The relative differences between the corrected DH10 $\xi_+^{\rm{unclip}}$ and the theoretical measurements are shown in the lower panel of Figure \ref{fig:DH10_Bias} for the five $S_8$ bins, and can be compared to the uncertainty included in the clipped predictions $\xi_+^{\rm{clip}}$ (shaded grey), which is incorporated in our cosmological parameter constraints.

\medskip

As was the case with the finite box effect (Section \ref{subsec:DH10_finite_box}), it is not possible to ascertain the extent to which the clipped predictions are affected by the cosmological bias in DH10, owing to the lack of a theoretical clipped statistic. Hence we again assume that the $\xi_+^{\rm{clip}}$ is biased in the same way as the corresponding $\xi_+^{\rm{unclip}}$ measurement. We find that our conclusions are not significantly changed however if the cosmological bias is unaccounted for; the combined clipped-and-unclipped analysis increases the constraining power by 20\%, instead of 15\% when the bias is corrected. We note that this bias was unaccounted for in the peak statistics analyses of DH10, \cite{Kacprzak_etal_2016} and \cite{Martinet_etal_2018}, and their results will likely be affected.

\subsection{Cosmological dependence of clipping} \label{subsec:Cosmol_Depend}

Although there is a large number of shear catalogues for the fiducial DH10 cosmology (35 independent simulations $\times$ 5 pseudo-independent catalogues corresponding to 5 different ray-tracing angles), there exist only 5 catalogues for the remaining 157 cosmologies. The average $\xi_\pm^{\rm{clip}}$, measured across each set of  non-fiducial DH10 cosmologies, is therefore more significantly impacted by shot noise in comparison to the fiducial set.  In the case of the unclipped correlation functions one can simply turn off the noisy galaxy ellipticities.  However, as is discussed in Section \ref{subsec:Choose_SS_Kc}, we find that the clipped correlation functions are critically dependent on the shape noise. This necessitates the inclusion of shape noise such that the noise properties of the mocks match the data.

\medskip

In order to reduce the impact of the shot noise whilst still including the effects of the galaxy shape noise, we determine the clipped correlation functions from DH10 with different realisations of the shape noise. We find that averaging $\xi_+^{\rm{clip}}$ across 75 or more noise realisations is sufficient for the measurement from each of the individual catalogues of the fiducial DH10 cosmology to stabilise. This averages away any bias in the measurement caused by a single realisation of the shape noise. We proceed to compute 75 noise realisations per catalogue for all of the DH10 cosmologies; the $\xi_+^{\rm{clip}}$ for each cosmology appearing in the likelihood analysis is the average over these. The remaining source of noise in the DH10 mocks is then the sampling variance across different catalogues of a given cosmology. In order to include this source of uncertainty in our likelihood analysis, we measure the covariance across the 175 clipped and unclipped $\xi_+$ from the fiducial DH10 cosmology, each of which is averaged across 75 noise realisations, via equation \ref{eqn:CovMat}. These covariance matrices, which are at the level of 5\% in the first angular separation bin (0.8 arcmin), increasing to $\sim$100\% in the last bin (220 arcmin; see Figure \ref{fig:DH10_Bias}), encompass our uncertainty on the model, both in terms of sampling variance and cosmological bias (see Section \ref{subsec:DH10_Bias}). We add this error in quadrature to the error measured from the SLICS which describes the uncertainty in the data itself. This is discussed in more detail in Section \ref{subsec:Lhd}. 

\medskip

\begin{figure}
\begin{center}
\includegraphics[width=0.5\textwidth]{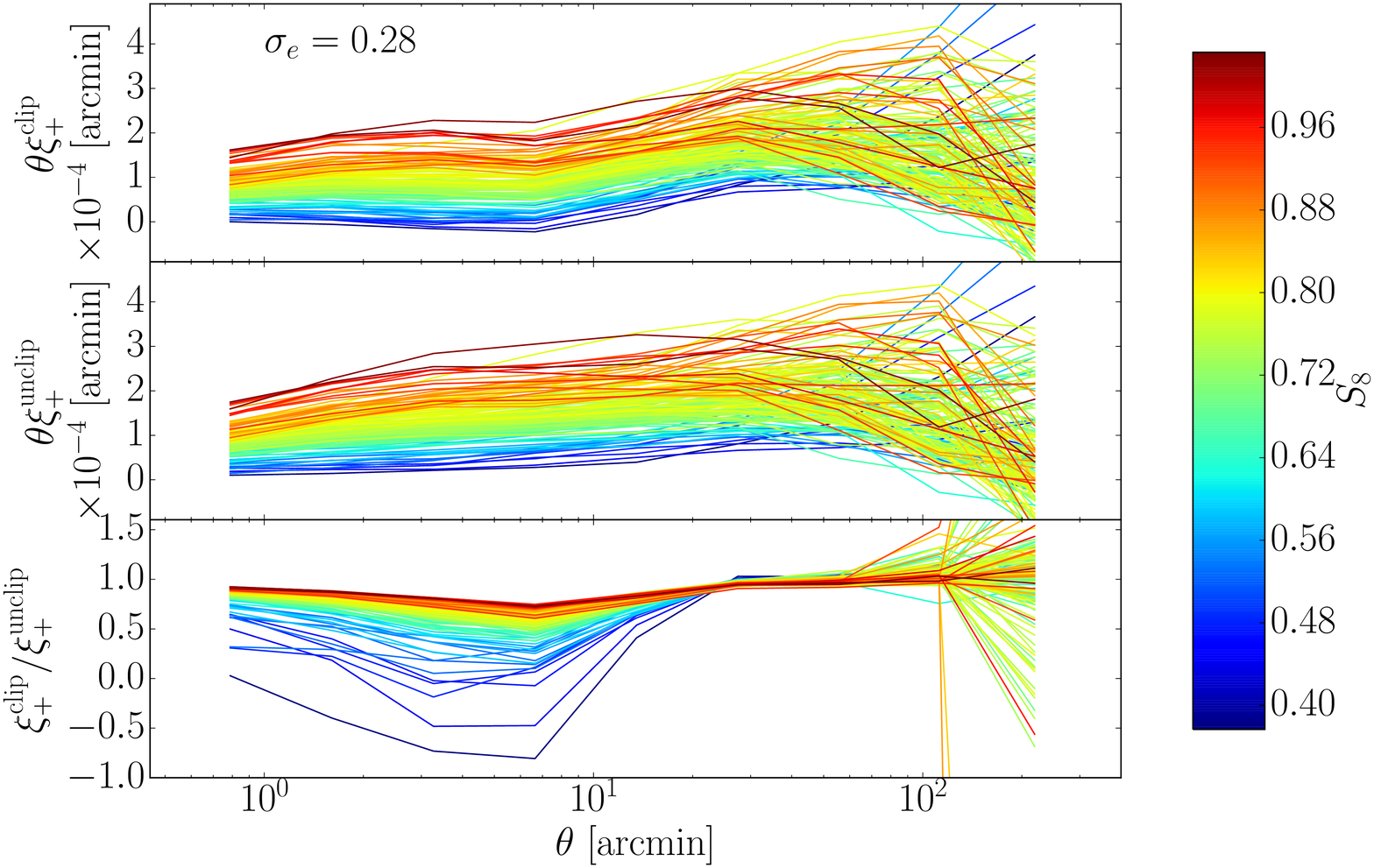}\\ 
\includegraphics[width=0.5\textwidth]{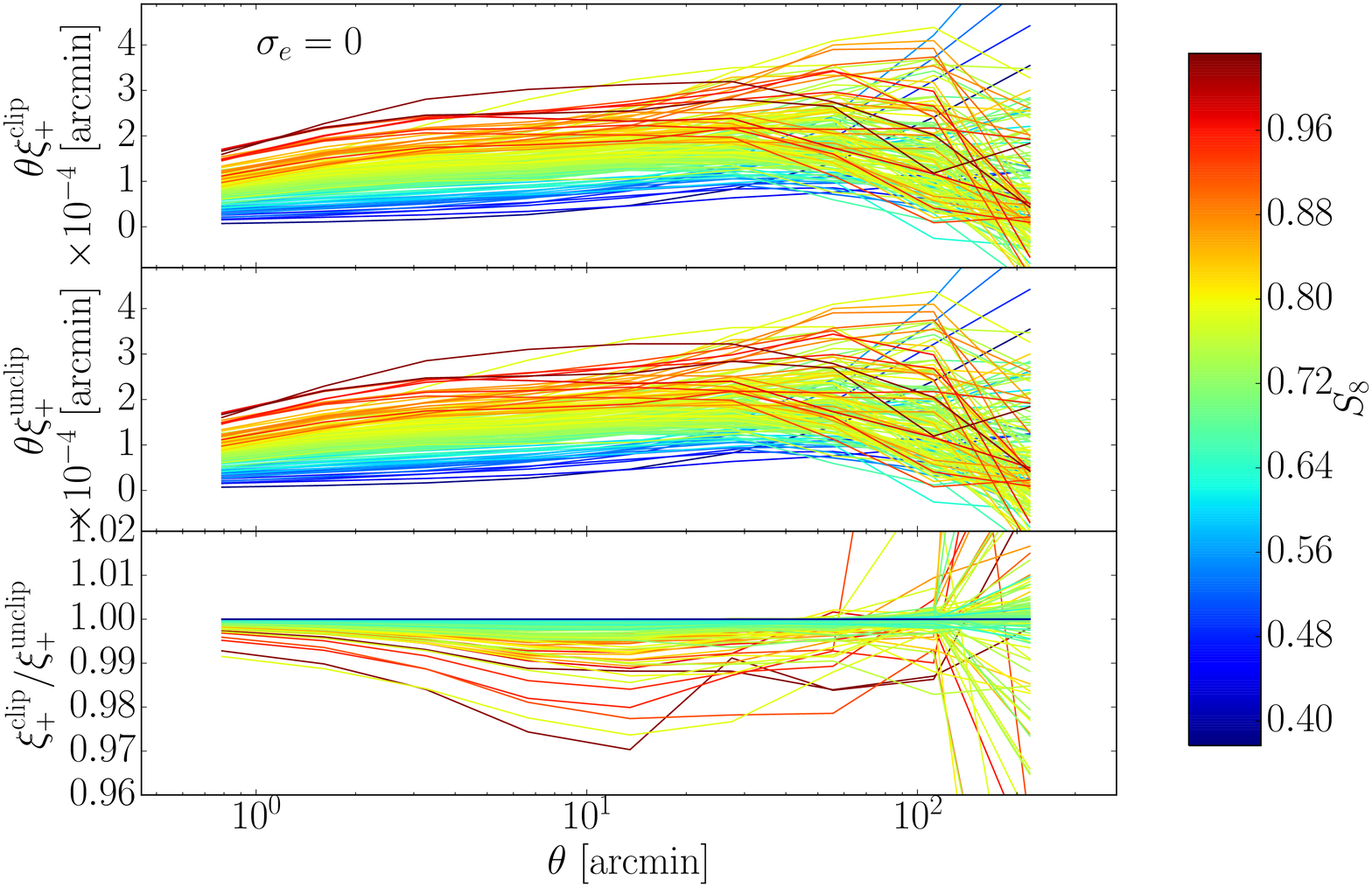}\\ 
\end{center}
\caption{\textit{Upper panels}: $\theta\xi_+^{\rm{clip}}$, $\theta\xi_+^{\rm{unclip}}$ and the ratio for all of the DH10 cosmologies, each of which is averaged over 75 shape noise realisations, colour-coded by $S_8$. The clipping threshold and smoothing scale are $\kappa^c=0.010, \sigma_{\rm{s}}=6.6$ arcmin respectively, selected via the methodology in Section \ref{subsec:Choose_SS_Kc}. The low clipped-to-unclipped ratios seen at $< 10$ arcmin for low $S_8$ cosmologies are brought about by clipping shape noise only. \textit{Lower panels}: the same measurements but with zero shape noise. The low $S_8$ cosmologies are not subject to clipping in this case, and the clipped and unclipped $\xi_+$ converge at all angular scales. All measurements have been corrected for the finite box effect and cosmological bias (see Sections \ref{subsec:DH10_finite_box}-\ref{subsec:DH10_Bias}).} \label{fig:DH10_CFs}
\end{figure}

In the upper panel of Figure \ref{fig:DH10_CFs} we present the clipped (upper), unclipped (middle), and the ratio (lower) for all of the DH10 cosmologies, each of which is averaged over the 75 realisations of the shape noise, with $\kappa^c=0.010$ and $\sigma_{\rm{s}}=6.6$ arcmin. All measurements have been corrected for the finite box effect and the cosmological bias (Sections \ref{subsec:DH10_finite_box} and \ref{subsec:DH10_Bias}). In general the power in the $\xi_+^{\rm{clip}}$ increases with $S_8$ in a similar capacity to the $\xi_+^{\rm{unclip}}$. The prominent reduction in power at angular scales $\sim$5 arcmin is also a common feature for all of the cosmologies. We observe a number of the low $S_8$ cosmologies with small or negative ratios at small angular separations. This effect is not caused by these cosmologies experiencing a greater degree of clipping; indeed we see that in general less of the field is clipped for lower $S_8$ cosmologies as expected. Rather, this is the result of these fields being dominated by shape noise. Smoothing these fields correlates the shape noise, and clipping then leads to a reduction in power and even anticorrelations to be seen in the $\xi_+^{\rm{clip}}$ for these low $S_8$ cosmologies. This is not observed in the higher $S_8$ measurements which have higher signal to noise, and consequently maintain larger power in the correlations throughout the clipping pipeline. In the case of the low $S_8$ cosmologies, smaller values of $\sigma_{\rm{s}}$ and $\kappa^c$ would have been more suitable for the clipped analysis. 

\medskip

\begin{figure*}
\begin{center}
\includegraphics[width=\textwidth]{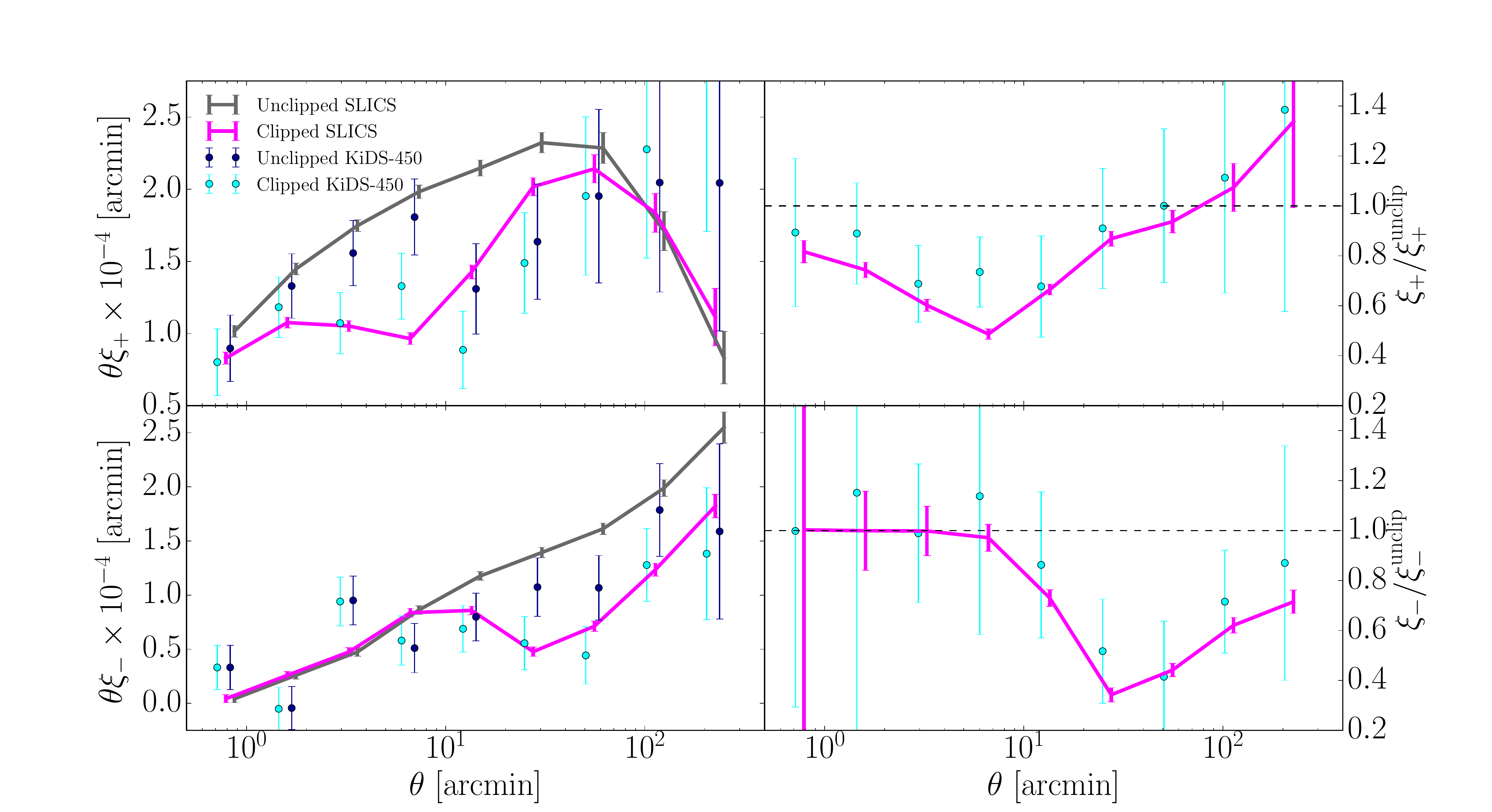}
\caption{The clipped and unclipped $\xi_{+}$ (\textit{upper)} and $\xi_{-}$ (\textit{lower}) for KiDS-450 (data points) relative to those from the fiducial cosmology of SLICS. The left hand panels display $\theta \xi_\pm$, the right hand the ratio of the clipped and unclipped measurements. The errors plotted for SLICS are that of the mean measurement. The error bars on KiDS-450 are equal to those of SLICS scaled by the ratio of the effective unmasked areas. The mock and the data were clipped with the fiducial parameters $\kappa^c=0.010$ and $\sigma_{\rm{s}}=6.6$ arcmin.} \label{fig:KiDSBigPlot}
\end{center}
\end{figure*}

\medskip

The lower panel of Figure \ref{fig:DH10_CFs} shows these measurements in the absence of shape noise, to verify that in this case the low $S_8$ cosmologies experience no clipping, and the $\xi_+^{\rm{clip}}$ and $\xi_+^{\rm{unclip}}$ converge at all scales. Intuitively we see lower clipped-to-unclipped ratios as $S_8$ increases, due to the greater degree of clipping of the cosmological signal. As with the shape-noise-free $\xi_+^{\rm{clip}}$ from SLICS presented in Section \ref{subsec:Choose_SS_Kc}, we see once again that clipping the non-linear signal causes only a small change in the correlation functions relative to the unclipped, but this small effect is ample for considerably informing the parameter inference (see Section \ref{sec:results}). This highlights the importance of selecting a smoothing scale and clipping threshold which are well suited to the properties of the data, in order to clip the cosmological signal rather than just the noise. One need not alter $\kappa^c$ and $\sigma_{\rm{s}}$ for our analysis however; the cosmologies in the extreme $S_8$ tail, are flagged as ill-fitting cosmologies to the data in our likelihood analysis, by virtue of the low power correlations/anticorrelations brought about by clipping noise only. 

\medskip


\section{Results} \label{sec:results}

In Figure \ref{fig:KiDSBigPlot} we present the $\xi_\pm^{\rm{clip}}$ measured from KiDS-450 and SLICS produced with a clipping threshold of $\kappa^c=0.010$ and smoothing scale $\sigma_{\rm{s}}=6.6$ arcmin (see Section \ref{subsec:Choose_SS_Kc}). The left hand panels of this figure display $\theta \xi_\pm$, the right displays the measurements normalised to the unclipped. The error bars come from the SLICS covariance (rescaled to the effective area of KiDS-450 in the case of the data), and we include the cross-covariance between the clipped and unclipped in the error on the ratios. We see similar trends in the clipped measurements between the mock and the data, which is expected given that SLICS are tailored to reflect KiDS-450.

\medskip

\subsection{Likelihood analysis} \label{subsec:Lhd}

We proceed to the likelihood analysis to constrain 
$S_8=\sigma_8(\Omega_{\rm{m}}/0.3)^{0.5}$, with the other parameters fixed to the DH10 fiducial values, $\Omega_{\rm{b}}=0.04, n_{\rm{s}}=1.0$ and $h=0.7$. We use only the clipped and unclipped $\xi_+$, omitting the $\xi_-$ for the reasons argued in Section \ref{subsec:Mask_Correction}, and all nine $\theta$-bins, logarithmically spaced from $\sim$0.8 to $\sim$220 arcmin. The products required to constrain these cosmological parameters are the clipped and unclipped auto- and cross-covariance matrices from SLICS, which describe the uncertainty on the data, those measured from the fiducial cosmology of DH10, which describe the uncertainty on the predictions themselves, and the $\xi_+^{\rm{clip}}$ predictions from DH10. Rather than use the $\xi_+^{\rm{unclip}}$ from DH10 in the likelihood analysis, we use the more accurate theoretical predictions (see equation \ref{eqn:xi+-theory}) evaluated at the DH10 cosmologies, from {\sc{Nicaea}}, which are free of the noise and low-level cosmological bias (Section \ref{subsec:DH10_Bias}) present in the simulations. When constraining the cosmology of a test dataset from DH10 of known cosmology, we use the {\sc{Halofit}} model from \citet{Smith_etal_2003}, as this matches these simulations more closely than the {\sc{Halofit}} model from \citet{Takahashi_etal_2012}. However, when constraining the cosmology of KiDS-450 we use the latter model, since it better describes the $\xi_+^{\rm{unclip}}$ on small, non-linear angular scales. We find that the combined clipped-and-unclipped analyses improve cosmological parameter constraints over the unclipped alone, irrespective of whether we use the simulated or theoretical $\xi_+^{\rm{unclip}}$. This is discussed further in Appendix  \ref{subsec:UCPredictions_Appendix}. We also find that the combined constraints are an improvement upon the unclipped irrespective of which $\theta$-scales are used in the likelihood analysis. The improvements do however tend to zero when the angular scales are restricted to the range where the clipped and unclipped converge.

\medskip

 The Bayesian posterior probability distribution for a particular set of cosmological parameters $\boldsymbol{\pi}$ given a data vector $\textbf{d}$ is given by

\begin{equation}
p(\boldsymbol{\pi}|\boldsymbol{d}) = \frac{\mathcal{L}(\boldsymbol{d}|\boldsymbol{\pi})p(\boldsymbol{\pi})}{E} \,,
\end{equation}

\noindent where $\mathcal{L}(\boldsymbol{d}|\boldsymbol{\pi})$ is the likelihood, $p(\boldsymbol{\pi})$ is the prior probability of the cosmological parameter configuration $\boldsymbol{\pi}$ and $E$ is the evidence, which normalises the integral of the posterior over all possible values of $\boldsymbol{\pi}$ to unity. We adopt a wide tophat prior over $\boldsymbol{\pi}$ which goes to zero where the likelihood becomes very small. Hence, in this case the posterior probability is simply proportional to the likelihood given by

\begin{equation} \label{eqn:Lhd}
\mathcal{L}(\boldsymbol{d}|\boldsymbol{\pi}) \propto  \exp \left( -\frac{1}{2} \left[\boldsymbol{d} - \boldsymbol{m}(\boldsymbol{\pi}) \right]^\intercal \Sigma^{-1} \left[\boldsymbol{d} - \boldsymbol{m}(\boldsymbol{\pi}) \right] \right) \,,
\end{equation} 

\noindent where the model prediction $\boldsymbol{m}(\boldsymbol{\pi})$ represents either the theoretical $\xi_+^{\rm{unclip}}$ from equation \ref{eqn:xi+-theory}, or the $\xi_+^{\rm{clip}}$ from DH10. The data vector $\boldsymbol{d}$ of course takes the form of the clipped and unclipped $\xi_+$ from the data. $\Sigma$ is the true covariance matrix describing the uncertainties affecting statistical inference. When computing the combined clipped-and-unclipped constraints, $\Sigma$ is built out of the auto-covariance matrices for the unclipped and clipped $\xi_+$, as well as the cross-covariance between them. Typically, uncertainties arise from the sampling variance in the data; here, we approximate this with the covariance matrix, $C_{\rm{data}}$, measured from the SLICS and rescaled to the effective area of the data. However, in this analysis we also have uncertainty on the clipped model predictions owing to the noise in the DH10 simulations. We incorporate these two independent sources of error by assuming $\Sigma \simeq C = C_{\rm{data}} + C_{\rm{model}}$, where $C_{\rm{model}}$ describes the covariance of the predictions $\boldsymbol{m(\boldsymbol{\pi})}$. The clipped auto-covariance component of $C_{\rm{model}}$ is measured across the various noise realisations for each of the catalogues for the fiducial DH10 cosmology, as is discussed in Section \ref{subsec:Cosmol_Depend}. Using the theoretical predictions from equation \ref{eqn:xi+-theory} for $\xi_+^{\rm{unclip}}$, which are free of noise, causes the unclipped auto-covariance, as well as the clipped-unclipped cross-covariance components within $C_{\rm{model}}$ to be zero. If we were to use the DH10 unclipped predictions instead, these elements are non-zero, and are again measured from these mocks (see Appendix \ref{subsec:UCPredictions_Appendix}). In this case, comparison of the diagonal elements of the clipped and unclipped parts of $C_{\rm{data}}$ and $C_{\rm{model}}$, reveals that $C_{\rm{data}}$, and hence the survey size of KiDS-450, is the dominant source of uncertainty, by a factor of $\sim$20 in the lowest angular separation bin, decreasing to $\sim$2 in the largest bin.


\medskip

Although the approximated covariance, $C$, is assumed to be an unbiased estimate of the true covariance, $\Sigma$, since it is calculated from simulations featuring noise, its inverse, $C^{-1}$, is a biased estimate of $\Sigma^{-1}$ which appears in equation \ref{eqn:Lhd}. This means that one cannot readily substitute $C^{-1}$ into this expression. \citet{Hartlap_etal_2007} advocate a correction whereby the inverse covariance is rescaled\footnote{Although see \citet{Sellentin_Heavens_2016} for a more rigorous correction scheme.} according to, 

\begin{equation}
\widehat{C^{-1}} = \frac{N-D-2}{N-2}C^{-1}.
\end{equation} 

\noindent Here $N$ is the number of simulations employed in estimating the covariance matrix $C$ containing $D \times D$ elements. In our analysis $C$ is the summation of $C_{\rm{data}}$ and $C_{\rm{model}}$, each of which have different Hartlap correction factors. This complicates efforts to obtain an unbiased estimate of the inverse covariance. However, the number of realisations, $N$, used to calculate the two matrices (906 for the data\footnote{After our clipping pipeline was run on these 906  SLICS realisations, 26 more where added to the ensemble presented in \citet{Harnois_etal_2018}. Given the negligible impact this would have on our analysis, we did not include them.}, 175 for the clipped model) greatly exceeds $D$, the number of $\theta$ bins in our correlation functions, (equal to 9 in the case of the separate clipped and unclipped analyses, and 18 for the combined). Thus $C_{\rm{data}}$ and $C_{\rm{model}}$ are sufficiently well estimated for us to safely neglect the Hartlap correction  in our likelihood analysis.

\medskip

Our cosmological constraints derive from an evaluation on a fine grid within the parameter space. In the case of the clipped analysis, we obtain 2D likelihood surfaces by interpolating from the DH10 cosmologies onto $\Omega_{\rm{m}}$-$\sigma_8$ and $\Omega_{\rm{m}}$-$S_8$ grids. Our 1D constraints on $S_8$ are then obtained by marginalising in the $\Omega_{\rm{m}}$-$S_8$ space. Although we have a theoretical prescription for the $\xi_+^{\rm{unclip}}$ as a function of cosmology (equations \ref{eqn:xi+-theory}--\ref{eqn:Pkappa}), we chose to also interpolate the theoretical unclipped model from the DH10 cosmologies in order to facilitate a direct comparison between the clipped and unclipped results. 

\medskip

An open question is whether this interpolation should be performed at the level of the clipped and unclipped $\xi_+$ or at the level of the likelihoods. If one interpolates the model, the cosmological parameter constraints are dependent on the square of any systematic bias which could potentially reside in the interpolation, whereas the dependence is only linear if one interpolates the likelihoods. We try both methods and find that extrapolating the likelihoods outside of the range of the DH10 cosmologies, is more reliable than extrapolating the model. Thus in this Section, we present the results having interpolated the DH10 likelihoods. We demonstrate in Appendix \ref{subsec:Appendix_Interp} however, that overall our conclusions are unchanged for a range of different interpolation schemes. We follow \citet{Martinet_etal_2018} and interpolate from the DH10 cosmologies using radial basis functions, employing the \textit{scipy.interpolate.Rbf} Python function set to the multiquadratic model\footnote{\url{https://docs.scipy.org/doc/scipy/reference/generated/scipy.interpolate.Rbf.html}}. Whereas the unclipped predictions are noise-less and come from theory, the clipped predictions, from DH10 have added uncertainty. Consequently, the interpolated clipped and combined likelihoods are fairly noisy, featuring spurious spikes which fracture the 68\% and 95\% clipped and combined contours. We apply a small amount of smoothing in the interpolation to reduce this effect and obtain cohesive contours. So that the clipped and combined contours can be directly compared to the unclipped, we apply the same level of smoothing when interpolating the unclipped predictions also. We verify with the unclipped statistic that the interpolation does not considerably affect the recovered cosmology relative to a standard grid-based likelihood method without interpolation.

\subsection{Cosmological constraints}

\subsubsection{DH10 constraints} \label{subsubsec:DH10_constraints}

Before constraining the cosmology of the KiDS-450 data, we investigate the power of combining the clipped and unclipped $\xi_+$ statistics for a case where the cosmology is already known. Since we only have clipped measurements at the cosmologies of the various simulations at our disposal, the natural choice for the ``data" in this test is the clipped and unclipped $\xi_+$ corresponding to the fiducial DH10 cosmology. Specifically, we take a subset of the simulations with this cosmology spanning 360 deg$^2$, the unmasked area of KiDS-450, as the data. We also omit the $\xi_+^{\rm{clip}}$ and $\xi_+^{\rm{unclip}}$ with the fiducial cosmology from the predictions, such that there is no ``perfect match" between the predictions and the data we are constraining the cosmology of, as is the case when working with real data. All cosmological constraints presented hereafter have the corrections for the finite box size and cosmological bias applied, not only to the predictions, but also to the data from DH10. We have verified that we better recover the known input cosmology with these calibrations included. 

\medskip

\begin{figure}
\begin{center}
\includegraphics[width=0.5\textwidth]{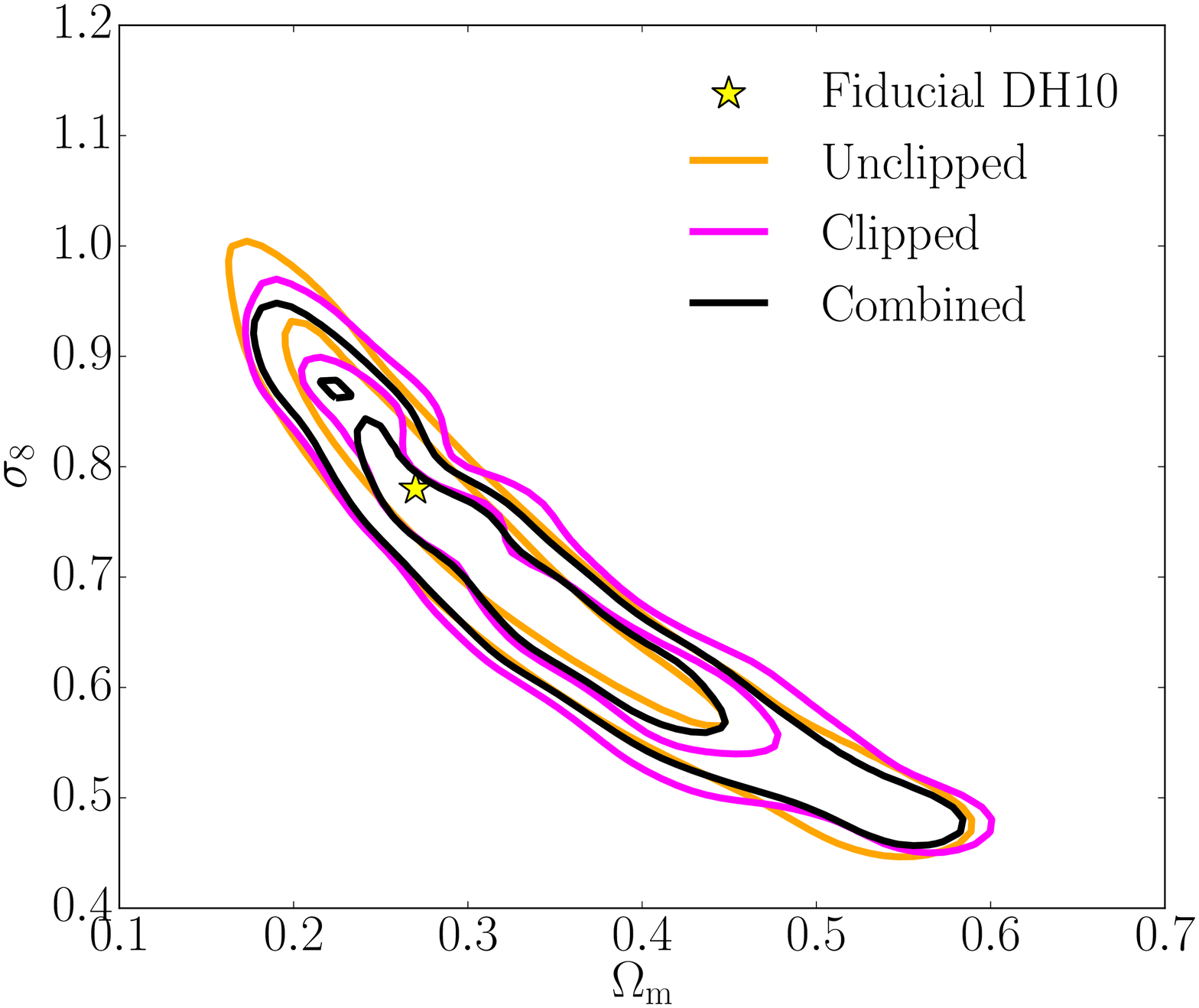}\\ 
\includegraphics[width=0.5\textwidth]{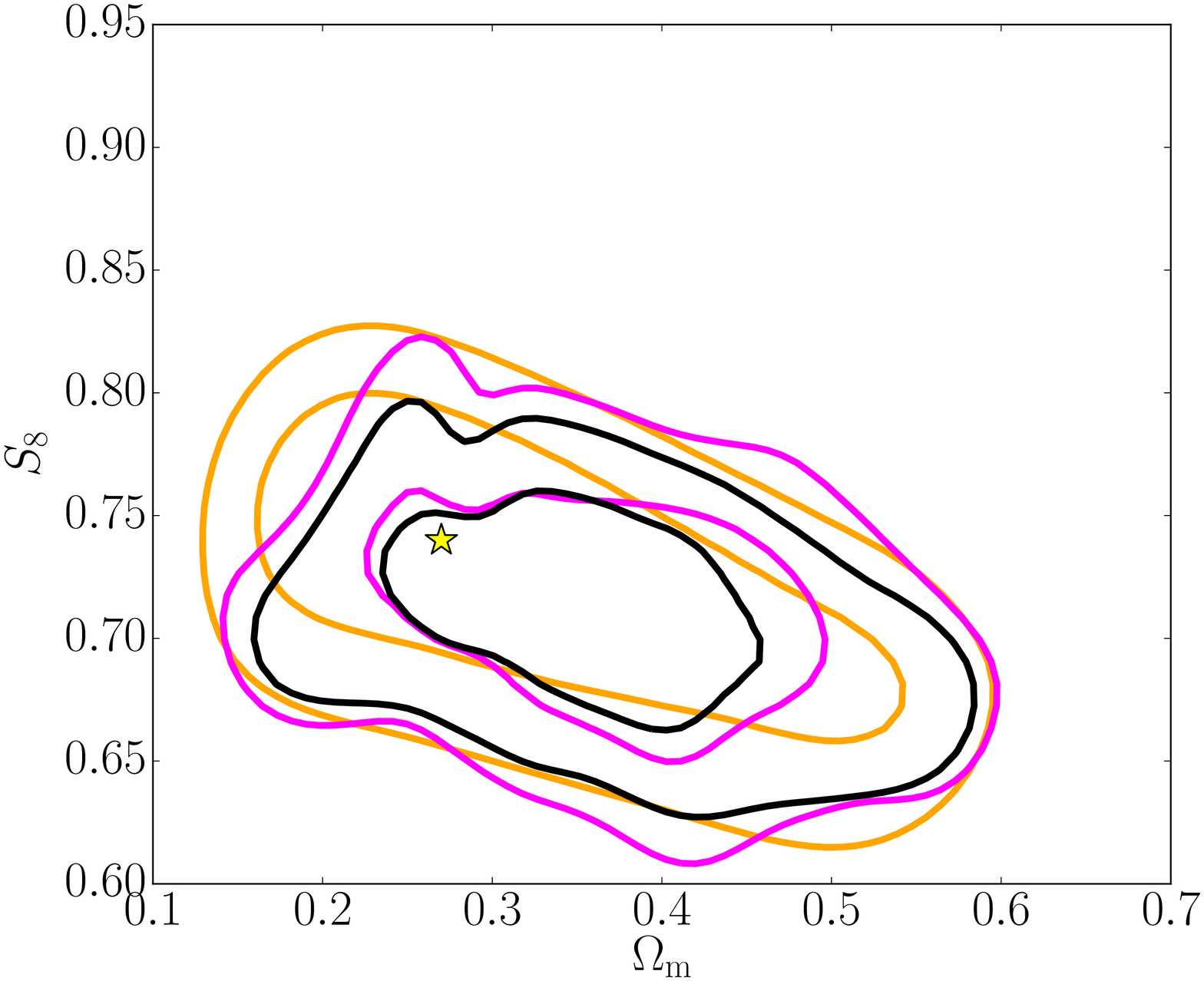}
\end{center}
\caption{The unclipped (orange), clipped (magenta) and combined (black) 68\% and 95\% confidence intervals for the fiducial cosmology from DH10 (shown by the yellow star) in the $\Omega_{\rm{m}}$-$\sigma_8$ and $\Omega_{\rm{m}}$-$S_8$ parameter spaces. We use only a subset of the fiducial cosmology simulations for the data vector in this test, corresponding to a KiDS-450-like survey of 360 deg$^2$. The unclipped contours are smooth as their theoretical expectation value is noise-free.  In contrast the clipped likelihood is interpolated across sparse measurements from DH10.  The resulting clipped and combined contours are therefore noisy in comparison to the unclipped constraints.} \label{fig:DH10_Constraints}
\end{figure}

\medskip

The upper panel of Figure \ref{fig:DH10_Constraints} shows the clipped (magenta), unclipped (orange) and combined (black) 68\% and 95\% constraints on the fiducial DH10 cosmology, in the $\Omega_{\rm{m}}$-$\sigma_8$ parameter space. The lower panel of this Figure shows the constraints in the $\Omega_{\rm{m}}$-$S_8$ parameter space, where $S_8=\sigma_8(\Omega_{\rm{m}}/0.3)^{0.5}$. We note first of all that we do not see the clockwise rotation of the clipped contours relative to the unclipped, predicted by S15. In answer to this, we remind the reader that this prediction was for a Euclid-like 5000 deg$^2$ survey, whereas our constraints correspond to a 360 deg$^2$ survey. It is possible that a rotation becomes evident given smaller error bars. If we were to scale the covariance on the data, $C_{\rm{data}}$, so as to correspond to a survey of Euclid-like proportions, the covariance on the clipped predictions from DH10, $C_{\rm{model}}$, becomes the dominant source of uncertainty. This prevents us from making a meaningful prediction for the cosmological constraints for a survey of this size. In the future, larger simulation suites will facilitate interesting predictions for larger size surveys.

\medskip

\begin{table}
  \begin{center}
    \caption{The marginalised means and 68\% confidence intervals on $S_8 = \sigma_8(\Omega_{\rm{m}}/0.3)^{0.5}$ for a subset of independent DH10 simulations with the fiducial cosmology spanning 360 deg$^2$. The improvements in the constraint over the unclipped are presented in bold to the nearest percentage. } \label{tab:Constraints_Table}
  \begin{tabular}{|cc|}
\hline
\hline
\textbf{Input} $\boldsymbol{S_8=0.740}$ & \\
Unclipped & $0.725 \pm 0.042$ \\
Clipped & $0.710 \pm 0.037$ $\boldsymbol{(11\%)}$ \\
Combined& $0.710 \pm 0.033$ $\boldsymbol{(22\%)}$ \\
\hline \hline
  \end{tabular}
  \end{center}
\end{table}

\medskip

The combined constraints shown in Figure \ref{fig:DH10_Constraints} recover the input cosmology, offering a significant improvement on the unclipped constraints. For example, the combined 95\% confidence intervals are 18\% and 29\% smaller in area than those of the unclipped, in the $\Omega_{\rm{m}}$-$\sigma_8$ and $\Omega_{\rm{m}}$-$S_8$ parameter spaces, respectively. In comparison, the clipped contours are of comparable size to the unclipped in either parameter space.

\medskip

Table \ref{tab:Constraints_Table} displays the marginalised mean and 68\% confidence intervals on $S_8$ from the clipped, unclipped and combined contours in the $\Omega_{\rm{m}}$-$S_8$ plane. The improvement in the size of the confidence intervals offered by the combined analysis relative to the unclipped is $22\%$. This improvement, which is not changed considerably by the corrections for the finite box effect and cosmological bias in DH10, is indicative of the independent information in the clipped and unclipped statistics. Indeed, this is evidenced by the cross-correlation coefficient matrices presented in Appendix \ref{sec:Appendix_Covariance}.

\medskip

The clipped analysis alone yields $S_8$ constraints which are 11\% tighter than the unclipped. For the clipped analysis to outperform the unclipped, the loss of information in clipping must be outweighed, by either the gain of phase information on the peaks, or the improvement in the clipped statistic for probing the more linear, clipped field. In this test however, we find that the success of the clipped analysis relative to the unclipped depends on the details of our interpolation scheme (see Appendix \ref{subsec:Appendix_Interp}). The combined analysis consistently outperforms the unclipped in constraining the cosmology of the DH10 dataset however, with all interpolations considered.

\medskip

\subsubsection{KiDS-450 constraints} \label{subsubsec:KiDS450_constraints}

\begin{figure}
\begin{center}
\includegraphics[width=0.5\textwidth]{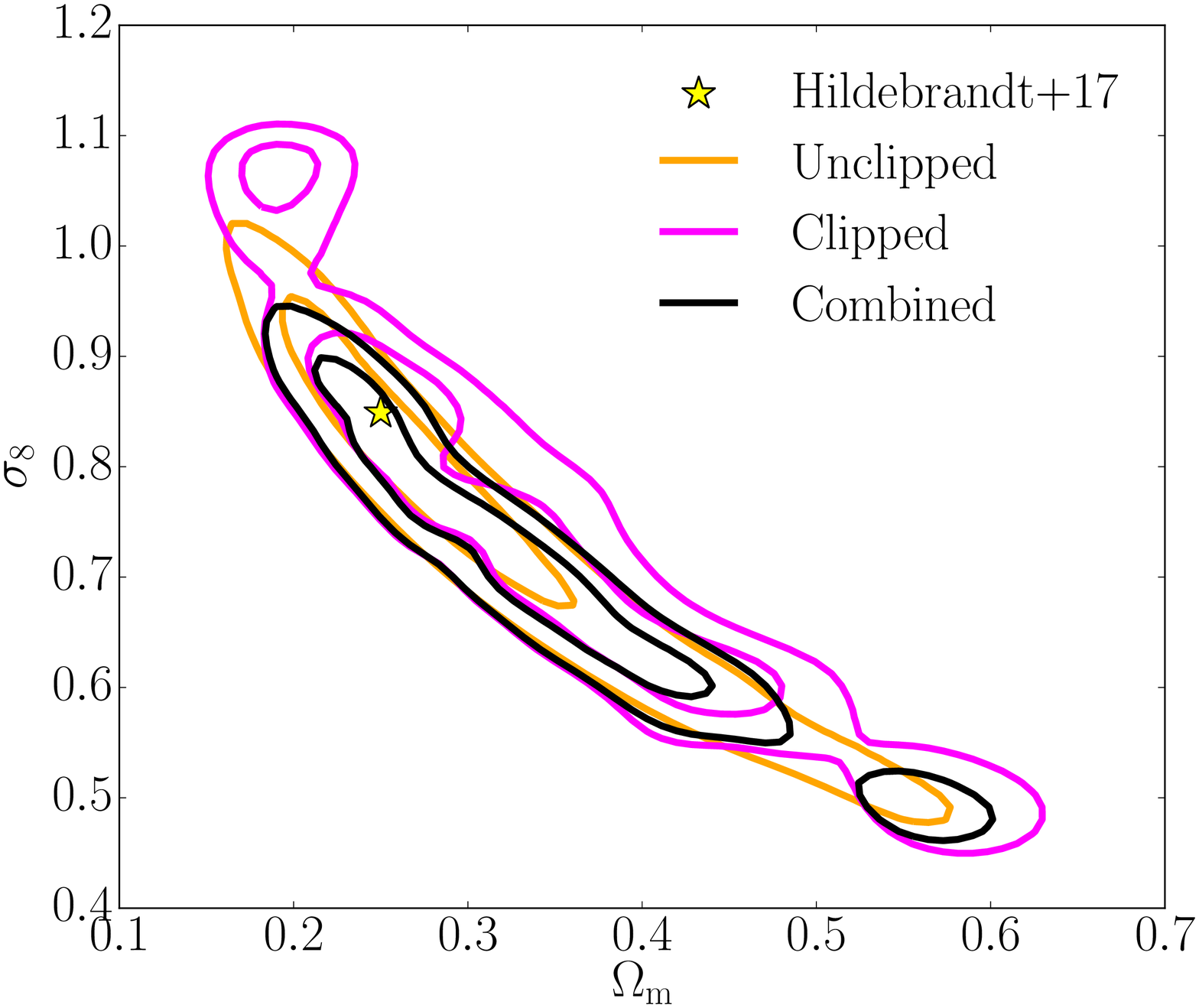}\\ 
\includegraphics[width=0.5\textwidth]{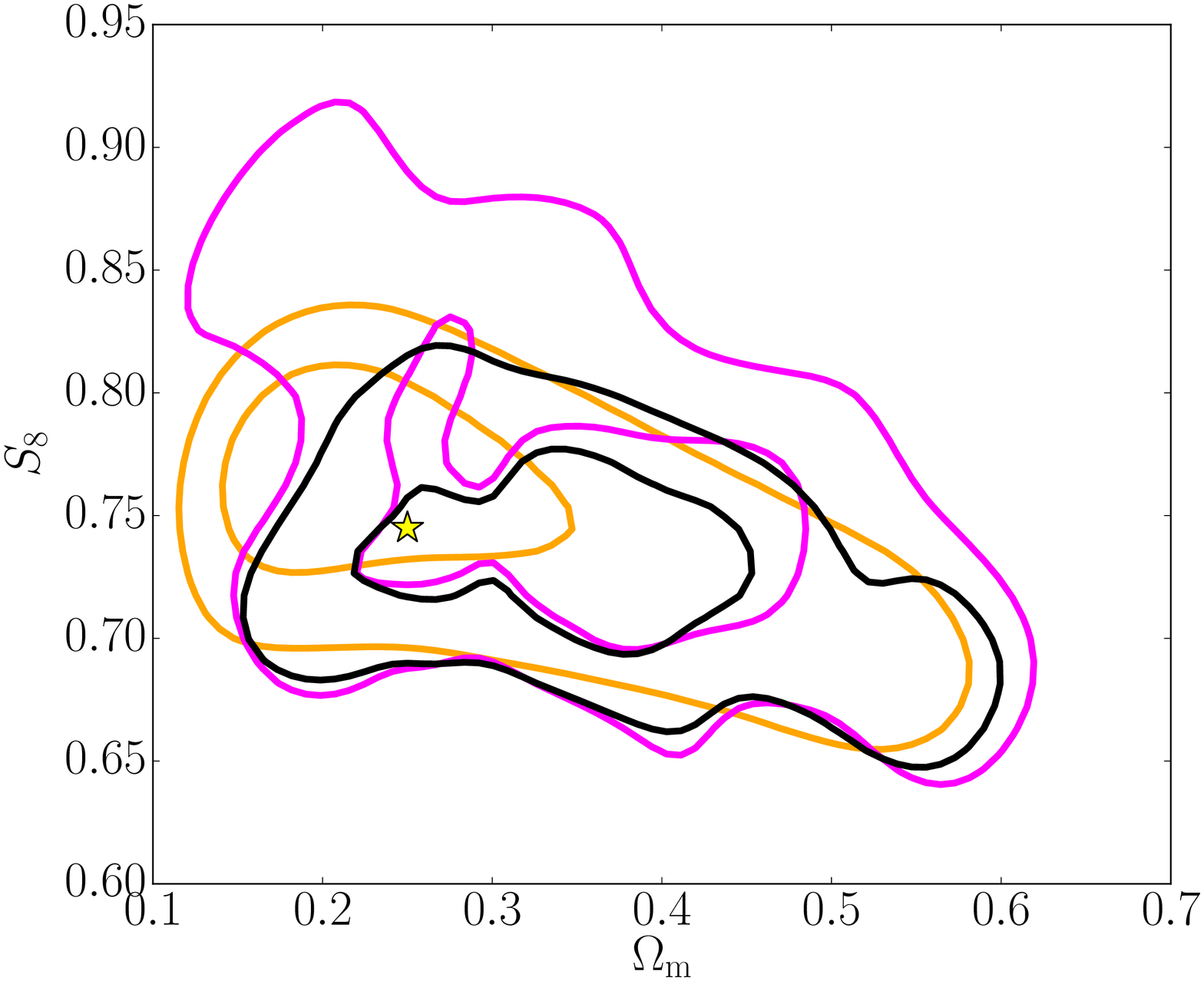}
\end{center}
\caption{The unclipped (orange), clipped (magenta) and combined (black) 68\% and 95\% confidence intervals for the KiDS-450 data in the $\Omega_{\rm{m}}$-$\sigma_8$ and $\Omega_{\rm{m}}$-$S_8$ parameter spaces. The yellow star depicts the best-fit cosmological parameters from the H17 cosmic shear analysis. The unclipped contours are smooth as their theoretical expectation value is noise-free.  In contrast the clipped likelihood is interpolated across sparse measurements from DH10.  The resulting clipped and combined contours are therefore noisy in comparison to the unclipped constraints.} \label{fig:KiDS450_Constraints}
\end{figure}

\medskip

After verifying that the combined clipped-and-unclipped analyses improve cosmological parameter constraints with a mock dataset, we proceed to constrain the cosmology of the KiDS-450 data. Figure \ref{fig:KiDS450_Constraints} displays the 68\% and 95\% confidence regions in the $\Omega_{\rm{m}}$-$\sigma_8$ and $\Omega_{\rm{m}}$-$S_8$ parameter spaces for this dataset. The best-fit cosmology from the H17 cosmic shear analysis is designated by the yellow star. Once again we have applied the finite box and cosmological bias calibration corrections to the clipped predictions from DH10. We have also interpolated from the DH10 cosmologies with radial basis functions, and applied the same degree of smoothing as in Figure \ref{fig:DH10_Constraints}. The slight discontinuities in the tails of the clipped and combined contours in the $\Omega_{\rm{m}}$-$\sigma_8$ space, seen also by \cite{Martinet_etal_2018} in their analysis involving the DH10 mocks, are a product of the sparsity of the simulated cosmologies, and disappear if we apply a greater degree of smoothing.

\medskip

In both the $\Omega_{\rm{m}}$-$\sigma_8$ and $\Omega_{\rm{m}}$-$S_8$ parameter spaces, we see that the clipped and combined contours are consistent with the best fit cosmological parameters from H17, despite the differences in the analyses. In addition to accounting for more systematics, the H17 result was obtained using four tomographic bins in the 0.1--0.9 photometric redshift range, as opposed to our single 0.5--0.9 bin. H17 also used both the $\xi_+^{\rm{unclip}}$ and $\xi_-^{\rm{unclip}}$, but omitted the largest two and smallest three $\theta$ bins for these statistics, respectively. In the $\Omega_{\rm{m}}$-$\sigma_8$ and $\Omega_{\rm{m}}$-$S_8$ parameter spaces shown in Figure \ref{fig:KiDS450_Constraints}, the 95\% confidence intervals from the combined analysis are about 13\% and 10\% smaller than the unclipped, respectively, whereas those of the clipped analysis are considerably larger. There are a number of extra sources of noise when working with the KiDS-450 data, which could cause the clipped contours to inflate relative to the unclipped, in contrast to what was observed when working with the DH10 data vector. These include galaxy shape measurement, baryonic effects and $n(z)$ uncertainties; this is discussed further in Appendix \ref{subsubsec:K450_constraints_Interp}.  

\medskip

The marginalised constraints on $S_8$ from the $\Omega_{\rm{m}}$-$S_8$ plane are shown in Table \ref{tab:Constraints_Table2}; again bold percentages detail improvements in the confidence intervals relative to the unclipped. As we saw with the DH10 data vector in Section \ref{subsubsec:DH10_constraints}, the combined analysis offers improvements on the unclipped constraint, by 17\%. This is comparable to the $\sim$20\% improvement in $S_8$ found by \cite{Martinet_etal_2018} when constraining the KiDS-450 cosmology with combined peak statistics and standard shear correlation functions.  

\medskip

\begin{table}
  \begin{center}
    \caption{The marginalised means and 68\% confidence intervals on $S_8 = \sigma_8(\Omega_{\rm{m}}/0.3)^{0.5}$ for KiDS-450. The improvement in the constraint over the unclipped are presented in bold to the nearest percentage. We remind the reader that the results of this work are not directly comparable to the H17 result, owing to the differences in the analyses discussed in the text. } \label{tab:Constraints_Table2}
  \begin{tabular}{|cc|}
\hline \hline
\textbf{H17} $\boldsymbol{S_8=0.745^{+0.038}_{-0.038}}$ & \\
Unclipped & $0.754 \pm 0.036$ \\
Clipped & $0.760 \pm 0.051$ \\
Combined& $0.734 \pm 0.030$ $\boldsymbol{(17\%)}$ \\ 
\hline  
\hline
  \end{tabular}
  \end{center}
\end{table}


\section{Conclusions} \label{sec:conclusions}

In this paper we have performed a proof-of-concept analysis demonstrating that clipping transformations, which suppress the contribution from overdense regions to the weak lensing signal, can be used alongside a conventional ``unclipped" cosmic shear analysis to improve cosmological parameter constraints. Our pipeline reconstructs the projected surface mass density, performs clipping, determines the shear corresponding to the overdensities, and obtains ``clipped" shear correlation functions. We have experimented with the threshold controlling the severity of the clipping transformation, and the smoothing employed in mass reconstruction, and found values well suited to the KiDS-450 dataset. 

\medskip

There is currently no analytical prediction for clipped statistics as a function of cosmology, and so we calibrate the clipped shear correlation functions with numerical simulations spanning a broad range of $\Omega_{\rm{m}}$ and $\sigma_8$. Consequently, we show that the combined clipped-and-unclipped analysis facilitates tighter constraints on $S_8=\sigma_8(\Omega_{\rm{m}}/0.3)^{0.5}$, at fixed values of $\Omega_{\rm{b}}, n_{\rm{s}}$ and $h$, than the conventional unclipped analysis alone. For a mock dataset with known cosmology, we find that the 68\% confidence interval on $S_8$ is improved upon the unclipped by 22\%. In the case of the KiDS-450 data, the improvement is 17\%. The combined constraints from clipping could improve further given optimisation for the clipping threshold and mass reconstruction smoothing scale, though we leave this for future work on account of the computational cost. 

\medskip

The DH10 mocks with the calibration corrections are sufficiently accurate for modelling in this work. However, the limitations of the mocks that we have examined here do impact the improvement reported for clipping and are likely to affect peak statistic studies also, reinforcing our conclusion that the success of these new statistics is intimately linked with the future accuracy and abundance of cosmological simulations. With new suites of simulations, the level of improvement seen in our joint analysis will increase in the future, as it will no longer be limited by the $\sim$7--100\% uncertainty that we currently include with DH10 predictions. We note that a joint analysis of peak counts, cosmic shear and clipping both peaks and voids, also poses an interesting topic for further investigation.

\medskip

Our best-fit $S_8=0.734\pm 0.030$ for the KiDS-450 data, inferred from a single photometric redshift bin in the range 0.5--0.9, is in good agreement with the tomographic cosmic shear analysis of H17, who found $S_8=0.745 \pm 0.038$. We note that H17 marginalised over $\Omega_{\rm{b}}, n_{\rm{s}}, h$ whereas our constraints are made at fixed values of these parameters. In the future, larger suites of numerical simulations will permit investigation of how clipped statistics vary with these cosmological parameters. H17 also marginalise over photometric redshift uncertainties, the effects of baryons and intrinsic alignments, which we have not contended with here. In order for clipping to become a standard tool for constraining cosmology, work must be done to fold these extra systematic uncertainties into the clipped analysis. Finally, mass reconstruction methods with a more sophisticated handling of the masks are needed to reduce the bias imposed by this essential part of the clipping pipeline. Nevertheless, the results obtained in this work robustly demonstrate that clipping improves constraining power and should be explored in future cosmic shear analyses.

\section*{Acknowledgements}

We thank Matthias Bartelmann for helpful comments on this analysis and for acting as the external blinder for the KiDS-450 data. We are also very grateful to Mike Jarvis for the {\sc{TreeCorr}} correlation function measurement software and Martin Kilbinger for the {\sc{Nicaea}} theoretical modelling software. We are thankful to Vasiliy Demchenko, Naomi Robertson, Shahab Joudaki, Laura Jean Wells and members of the KiDS consortium for useful discussions. We extend our gratitude to the referee for their comments which improved this paper. 

\medskip

BG, CH, JHD, AA, MA and TT acknowledge support from the European Research Council under grant number 647112. JHD also acknowledges support from the European Commission under a Marie-Sklodwoska-Curie European Fellowship
(EU project 656869). LvW is supported by NSERC of Canada. HH is supported by an Emmy Noether grant (No. Hi 1495/2-1) of the Deutsche Forschungsgemeinschaft. KK acknowledges support from the Alexander von Humboldt Foundation. PS acknowledges support from the Deutsche Forschungsgemeinschaft in the framework of the TR33 `The Dark Universe'. TT acknowledges funding from the European Union’s Horizon 2020 research and innovation programme under the Marie Sk{l}odowska-Curie grant agreement No 797794. 

\medskip

This KiDS data are based on data products from observations made with ESO Telescopes at the La Silla Paranal 
Observatory under programme IDs 177.A-3016, 177.A-3017 and 177.A-3018. Computations for the SLICS $N$-body simulations (now publicly available; \url{http://slics.roe.ac.uk}) were performed in part on the Orcinus supercomputer at the WestGrid HPC consortium (\url{www.westgrid.ca}),
in part on the GPC supercomputer at the SciNet HPC Consortium. SciNet is funded by: the Canada Foundation for Innovation under the auspices of Compute Canada;
the Government of Ontario; Ontario Research Fund - Research Excellence; and the University of Toronto. 

\medskip

{\footnotesize{\textit{Author Contributions:} All authors contributed to the development and writing of this paper. The authorship list is given in three groups: the lead authors (BG, CH, JHD, FS), followed by two alphabetical groups. The first alphabetical group includes those who are key contributors to both the scientific analysis and the data products. The second group covers those who have either made a significant contribution to the data products, or to the scientific analysis.}}

\bibliographystyle{mnras}
\bibliography{references}

\begin{thebibliography}{}
\makeatletter
\relax
\def\mn@urlcharsother{\let\do\@makeother \do\$\do\&\do\#\do\^\do\_\do\%\do\~}
\def\mn@doi{\begingroup\mn@urlcharsother \@ifnextchar [ {\mn@doi@}
  {\mn@doi@[]}}
\def\mn@doi@[#1]#2{\def\@tempa{#1}\ifx\@tempa\@empty \href
  {http://dx.doi.org/#2} {doi:#2}\else \href {http://dx.doi.org/#2} {#1}\fi
  \endgroup}
\def\mn@eprint#1#2{\mn@eprint@#1:#2::\@nil}
\def\mn@eprint@arXiv#1{\href {http://arxiv.org/abs/#1} {{\tt arXiv:#1}}}
\def\mn@eprint@dblp#1{\href {http://dblp.uni-trier.de/rec/bibtex/#1.xml}
  {dblp:#1}}
\def\mn@eprint@#1:#2:#3:#4\@nil{\def\@tempa {#1}\def\@tempb {#2}\def\@tempc
  {#3}\ifx \@tempc \@empty \let \@tempc \@tempb \let \@tempb \@tempa \fi \ifx
  \@tempb \@empty \def\@tempb {arXiv}\fi \@ifundefined
  {mn@eprint@\@tempb}{\@tempb:\@tempc}{\expandafter \expandafter \csname
  mn@eprint@\@tempb\endcsname \expandafter{\@tempc}}}

\bibitem[\protect\citeauthoryear{{Bartelmann} \& {Schneider}}{{Bartelmann} \&
  {Schneider}}{2001}]{Bartelmann_Schneider_2001}
{Bartelmann} M.,  {Schneider} P.,  2001, \mn@doi [\physrep]
  {10.1016/S0370-1573(00)00082-X}, \href
  {http://adsabs.harvard.edu/abs/2001PhR...340..291B} {340, 291}

\bibitem[\protect\citeauthoryear{{Ben{\'{\i}}tez}}{{Ben{\'{\i}}tez}}{2000}]{Benitez_2000}
{Ben{\'{\i}}tez} N.,  2000, \mn@doi [\apj] {10.1086/308947}, \href
  {http://adsabs.harvard.edu/abs/2000ApJ...536..571B} {536, 571}

\bibitem[\protect\citeauthoryear{{Benjamin} et~al.,}{{Benjamin}
  et~al.}{2013}]{Benjamin_et_al_2013}
{Benjamin} J.,  et~al., 2013, \mn@doi [\mnras] {10.1093/mnras/stt276}, \href
  {http://adsabs.harvard.edu/abs/2013MNRAS.431.1547B} {431, 1547}

\bibitem[\protect\citeauthoryear{{Bernardeau}}{{Bernardeau}}{2005}]{Bernardeau_2005}
{Bernardeau} F.,  2005, \mn@doi [\aap] {10.1051/0004-6361:20053440}, \href
  {http://adsabs.harvard.edu/abs/2005A%26A...441..873B} {441, 873}

\bibitem[\protect\citeauthoryear{{Bridle} \& {King}}{{Bridle} \&
  {King}}{2007}]{Bridle_King_2007}
{Bridle} S.,  {King} L.,  2007, \mn@doi [New Journal of Physics]
  {10.1088/1367-2630/9/12/444}, \href
  {http://adsabs.harvard.edu/abs/2007NJPh....9..444B} {9, 444}

\bibitem[\protect\citeauthoryear{{Chang} et~al.,}{{Chang}
  et~al.}{2017}]{Chang_etal_2017}
{Chang} C.,  et~al., 2017, preprint, \href
  {http://adsabs.harvard.edu/abs/2017arXiv170801535C} {} (\mn@eprint {arXiv}
  {1708.01535})

\bibitem[\protect\citeauthoryear{{DES Collaboration} et~al.,}{{DES
  Collaboration} et~al.}{2017}]{DES_Clustering_Cosmic_Shear_Yr1}
{DES Collaboration} et~al., 2017, preprint, \href
  {http://adsabs.harvard.edu/abs/2017arXiv170801530D} {} (\mn@eprint {arXiv}
  {1708.01530})

\bibitem[\protect\citeauthoryear{{Dietrich} \& {Hartlap}}{{Dietrich} \&
  {Hartlap}}{2010}]{Dietrich_Hartlap_2010}
{Dietrich} J.~P.,  {Hartlap} J.,  2010, \mn@doi [\mnras]
  {10.1111/j.1365-2966.2009.15948.x}, \href
  {http://adsabs.harvard.edu/abs/2010MNRAS.402.1049D} {402, 1049}

\bibitem[\protect\citeauthoryear{{Eifler}, {Schneider}  \& {Hartlap}}{{Eifler}
  et~al.}{2009}]{Eifler_etal_2009}
{Eifler} T.,  {Schneider} P.,   {Hartlap} J.,  2009, \mn@doi [\aap]
  {10.1051/0004-6361/200811276}, \href
  {http://adsabs.harvard.edu/abs/2009A%26A...502..721E} {502, 721}

\bibitem[\protect\citeauthoryear{{Eisenstein} \& {Hu}}{{Eisenstein} \&
  {Hu}}{1998}]{Eisenstein_Hu_1998}
{Eisenstein} D.~J.,  {Hu} W.,  1998, \mn@doi [\apj] {10.1086/305424}, \href
  {http://adsabs.harvard.edu/abs/1998ApJ...496..605E} {496, 605}

\bibitem[\protect\citeauthoryear{{Fenech Conti}, {Herbonnet}, {Hoekstra},
  {Merten}, {Miller}  \& {Viola}}{{Fenech Conti}
  et~al.}{2017}]{Fenech_Conti_et_al_2017}
{Fenech Conti} I.,  {Herbonnet} R.,  {Hoekstra} H.,  {Merten} J.,  {Miller} L.,
    {Viola} M.,  2017, \mn@doi [\mnras] {10.1093/mnras/stx200}, \href
  {http://adsabs.harvard.edu/abs/2017MNRAS.467.1627F} {467, 1627}

\bibitem[\protect\citeauthoryear{{Fu} et~al.,}{{Fu}
  et~al.}{2014}]{Fu_etal_2014}
{Fu} L.,  et~al., 2014, \mn@doi [\mnras] {10.1093/mnras/stu754}, \href
  {http://adsabs.harvard.edu/abs/2014MNRAS.441.2725F} {441, 2725}

\bibitem[\protect\citeauthoryear{{Harnois-D{\'e}raps} \& {van
  Waerbeke}}{{Harnois-D{\'e}raps} \& {van
  Waerbeke}}{2015}]{Harnois_Waerbeke_2015}
{Harnois-D{\'e}raps} J.,  {van Waerbeke} L.,  2015, \mn@doi [\mnras]
  {10.1093/mnras/stv794}, \href
  {http://adsabs.harvard.edu/abs/2015MNRAS.450.2857H} {450, 2857}

\bibitem[\protect\citeauthoryear{{Harnois-D{\'e}raps}, {Pen}, {Iliev}, {Merz},
  {Emberson}  \& {Desjacques}}{{Harnois-D{\'e}raps}
  et~al.}{2013}]{Harnois_etal_2013}
{Harnois-D{\'e}raps} J.,  {Pen} U.-L.,  {Iliev} I.~T.,  {Merz} H.,  {Emberson}
  J.~D.,   {Desjacques} V.,  2013, \mn@doi [\mnras] {10.1093/mnras/stt1591},
  \href {http://adsabs.harvard.edu/abs/2013MNRAS.436..540H} {436, 540}

\bibitem[\protect\citeauthoryear{{Harnois-D{\'e}raps}
  et~al.,}{{Harnois-D{\'e}raps} et~al.}{2018}]{Harnois_etal_2018}
{Harnois-D{\'e}raps} J.,  et~al., 2018, preprint, \href
  {http://adsabs.harvard.edu/abs/2018arXiv180504511H} {} (\mn@eprint {arXiv}
  {1805.04511})

\bibitem[\protect\citeauthoryear{{Hartlap}, {Simon}  \& {Schneider}}{{Hartlap}
  et~al.}{2007}]{Hartlap_etal_2007}
{Hartlap} J.,  {Simon} P.,   {Schneider} P.,  2007, \mn@doi [\aap]
  {10.1051/0004-6361:20066170}, \href
  {http://adsabs.harvard.edu/abs/2007A%26A...464..399H} {464, 399}

\bibitem[\protect\citeauthoryear{{Heitmann} et~al.,}{{Heitmann}
  et~al.}{2016}]{Heitmann_etal_2016}
{Heitmann} K.,  et~al., 2016, \mn@doi [\apj] {10.3847/0004-637X/820/2/108},
  \href {http://adsabs.harvard.edu/abs/2016ApJ...820..108H} {820, 108}

\bibitem[\protect\citeauthoryear{{Heymans} et~al.,}{{Heymans}
  et~al.}{2013}]{Heymans_etal_2013}
{Heymans} C.,  et~al., 2013, \mn@doi [\mnras] {10.1093/mnras/stt601}, \href
  {http://adsabs.harvard.edu/abs/2013MNRAS.432.2433H} {432, 2433}

\bibitem[\protect\citeauthoryear{{Hildebrandt} et~al.,}{{Hildebrandt}
  et~al.}{2012}]{Hildebrandt_et_al_2012}
{Hildebrandt} H.,  et~al., 2012, \mn@doi [\mnras]
  {10.1111/j.1365-2966.2012.20468.x}, \href
  {http://adsabs.harvard.edu/abs/2012MNRAS.421.2355H} {421, 2355}

\bibitem[\protect\citeauthoryear{{Hildebrandt} et~al.,}{{Hildebrandt}
  et~al.}{2017}]{Hildebrandtetal_2016}
{Hildebrandt} H.,  et~al., 2017, \mn@doi [\mnras] {10.1093/mnras/stw2805},
  \href {http://adsabs.harvard.edu/abs/2017MNRAS.465.1454H} {465, 1454}

\bibitem[\protect\citeauthoryear{{Hinshaw} et~al.,}{{Hinshaw}
  et~al.}{2013}]{Hinshaw_etal_2013}
{Hinshaw} G.,  et~al., 2013, \mn@doi [\apjs] {10.1088/0067-0049/208/2/19},
  \href {http://adsabs.harvard.edu/abs/2013ApJS..208...19H} {208, 19}

\bibitem[\protect\citeauthoryear{{Jain} \& {Van Waerbeke}}{{Jain} \& {Van
  Waerbeke}}{2000}]{Jain_LvW_2000}
{Jain} B.,  {Van Waerbeke} L.,  2000, \mn@doi [\apjl] {10.1086/312480}, \href
  {http://adsabs.harvard.edu/abs/2000ApJ...530L...1J} {530, L1}

\bibitem[\protect\citeauthoryear{{Jarvis}}{{Jarvis}}{2015}]{TreeCorr}
{Jarvis} M.,  2015, {TreeCorr: Two-point correlation functions}, Astrophysics
  Source Code Library (\mn@eprint {ascl} {1508.007})

\bibitem[\protect\citeauthoryear{{Joudaki} et~al.,}{{Joudaki}
  et~al.}{2017}]{Joudaki_etal_2016}
{Joudaki} S.,  et~al., 2017, \mn@doi [\mnras] {10.1093/mnras/stx998}, \href
  {http://adsabs.harvard.edu/abs/2017MNRAS.471.1259J} {471, 1259}

\bibitem[\protect\citeauthoryear{{Jullo}, {Pires}, {Jauzac}  \&
  {Kneib}}{{Jullo} et~al.}{2014}]{Jullo_etal_2014}
{Jullo} E.,  {Pires} S.,  {Jauzac} M.,   {Kneib} J.-P.,  2014, \mn@doi [\mnras]
  {10.1093/mnras/stt2207}, \href
  {http://adsabs.harvard.edu/abs/2014MNRAS.437.3969J} {437, 3969}

\bibitem[\protect\citeauthoryear{{Kacprzak} et~al.,}{{Kacprzak}
  et~al.}{2016}]{Kacprzak_etal_2016}
{Kacprzak} T.,  et~al., 2016, \mn@doi [\mnras] {10.1093/mnras/stw2070}, \href
  {http://adsabs.harvard.edu/abs/2016MNRAS.463.3653K} {463, 3653}

\bibitem[\protect\citeauthoryear{{Kaiser}}{{Kaiser}}{1992}]{Kaiser92}
{Kaiser} N.,  1992, \mn@doi [\apj] {10.1086/171151}, \href
  {http://adsabs.harvard.edu/abs/1992ApJ...388..272K} {388, 272}

\bibitem[\protect\citeauthoryear{{Kaiser} \& {Squires}}{{Kaiser} \&
  {Squires}}{1993}]{KSB}
{Kaiser} N.,  {Squires} G.,  1993, \mn@doi [\apj] {10.1086/172297}, \href
  {http://adsabs.harvard.edu/abs/1993ApJ...404..441K} {404, 441}

\bibitem[\protect\citeauthoryear{{Kilbinger} \& {Schneider}}{{Kilbinger} \&
  {Schneider}}{2005}]{Kilbinger_Schneider_2005}
{Kilbinger} M.,  {Schneider} P.,  2005, \mn@doi [\aap]
  {10.1051/0004-6361:20053531}, \href
  {http://adsabs.harvard.edu/abs/2005A%26A...442...69K} {442, 69}

\bibitem[\protect\citeauthoryear{{Kilbinger} et~al.,}{{Kilbinger}
  et~al.}{2009}]{Kilbinger_etal_2009}
{Kilbinger} M.,  et~al., 2009, \mn@doi [\aap] {10.1051/0004-6361/200811247},
  \href {http://adsabs.harvard.edu/abs/2009A%26A...497..677K} {497, 677}

\bibitem[\protect\citeauthoryear{{Kilbinger} et~al.,}{{Kilbinger}
  et~al.}{2017}]{Kilbinger_etal_2018}
{Kilbinger} M.,  et~al., 2017, \mn@doi [\mnras] {10.1093/mnras/stx2082}, \href
  {http://adsabs.harvard.edu/abs/2017MNRAS.472.2126K} {472, 2126}

\bibitem[\protect\citeauthoryear{{K{\"o}hlinger} et~al.,}{{K{\"o}hlinger}
  et~al.}{2017}]{Kohlinger_etal_2017}
{K{\"o}hlinger} F.,  et~al., 2017, preprint, \href
  {http://adsabs.harvard.edu/abs/2017arXiv170602892K} {} (\mn@eprint {arXiv}
  {1706.02892})

\bibitem[\protect\citeauthoryear{{Kuijken} et~al.,}{{Kuijken}
  et~al.}{2015}]{Kuijken_etal_2015}
{Kuijken} K.,  et~al., 2015, \mn@doi [\mnras] {10.1093/mnras/stv2140}, \href
  {http://adsabs.harvard.edu/abs/2015MNRAS.454.3500K} {454, 3500}

\bibitem[\protect\citeauthoryear{{Lewis}, {Challinor}  \& {Lasenby}}{{Lewis}
  et~al.}{2000}]{Lewis_etal_2000}
{Lewis} A.,  {Challinor} A.,   {Lasenby} A.,  2000, \mn@doi [\apj]
  {10.1086/309179}, \href {http://adsabs.harvard.edu/abs/2000ApJ...538..473L}
  {538, 473}

\bibitem[\protect\citeauthoryear{{Lima}, {Cunha}, {Oyaizu}, {Frieman}, {Lin}
  \& {Sheldon}}{{Lima} et~al.}{2008}]{Lima_etal_2008}
{Lima} M.,  {Cunha} C.~E.,  {Oyaizu} H.,  {Frieman} J.,  {Lin} H.,   {Sheldon}
  E.~S.,  2008, \mn@doi [\mnras] {10.1111/j.1365-2966.2008.13510.x}, \href
  {http://adsabs.harvard.edu/abs/2008MNRAS.390..118L} {390, 118}

\bibitem[\protect\citeauthoryear{{Liu}, {Wang}, {Pan}  \& {Fan}}{{Liu}
  et~al.}{2014}]{Liu_etal_2014}
{Liu} X.,  {Wang} Q.,  {Pan} C.,   {Fan} Z.,  2014, \mn@doi [\apj]
  {10.1088/0004-637X/784/1/31}, \href
  {http://adsabs.harvard.edu/abs/2014ApJ...784...31L} {784, 31}

\bibitem[\protect\citeauthoryear{{Lombriser}, {Simpson}  \& {Mead}}{{Lombriser}
  et~al.}{2015}]{Lombriser_etal_2015}
{Lombriser} L.,  {Simpson} F.,   {Mead} A.,  2015, \mn@doi [Physical Review
  Letters] {10.1103/PhysRevLett.114.251101}, \href
  {http://adsabs.harvard.edu/abs/2015PhRvL.114y1101L} {114, 251101}

\bibitem[\protect\citeauthoryear{{Martinet} et~al.,}{{Martinet}
  et~al.}{2018}]{Martinet_etal_2018}
{Martinet} N.,  et~al., 2018, \mn@doi [\mnras] {10.1093/mnras/stx2793}, \href
  {http://adsabs.harvard.edu/abs/2018MNRAS.474..712M} {474, 712}

\bibitem[\protect\citeauthoryear{{Miller} et~al.,}{{Miller}
  et~al.}{2013}]{Miller_et_al_2013}
{Miller} L.,  et~al., 2013, \mn@doi [\mnras] {10.1093/mnras/sts454}, \href
  {http://adsabs.harvard.edu/abs/2013MNRAS.429.2858M} {429, 2858}

\bibitem[\protect\citeauthoryear{Neyrinck, Szapudi  \& Szalay}{Neyrinck
  et~al.}{2009}]{Neyrinck_etal_2009}
Neyrinck M.~C.,  Szapudi I.,   Szalay A.~S.,  2009, The Astrophysical Journal
  Letters, 698, L90

\bibitem[\protect\citeauthoryear{{Planck Collaboration} et~al.,}{{Planck
  Collaboration} et~al.}{2018}]{Planck_2018}
{Planck Collaboration} et~al., 2018, preprint, \href
  {http://adsabs.harvard.edu/abs/2018arXiv180706209P} {} (\mn@eprint {arXiv}
  {1807.06209})

\bibitem[\protect\citeauthoryear{{Schneider}, {van Waerbeke}  \&
  {Mellier}}{{Schneider} et~al.}{2002a}]{Schneider_etal_2002b}
{Schneider} P.,  {van Waerbeke} L.,   {Mellier} Y.,  2002a, \mn@doi [\aap]
  {10.1051/0004-6361:20020626}, \href
  {http://adsabs.harvard.edu/abs/2002A%26A...389..729S} {389, 729}

\bibitem[\protect\citeauthoryear{{Schneider}, {van Waerbeke}, {Kilbinger}  \&
  {Mellier}}{{Schneider} et~al.}{2002b}]{Schneider_et_al_2002}
{Schneider} P.,  {van Waerbeke} L.,  {Kilbinger} M.,   {Mellier} Y.,  2002b,
  \mn@doi [\aap] {10.1051/0004-6361:20021341}, \href
  {http://cdsads.u-strasbg.fr/abs/2002A%26A...396....1S} {396, 1}

\bibitem[\protect\citeauthoryear{{Seitz} \& {Schneider}}{{Seitz} \&
  {Schneider}}{1996}]{Seitz_Schneider_1996}
{Seitz} S.,  {Schneider} P.,  1996, \aap, \href
  {http://adsabs.harvard.edu/abs/1996A%26A...305..383S} {305, 383}

\bibitem[\protect\citeauthoryear{{Sellentin} \& {Heavens}}{{Sellentin} \&
  {Heavens}}{2016}]{Sellentin_Heavens_2016}
{Sellentin} E.,  {Heavens} A.~F.,  2016, \mn@doi [\mnras]
  {10.1093/mnrasl/slv190}, \href
  {http://adsabs.harvard.edu/abs/2016MNRAS.456L.132S} {456, L132}

\bibitem[\protect\citeauthoryear{{Sellentin}, {Heymans}  \&
  {Harnois-D{\'e}raps}}{{Sellentin} et~al.}{2017}]{Sellentin_etal_2017}
{Sellentin} E.,  {Heymans} C.,   {Harnois-D{\'e}raps} J.,  2017, preprint,
  \href {http://adsabs.harvard.edu/abs/2017arXiv171204923S} {} (\mn@eprint
  {arXiv} {1712.04923})

\bibitem[\protect\citeauthoryear{{Semboloni}, {Schrabback}, {van Waerbeke},
  {Vafaei}, {Hartlap}  \& {Hilbert}}{{Semboloni}
  et~al.}{2011a}]{Semboloni_etal_2011}
{Semboloni} E.,  {Schrabback} T.,  {van Waerbeke} L.,  {Vafaei} S.,  {Hartlap}
  J.,   {Hilbert} S.,  2011a, \mn@doi [\mnras]
  {10.1111/j.1365-2966.2010.17430.x}, \href
  {http://adsabs.harvard.edu/abs/2011MNRAS.410..143S} {410, 143}

\bibitem[\protect\citeauthoryear{{Semboloni}, {Hoekstra}, {Schaye}, {van
  Daalen}  \& {McCarthy}}{{Semboloni} et~al.}{2011b}]{Semboloni_etal_2011b}
{Semboloni} E.,  {Hoekstra} H.,  {Schaye} J.,  {van Daalen} M.~P.,   {McCarthy}
  I.~G.,  2011b, \mn@doi [\mnras] {10.1111/j.1365-2966.2011.19385.x}, \href
  {http://adsabs.harvard.edu/abs/2011MNRAS.417.2020S} {417, 2020}

\bibitem[\protect\citeauthoryear{{Seo}, {Sato}, {Dodelson}, {Jain}  \&
  {Takada}}{{Seo} et~al.}{2011}]{Seo_et_al_2011}
{Seo} H.-J.,  {Sato} M.,  {Dodelson} S.,  {Jain} B.,   {Takada} M.,  2011,
  \mn@doi [\apjl] {10.1088/2041-8205/729/1/L11}, \href
  {http://adsabs.harvard.edu/abs/2011ApJ...729L..11S} {729, L11}

\bibitem[\protect\citeauthoryear{{Simpson}, {James}, {Heavens}  \&
  {Heymans}}{{Simpson} et~al.}{2011}]{Simpson_etal_2011}
{Simpson} F.,  {James} J.~B.,  {Heavens} A.~F.,   {Heymans} C.,  2011, \mn@doi
  [Physical Review Letters] {10.1103/PhysRevLett.107.271301}, \href
  {http://adsabs.harvard.edu/abs/2011PhRvL.107A1301S} {107, 271301}

\bibitem[\protect\citeauthoryear{{Simpson}, {Heavens}  \& {Heymans}}{{Simpson}
  et~al.}{2013}]{Simpson_etal_2013}
{Simpson} F.,  {Heavens} A.~F.,   {Heymans} C.,  2013, \mn@doi [\prd]
  {10.1103/PhysRevD.88.083510}, \href
  {http://adsabs.harvard.edu/abs/2013PhRvD..88h3510S} {88, 083510}

\bibitem[\protect\citeauthoryear{{Simpson} et~al.,}{{Simpson}
  et~al.}{2016a}]{Simpson_etal_2016}
{Simpson} F.,  et~al., 2016a, \mn@doi [\prd] {10.1103/PhysRevD.93.023525},
  \href {http://adsabs.harvard.edu/abs/2016PhRvD..93b3525S} {93, 023525}

\bibitem[\protect\citeauthoryear{{Simpson}, {Harnois-D{\'e}raps}, {Heymans},
  {Jimenez}, {Joachimi}  \& {Verde}}{{Simpson}
  et~al.}{2016b}]{Simpson_et_al_2015}
{Simpson} F.,  {Harnois-D{\'e}raps} J.,  {Heymans} C.,  {Jimenez} R.,
  {Joachimi} B.,   {Verde} L.,  2016b, \mn@doi [\mnras]
  {10.1093/mnras/stv2474}, \href
  {http://adsabs.harvard.edu/abs/2016MNRAS.456..278S} {456, 278}

\bibitem[\protect\citeauthoryear{{Smith} et~al.,}{{Smith}
  et~al.}{2003}]{Smith_etal_2003}
{Smith} R.~E.,  et~al., 2003, \mn@doi [\mnras]
  {10.1046/j.1365-8711.2003.06503.x}, \href
  {http://adsabs.harvard.edu/abs/2003MNRAS.341.1311S} {341, 1311}

\bibitem[\protect\citeauthoryear{{Springel}}{{Springel}}{2005}]{Springel2005}
{Springel} V.,  2005, \mn@doi [\mnras] {10.1111/j.1365-2966.2005.09655.x},
  \href {http://adsabs.harvard.edu/abs/2005MNRAS.364.1105S} {364, 1105}

\bibitem[\protect\citeauthoryear{{Takada} \& {Jain}}{{Takada} \&
  {Jain}}{2002}]{Takada_Jain_2002}
{Takada} M.,  {Jain} B.,  2002, \mn@doi [\mnras]
  {10.1046/j.1365-8711.2002.05972.x}, \href
  {http://adsabs.harvard.edu/abs/2002MNRAS.337..875T} {337, 875}

\bibitem[\protect\citeauthoryear{{Takahashi}, {Sato}, {Nishimichi}, {Taruya}
  \& {Oguri}}{{Takahashi} et~al.}{2012}]{Takahashi_etal_2012}
{Takahashi} R.,  {Sato} M.,  {Nishimichi} T.,  {Taruya} A.,   {Oguri} M.,
  2012, \mn@doi [\apj] {10.1088/0004-637X/761/2/152}, \href
  {http://adsabs.harvard.edu/abs/2012ApJ...761..152T} {761, 152}

\bibitem[\protect\citeauthoryear{{Troxel} et~al.,}{{Troxel}
  et~al.}{2017}]{DES_Cosmic_Shear_Yr1}
{Troxel} M.~A.,  et~al., 2017, preprint, \href
  {http://adsabs.harvard.edu/abs/2017arXiv170801538T} {} (\mn@eprint {arXiv}
  {1708.01538})

\bibitem[\protect\citeauthoryear{{Troxel} et~al.,}{{Troxel}
  et~al.}{2018}]{Troxel_etal_2018}
{Troxel} M.~A.,  et~al., 2018, preprint, \href
  {http://adsabs.harvard.edu/abs/2018arXiv180410663T} {} (\mn@eprint {arXiv}
  {1804.10663})

\bibitem[\protect\citeauthoryear{{Van Waerbeke} et~al.,}{{Van Waerbeke}
  et~al.}{2013}]{vanWaerbeke_et_al_2013}
{Van Waerbeke} L.,  et~al., 2013, \mn@doi [\mnras] {10.1093/mnras/stt971},
  \href {http://adsabs.harvard.edu/abs/2013MNRAS.433.3373V} {433, 3373}

\bibitem[\protect\citeauthoryear{{VanderPlas}, {Connolly}, {Jain}  \&
  {Jarvis}}{{VanderPlas} et~al.}{2012}]{VanderPlas_etal_2012}
{VanderPlas} J.~T.,  {Connolly} A.~J.,  {Jain} B.,   {Jarvis} M.,  2012,
  \mn@doi [\apj] {10.1088/0004-637X/744/2/180}, \href
  {http://adsabs.harvard.edu/abs/2012ApJ...744..180V} {744, 180}

\bibitem[\protect\citeauthoryear{{Wilson}}{{Wilson}}{2016}]{Wilson_2016}
{Wilson} M.~J.,  2016, preprint, \href
  {http://adsabs.harvard.edu/abs/2016arXiv161008362W} {} (\mn@eprint {arXiv}
  {1610.08362})

\bibitem[\protect\citeauthoryear{{Yoon}, {Jee}, {Tyson}, {Schmidt}, {Wittman}
  \& {Choi}}{{Yoon} et~al.}{2018}]{Yoon_etal_2018}
{Yoon} M.,  {Jee} M.~J.,  {Tyson} J.~A.,  {Schmidt} S.,  {Wittman} D.,   {Choi}
  A.,  2018, preprint, \href
  {http://adsabs.harvard.edu/abs/2018arXiv180709195Y} {} (\mn@eprint {arXiv}
  {1807.09195})

\bibitem[\protect\citeauthoryear{{de Jong} et~al.,}{{de Jong}
  et~al.}{2017}]{deJong_etal_2017}
{de Jong} J.~T.~A.,  et~al., 2017, \mn@doi [\aap]
  {10.1051/0004-6361/201730747}, \href
  {http://adsabs.harvard.edu/abs/2017A%26A...604A.134D} {604, A134}

\bibitem[\protect\citeauthoryear{{van Uitert} et~al.,}{{van Uitert}
  et~al.}{2018}]{vanUitert_etal_2017}
{van Uitert} E.,  et~al., 2018, \mn@doi [\mnras] {10.1093/mnras/sty551}, \href
  {http://adsabs.harvard.edu/abs/2018MNRAS.476.4662V} {476, 4662}

\makeatother
\end{thebibliography}

\medskip

\appendix

\section{SLICS covariance matrices} \label{sec:Appendix_Covariance}

Our likelihood analysis for cosmological parameters necessitates auto-covariance matrices for the clipped and unclipped statistics, as well as the cross-covariance between the two. Non-Gaussianity in cosmological density fields engenders correlations between the different angular scales probed by these measurements which are not well described by theory. Therefore we use the SLICS numerical simulations to model these covariance matrices. From the SLICS covariance matrices, defined in equation \ref{eqn:CovMat}, we calculate correlation coefficient matrices, defined as

\begin{equation} \label{eqn:CCC_Mat}
\mathcal{CC}_\pm(\theta_i,\theta_j) = \frac{C_\pm(\theta_i,\theta_j)}{\sqrt{C_\pm(\theta_i,\theta_i) \times C_\pm(\theta_j,\theta_j)}} \,,
\end{equation}   

\noindent where $C_\pm(\theta_i,\theta_j)$ represents either the auto-covariance matrices for the $\xi_\pm^{\rm{unclip}}$ or $\xi_\pm^{\rm{clip}}$ statistics, or the cross-covariance matrix between the clipped and unclipped statistics. In the correlation coefficient matrix, the covariance is normalised to a value of unity for the strongest positive correlations on the leading diagonal, and values between -1 and 1 for all other elements. 

\medskip

\begin{figure}
\centering
\includegraphics[width=0.5\textwidth]{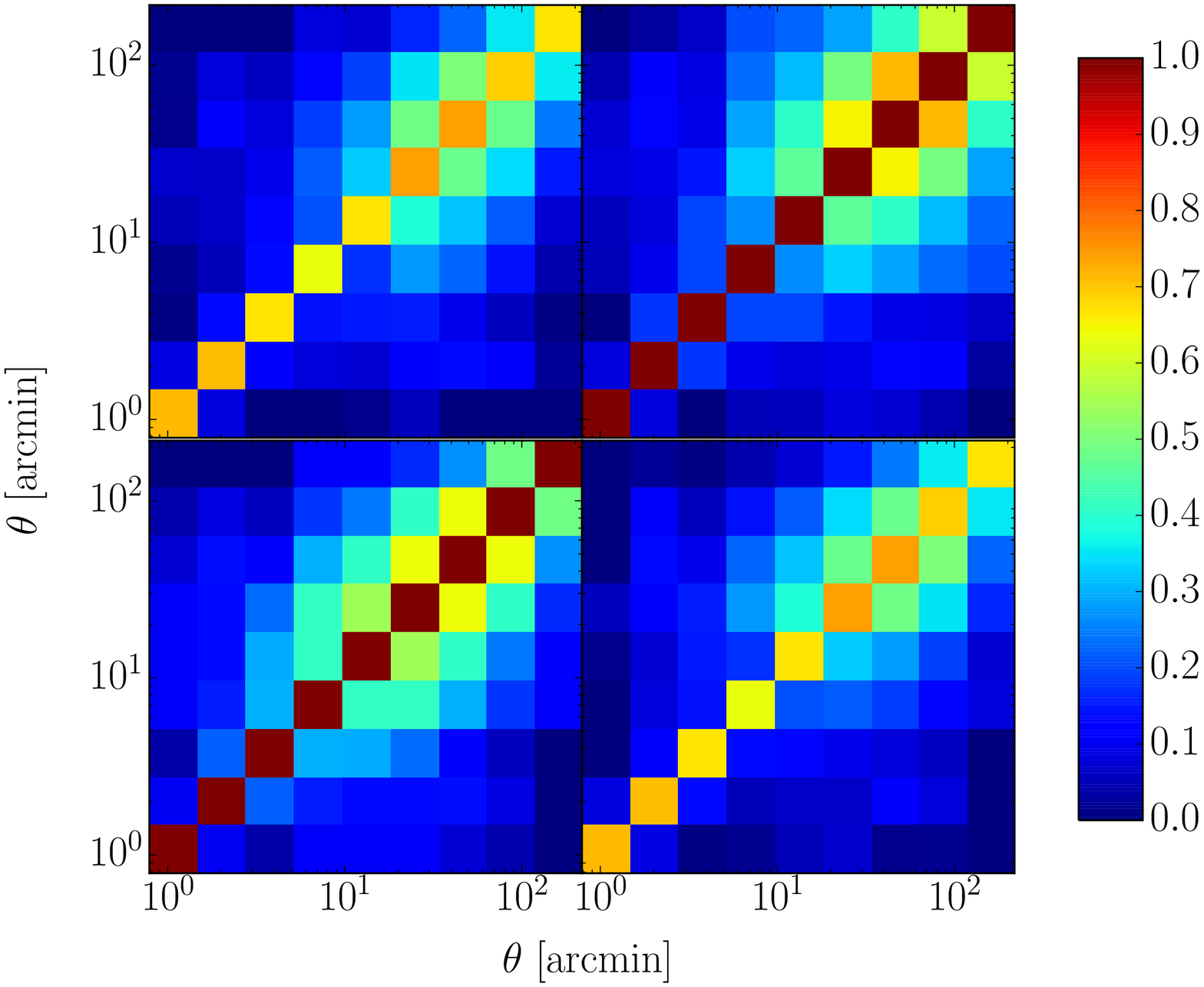} \\
\includegraphics[width=0.5\textwidth]{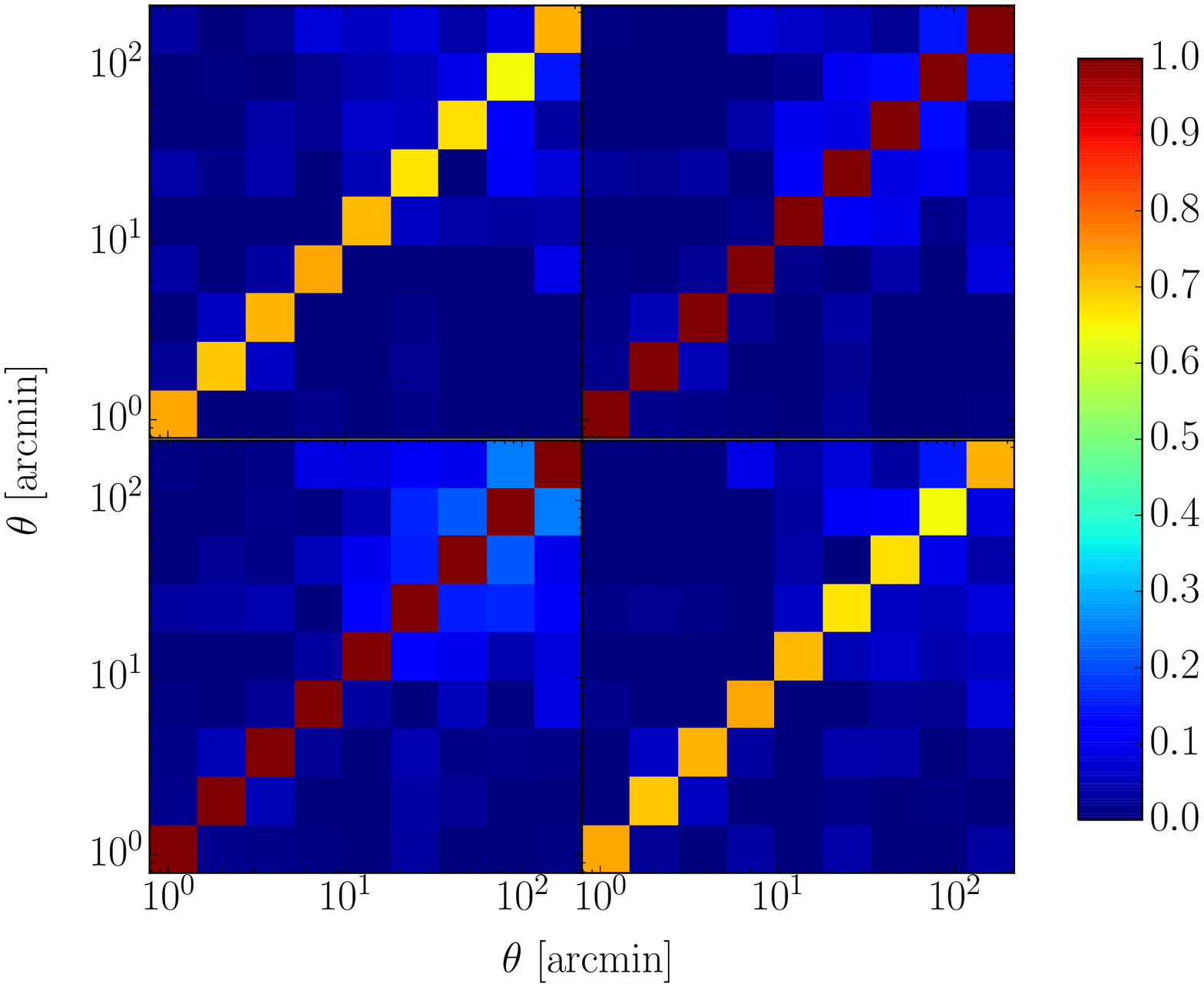} \\  
\caption{The correlation coefficient matrices measured from SLICS (featuring shape noise typical of KiDS-450) for the clipped and unclipped $\xi_+$ \textit{(upper panel)} and $\xi_-$ \textit{(lower panel)}. Each panel consists of the following components. \textit{Lower left:} the auto-correlations for the $\xi_\pm^{\rm{unclip}}$. \textit{Upper right:} the auto-correlations for the $\xi_\pm^{\rm{clip}}$. \textit{Upper-left} (and \textit{lower-right)}: the cross-correlations between $\xi_\pm^{\rm{unclip}}$ and $\xi_\pm^{\rm{clip}}$ (and its transpose).} \label{fig:CCCMat} 
\end{figure} 

\medskip

In Figure \ref{fig:CCCMat}, we display correlation coefficient matrices for the clipped and unclipped $\xi_+$ in the upper panel, and for the $\xi_-$ in the lower panel. Each of these matrices are built out of the following components. The auto-correlation coefficient matrix for the unclipped statistic is in the lower-left corner, the clipped is in the upper-right corner, the matrix describing the cross-correlation coefficients between these two statistics is in the upper-left, and its transpose is in the lower-right. 

\medskip

The fact that many of the off-diagonal elements of these matrices are non-zero (varying in the range -0.1 to 0.8 in either panel), indicates the need for simulations in order to model the correlations not only across angular scales, but also the correlations between the clipped and unclipped statistics. The cross-correlation matrices reveal that the clipped and unclipped statistics are not perfectly correlated and thus contain some independent information. It is also interesting to note that the clipped auto-correlation matrices seem to feature slightly weaker correlations between scales around $\sim$10 arcmin and $\sim$100 arcmin in the upper and lower panels respectively, than the unclipped auto-correlation matrices. This is consistent with the clipped field being more Gaussian than the unclipped. We note that the correlation between the clipped and unclipped measurements does not tend to unity on the largest scales probed in this analysis.  This is a reflection of that fact that the largest-scale clipped and unclipped measurements for our fiducial analysis also do not converge.   We find that for a less aggressive clipping threshold, (see the upper panel in Figure \ref{fig:BiggerPlot}), both the cross-correlation coefficients and the ratio between the clipped and unclipped measurements do however converge to unity as expected. For our fiducial set-up we would expect perfect correlation between the clipped and unclipped signals to occur on scales that are larger than we can currently test with the SLICS or KiDS-450 survey area.

\section{Cosmological constraints} \label{sec:Appendix_Constraints}

\subsection{Sensitivity to the unclipped predictions} \label{subsec:UCPredictions_Appendix}

In Section \ref{subsubsec:DH10_constraints}, we use the theoretical $\xi_+^{\rm{unclip}}$ from equation \ref{eqn:xi+-theory} to constrain the cosmology of the subset of DH10 simulations with the fiducial cosmology spanning 360 deg$^2$. We could alternatively have used the unclipped predictions from the simulations themselves, though these predictions are subject to the finite box effect, cosmological bias and additional uncertainty, as discussed in Sections \ref{subsec:DH10_finite_box}, \ref{subsec:DH10_Bias} and \ref{subsec:Cosmol_Depend} respectively. The noise-free theoretical predictions (e.g. from {\sc{Nicaea}}) are a more suitable choice for constraining cosmology where such predictions are available (which is of course not so, in the case of the clipped statistic). Nevertheless, we verify that one still obtains improved cosmological constraints in the combined analysis irrespective of whether we employ the theoretical or simulated $\xi_+^{\rm{unclip}}$.

\medskip

\begin{figure}
\begin{center}
\includegraphics[width=0.5\textwidth]{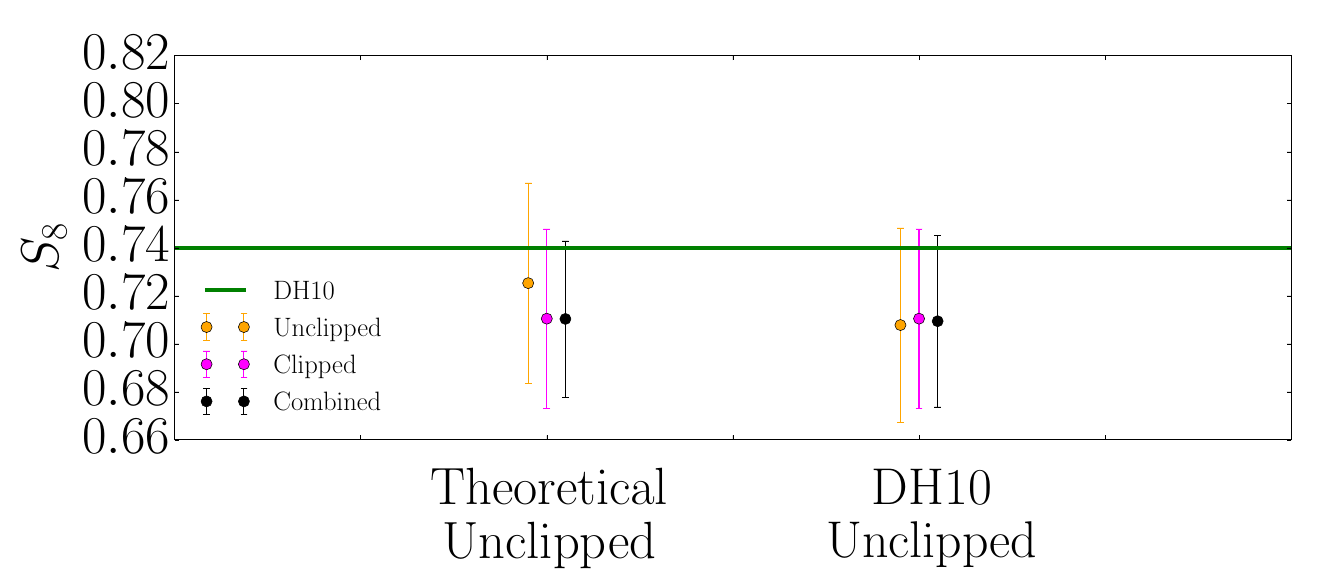}
\caption{The marginalised means and 68\% confidence intervals on $S_8$ from the $\Omega_{\rm{m}}$-$S_8$ plane for the DH10 fiducial cosmology data vector, depending on whether the $\xi_+^{\rm{unclip}}$ derive from equation \ref{eqn:xi+-theory} or from the DH10 mocks themselves. The input $S_8$ is designated by the horizontal green line. The corrections for the finite box size and cosmological bias have been applied to the predictions from DH10.} \label{fig:UCPredictions_DH10}
\end{center}
\end{figure}

Figure \ref{fig:UCPredictions_DH10} compares the marginalised means and 68\% confidence intervals on $S_8$ from the $\Omega_{\rm{m}}$-$S_8$ parameter space when we use the unclipped predictions from equation \ref{eqn:xi+-theory} and from DH10. These constraints are clearly consistent with one another and the input $S_8$, but differ in their details, as is shown in Table \ref{tab:UCPredictions_DH10}. The theoretical unclipped better recovers the input $S_8$ indicating again that they should be used over DH10 whenever possible. One consequence of this choice however, is that we find it leads to a $\sim 0.4 \sigma$ difference in the mean marginalised constraints on $S_8$ when comparing the clipped and unclipped analyses in Table \ref{tab:UCPredictions_DH10}.  Given the high correlation between these two statistics, shown in Figure \ref{fig:CCCMat}, we would expect better agreement, which we indeed find when using the DH10 measurements for both the clipped and unclipped predictions.  In this case the mean $S_8$ agree to within $0.05 \sigma$.   When using DH10 for both the clipped and unclipped predictions, our finding that the combined clipped-and-unclipped analyses improves cosmological parameter constraints holds but in this case the level of improved constraining power decreases to $12\%$. We find these conclusions are robust to different realisations of the DH10 data vector. 

\medskip

\begin{table}
  \begin{center}
    \caption{The marginalised means and 68\% confidence intervals on the DH10 data vector from Figure \ref{fig:UCPredictions_DH10} expressed in tabular form. Improvements over the unclipped confidence intervals are detailed in bold to the nearest percentage. The corrections for the finite box size and cosmological bias have been applied to the predictions from DH10.} \label{tab:UCPredictions_DH10}
  \begin{tabular}{|ccc|}
  & \multicolumn{1}{|c|}{Theoretical Unclipped} & \multicolumn{1}{|c|}{DH10 Unclipped}  \\  
\hline \hline
\textbf{Input} $\boldsymbol{S_8=0.740}$ & &\\
Unclipped & $0.725 \pm 0.042$  & $0.708 \pm 0.041$ \\
Clipped & $0.710 \pm 0.037$  $\boldsymbol{(11\%)}$ & $0.710 \pm 0.037$  $\boldsymbol{(8\%)}$\\
Combined & $0.710 \pm 0.033$  $\boldsymbol{(22\%)}$ & $0.709 \pm 0.036$  $\boldsymbol{(12\%)}$\\
\hline \hline
  \end{tabular}
  \end{center} 
\end{table}

\subsection{Sensitivity to the interpolation method} \label{subsec:Appendix_Interp}

Qualitatively our finding that the combined clipped-and-unclipped analyses improves cosmological parameter constraints holds irrespective of how we choose to interpolate from the DH10 cosmologies onto the $\Omega_{\rm{m}}$-$\sigma_8$ and $\Omega_{\rm{m}}$-$S_8$ grids.  Quantitatively however there is a dependence of the marginalised constraints on these choices, particularly for the highly degenerate $\Omega_{\rm{m}}$ and $\sigma_8$ parameters.  This is to be expected given the level of noise in the predictions and the sparsity with which the predictions are sampled across the parameter space.   We find that the measurement of $S_8$ is the least sensitive to the interpolation scheme adopted, motivating the use of this statistic to highlight the benefit of clipping throughout this paper.   In this Appendix we compare our marginalised $S_8$ constraints for KiDS-450 and the DH10 mock data for four different interpolation methods.

\subsubsection{DH10 constraints} \label{subsubsec:Dh10_constraints_Interp}

\begin{table*}
  \begin{center}
    \caption{The marginalised means and 68\% confidence intervals on $S_8$ for the DH10 data vector from Figure \ref{fig:Interp_Method_DH10} expressed in tabular form. Improvements over the unclipped confidence intervals are detailed to the nearest percentage in bold. } \label{tab:Interp_Method_DH10}
  \begin{tabular}{|ccccc|}
  & \multicolumn{1}{|c|}{RBF$+$Smooth} & \multicolumn{1}{|c|}{RBF} & \multicolumn{1}{|c|}{2D Lin Int} & \multicolumn{1}{|c|}{$\xi_+$-Int}  \\  
\hline \hline
\textbf{Input} $\boldsymbol{S_8=0.740}$ & & & &\\
Unclipped & $0.725 \pm 0.042$  & $0.725 \pm 0.043$  & $0.727 \pm 0.040$  & $0.727 \pm 0.040$ \\
Clipped & $0.710 \pm 0.037$  $\boldsymbol{(11\%)}$ & $0.717 \pm 0.039$  $\boldsymbol{(9\%)}$ & $0.718 \pm 0.039$  $\boldsymbol{(3\%)}$ & $0.724 \pm 0.041$ \\
Combined & $0.710 \pm 0.033$  $\boldsymbol{(22\%)}$ & $0.713 \pm 0.034$  $\boldsymbol{(21\%)}$ & $0.716 \pm 0.033$  $\boldsymbol{(17\%)}$ & $0.726 \pm 0.035$  $\boldsymbol{(12\%)}$\\
\hline \hline
  \end{tabular}
  \end{center} 
\end{table*}

\medskip

The first method we consider for interpolating the likelihoods from the DH10 simulations, is the interpolation with radial basis functions (RBFs), smoothing the contours as described in Section \ref{sec:results}. Secondly, we have the RBF interpolation with no contour smoothing. Thirdly, we have simple 2D linear interpolation. We also consider the results of interpolating the clipped and unclipped DH10 $\xi_+$ statistics, for each $\theta$ bin individually, rather than the likelihoods, onto the $\Omega_{\rm{m}}$-$S_8$ plane. We use the smoothed-RBF method when interpolating the correlation functions in this comparison. By comparing the theoretical $\xi_+^{\rm{unclip}}$ with those extrapolated outside of the range of the DH10 cosmologies, we find that the extrapolation of the correlation functions is inaccurate. Thus we impose a prior which sets the likelihoods calculated from the extrapolated clipped and unclipped $\xi_+$ to zero. Since we find good agreement between theory and the mocks when we extrapolate the unclipped likelihoods instead of the unclipped correlation functions, we do not impose this prior when performing the likelihood-interpolations. Indeed, we find it does not change our results significantly when it is imposed.

\medskip

\begin{figure}
\begin{center}
\includegraphics[width=0.5\textwidth]{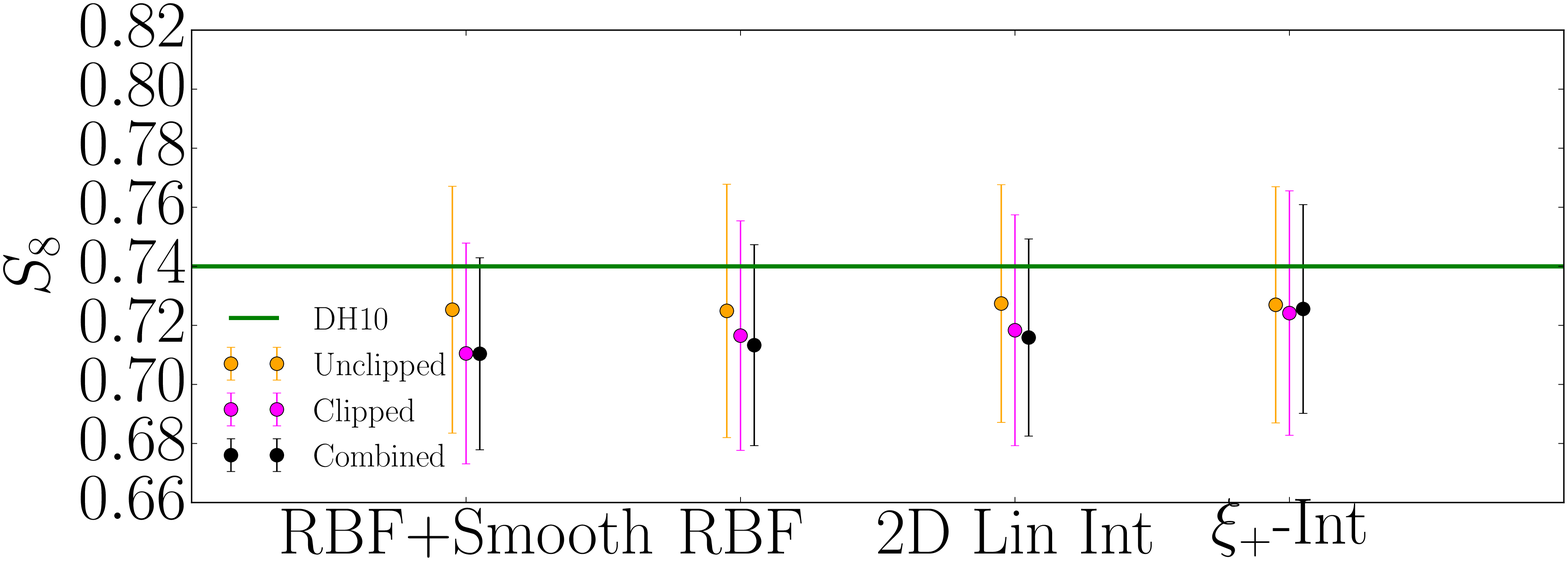}
\caption{The marginalised means and 68\% confidence intervals on $S_8$ from the $\Omega_{\rm{m}}$-$S_8$ plane for the DH10 fiducial cosmology data vector, via different intepolation methods listed on the horizontal axis. From the left-hand side, the first three methods are likelihood-interpolations. ``RBF$+$Smooth" refers to the likelihood-interpolation with radial basis functions and contour smoothing. ``RBF" refers to this interpolation with no smoothing, and ``2D Lin Int" designates simple 2D linear interpolation. ``$\xi_+$-Int" refers to interpolating the clipped and unclipped shear correlation functions, instead of the likelihoods, again with the smoothed-RBF method. The input $S_8$ is designated by the horizontal green line. The corrections for the finite box size and cosmological bias have been applied to the $\xi_+^{\rm{clip}}$ predictions from DH10. The $\xi_+^{\rm{unclip}}$ predictions come from equation \ref{eqn:xi+-theory} and are calculated using {\sc{Nicaea}}.} \label{fig:Interp_Method_DH10}
\end{center}
\end{figure}

Figure \ref{fig:Interp_Method_DH10} and Table \ref{tab:Interp_Method_DH10} present a comparison of the marginalised means and 68\% confidence intervals on $S_8$ for the DH10 dataset. Featured, are the three likelihood-interpolations methods and one $\xi_+$-interpolation method. Clearly all of the marginalised constraints from the different ways of interpolating are consistent with one another, and with the true cosmological parameters to $< 1\sigma$. We see that the combined analysis invariably is an improvement upon the unclipped, with 68\% confidence intervals that are between 12\% and 22\% tighter. The combined analysis also yields improvements on $\Omega_{\rm{m}}$ and $\sigma_8$, of 28\% and 24\% respectively, in the standard analysis with the DH10 dataset presented in Section \ref{sec:results}. Though the greater sensitivity of these results to the interpolation scheme means that we ascribe more confidence in our measurement of $S_8$. 

\medskip

\medskip

\subsubsection{KiDS-450 constraints} \label{subsubsec:K450_constraints_Interp}

The marginalised constraints on the KiDS-450 data fluctuate more than those on the DH10 data vector across the different interpolation schemes (described in Section \ref{subsubsec:Dh10_constraints_Interp}). This is to be expected given that KiDS-450 features extra sources of noise, such as galaxy shape measurement, baryonic effects and  $n(z)$ uncertainties which have not been accounted for in this proof-of-concept analysis. These may engender spurious peaks in the interpolated likelihoods which bias some interpolation methods more than others. What is more, the nuisance cosmological parameters $\Omega_{\rm{b}},n_{\rm{s}}$ and $h$ are almost certainly mismatched between the data and the predictions. In principle this could affect the $\xi_+^{\rm{clip}}$ differently than the $\xi_+^{\rm{unclip}}$ predictions.

\medskip

\begin{figure}
\begin{center}
\includegraphics[width=0.5\textwidth]{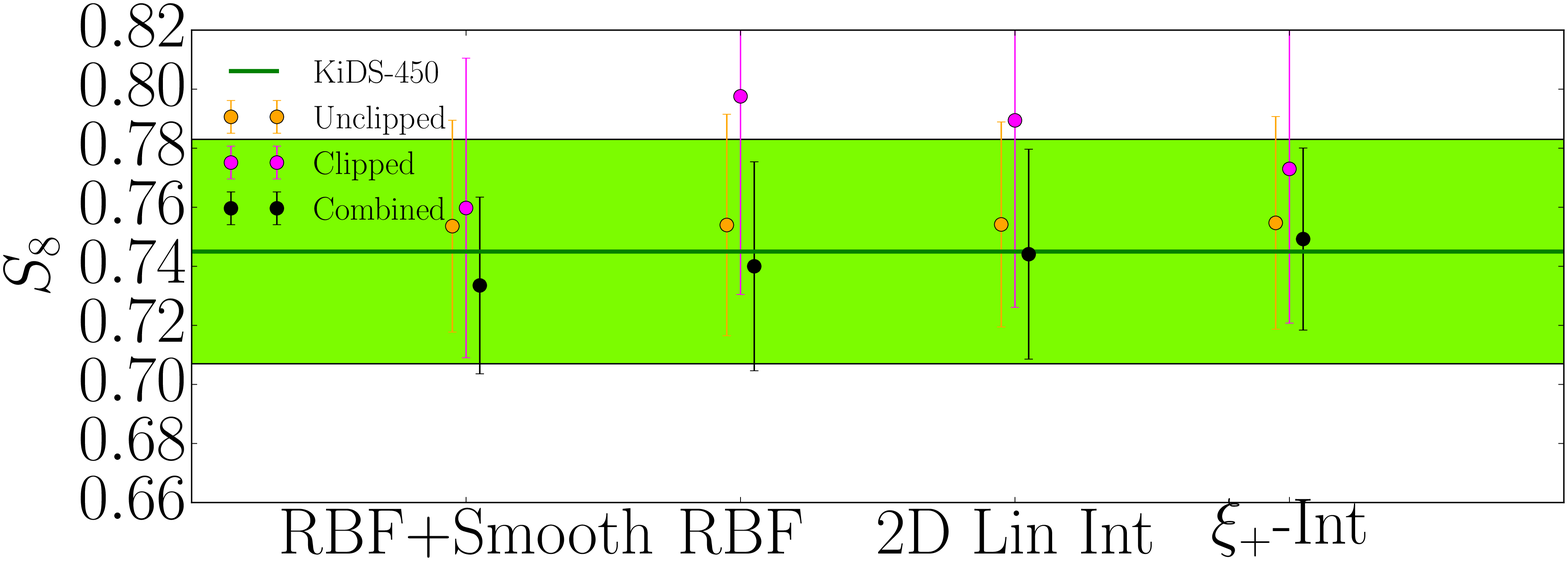}
 \caption{The same as Figure \ref{fig:Interp_Method_DH10} but for the KiDS-450 data. The light-green region corresponds to the 68\% confidence region from the H17 cosmic shear analysis.} \label{fig:Interp_Method_K450}
\end{center}
\end{figure}

We find that the improvements over the unclipped found in the combined marginalised $S_8$ constraints, displayed visually in Figure \ref{fig:Interp_Method_K450} and numerically in Table \ref{fig:Interp_Method_K450}, are consistent for the interpolation schemes which incorporate smoothing, ``RBF$+$Smooth" and ``$\xi_+$-Int", between 14\% and 17\%. The interpolation schemes without smoothing however, ``RBF" and ``2D Lin Int", yield little to no improvement in the combined constraints. This is because the interpolated clipped and combined likelihoods for the KiDS-450 dataset are reasonably noisy, and the methods without smoothing are more strongly affected by this. The smoothing reduces the impact of spurious noise spikes in the likelihoods biasing the parameter constraints. Thus we regard the constraints obtained with these interpolations as more accurate, and maintain that the improvement found by combining the clipped and unclipped analyses is around the 17\% level for the KiDS-450 data. 

\medskip

\medskip

\begin{table*}
  \begin{center}
    \caption{The same as Table \ref{tab:Interp_Method_DH10} but for the KiDS-450 data. Improvements over the unclipped confidence intervals are detailed to the nearest percentage in bold. We remind the reader that the results of this work are not directly comparable to the H17 result, owing to the differences in the analyses discussed in the text.} \label{tab:Interp_Method_K450}
  \begin{tabular}{|ccccc|}
  & \multicolumn{1}{|c|}{RBF$+$Smooth} & \multicolumn{1}{|c|}{RBF} & \multicolumn{1}{|c|}{2D Lin Int} & \multicolumn{1}{|c|}{$\xi_+$-Int}  \\
\hline \hline
\textbf{H17} $\boldsymbol{S_8=0.745^{+0.038}_{-0.038}}$ & & & &\\
Unclipped & $0.754 \pm 0.036$  & $0.754 \pm 0.038$  & $0.754 \pm 0.035$  & $0.755 \pm 0.036$ \\
Clipped & $0.760 \pm 0.051$  & $0.798 \pm 0.067$  & $0.789 \pm 0.063$  & $0.773 \pm 0.052$ \\
Combined & $0.734 \pm 0.030$  $\boldsymbol{(17\%)}$ & $0.740 \pm 0.035$  $\boldsymbol{(6\%)}$ & $0.744 \pm 0.036$  & $0.749 \pm 0.031$  $\boldsymbol{(14\%)}$\\
\hline \hline
  \end{tabular}
  \end{center} 
\end{table*}

\medskip

\section{KiDS-450 mass maps} \label{sec:Appendix_MassMaps}

\begin{figure*}
\centering
\includegraphics[width=\textwidth]{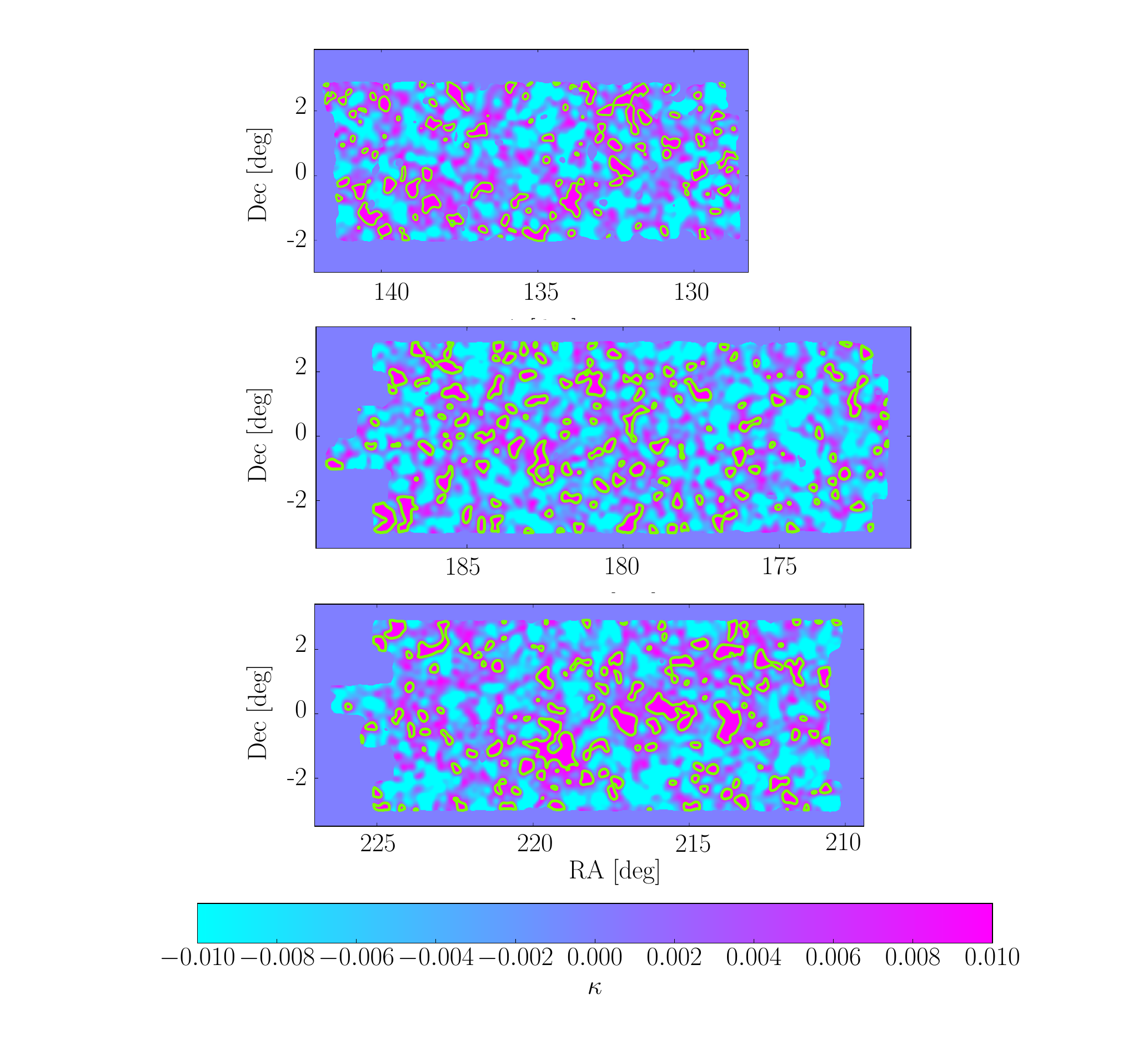}  
\caption{Maps of the convergence, $\kappa$, for the three KiDS-450 North patches, G9 (\textit{upper}), G12 (\textit{middle}) and G15 (\textit{lower}). The maps have been smoothed with a Gaussian filter with width $\sigma_{\rm{s}}=6.6$ arcmin. Unobserved/masked regions are given zero convergence, as is described in the text. The regions highlighted by the green contours, exceed the clipping threshold, $\kappa^c=0.010$, and are therefore clipped in our pipeline. The clipped regions make up $12 \pm 1\%$ of the effective area of the five KiDS-450 patches.} \label{fig:KiDS450_MassMaps} 
\end{figure*}

\begin{figure*}
\centering
\includegraphics[width=\textwidth]{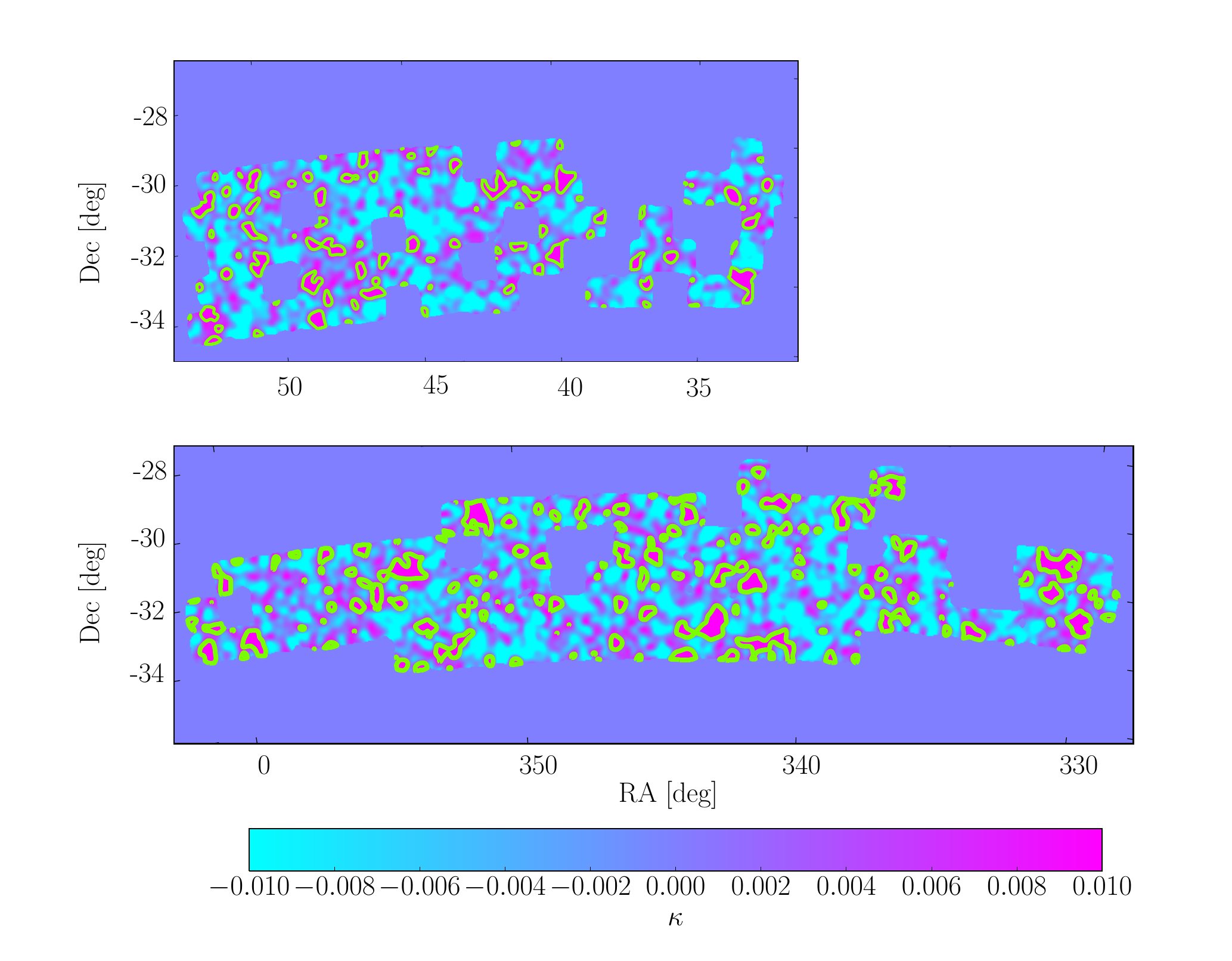}  
\caption{The same as Figure \ref{fig:KiDS450_MassMaps} but for the two KiDS-450 South patches, GS (\textit{upper}), and G23 (\textit{lower}).} \label{fig:KiDS450_MassMaps2} 
\end{figure*}

In Figures \ref{fig:KiDS450_MassMaps} and \ref{fig:KiDS450_MassMaps2} we present convergence maps for the North and South KiDS-450 patches respectively. In producing these maps, we follow the mass reconstruction methodology of \cite{KSB} as detailed in Section \ref{subsec:Mass_Recon}. The maps are smoothed with a Gaussian filter with width $\sigma_{\rm{s}}=6.6$ arcmin, and the regions exceeding the clipping threschold $\kappa^c=0.010$ are highlighted with the green contours. We follow \cite{vanWaerbeke_et_al_2013} and set the convergence to zero in regions where more than 50\% of the Gaussian smoothing window is centred on masked pixels.

\medskip

\end{document}